\documentclass[12pt,oneside]{article}

\usepackage{amsmath,amssymb,amsfonts,mathrsfs,undertilde,latexsym}
\usepackage{graphicx}
\usepackage{enumerate,bm}
\usepackage{dsfont,comment}
\usepackage{float,multirow}
\usepackage{mwe}
\usepackage{chngcntr}
\usepackage{epstopdf}
\usepackage{cleveref}
\counterwithout{figure}{section}
\counterwithout{table}{section}

\usepackage[authoryear]{natbib}
\usepackage{pdflscape}
\usepackage{rotating}

\usepackage{multibib}
\newcites{append}{Appendix References}

\newcommand{\GG}[1]{}

\def\1{\mathbf{1}}

\def\<{\langle}

\def\>{\rangle}

\newtheorem{lem}{Lemma}
\newtheorem{pro}{Proposition}

\newtheorem{cor}{Corollary}

\def\var{\mathrm{var}}
\def\cov{\mathrm{cov}}

\def\T{{ \mathrm{\scriptscriptstyle T} }}

\begin{document}

\begin{titlepage}

\begin{center}
{\bf Improved Estimation of Average Treatment Effects on the Treated: Local Efficiency, Double Robustness, and Beyond}

\vspace{.15in} Heng Shu \& Zhiqiang Tan\footnotemark[1]

\vspace{.05in}
\end{center}

\footnotetext[1]{Heng Shu is with JPMorgan Chase, New York, NY 10017 and Zhiqiang Tan is Professor,
Department of Statistics, Rutgers University, Piscataway, NJ 08854 (E-mail: ztan@stat.rutgers.edu).
An earlier version of this work was completed as part of the PhD thesis of Heng Shu at Rutgers University.
}

\vspace{-.3in}
\paragraph{Abstract.} Estimation of average treatment effects on the treated (ATT) is an important topic of causal inference in econometrics and statistics. This problem seems to be often treated as a simple modification or extension of that of estimating overall average treatment effects (ATE). However, the propensity score is no longer ancillary for estimation of ATT, in contrast with estimation of ATE. \
In this article, we review semiparametric theory for estimation of ATT and the use of efficient influence functions
to derive augmented inverse probability weighted (AIPW) estimators that are locally efficient and doubly robust. Moreover, we
discuss improved estimation over AIPW by developing calibrated regression and likelihood estimators that are not only locally efficient and doubly robust, but also intrinsically efficient in achieving smaller variances than AIPW estimators when a propensity score model is correctly specified but an outcome regression model may be misspecified. Finally, we present two simulation studies and an econometric application to demonstrate the advantage of the proposed methods when compared with existing methods.

\paragraph{Key words and phrases.} Average treatment effect; Average treatment effect on the treated; Causal inference; Double robustness; Inverse probability weighting; Intrinsic efficiency; Local efficiency; Semiparametric estimation.

\end{titlepage}

\section{Introduction} \label{intro}

A central problem in various social and behavioral studies is to evaluate average effects of treatments and actions {\it ceteris paribus} (with all other things being equal).
Such problems can be addressed by introducing potential outcomes that would be observed for each subject under
different treatments (\citealt{Neyman1923, Rubin1974}).
Average causal effects are then defined as statistical comparisons (e.g., mean differences) of potential outcomes over a population or subpopulation.
Two causal parameters commonly studied are the average treatment effect (ATE) and the average treatment effect on the treated (ATT). The ATE is defined as the mean difference of two potential outcomes under the active treatment and the control over the entire population, whereas ATT is defined as the mean difference of two potential outcomes over the subpopulation of individuals who received the active treatment.
As argued by \citet{Heckman1985} and \citet{Heckman1997} in the context of evaluating training programs, the ATT answers the question ``How much did persons participating in the programme
benefit compared to what they would have experienced without participating in the programme?" The ATT is relevant in
making forecasts when the same selection rule operates in the future as has operated in the past.

Drawing inferences about ATE and ATT is challenging because, in reality, all but one potential outcome are missing for each subject.
Nevertheless, under unconfoundedness (i.e., exogeneity) and overlap assumptions, the ATE and ATT are point identifiable from  observed data (e.g., \citealt{Imbens2004}).
There is an extensive collection of theory and methods developed for statistical estimation of ATE and ATT under exogeneity.
Let $Y$ be an observed outcome,
$T$ a treatment indicator, and $X$ a vector of covariates.
Semiparametric efficiency bounds for estimation of both ATE and ATT are obtained by \citet{Hahn1998}, and can be seen as special cases of semiparametric theory
in \citet{Robins1994} and \citet{Chen2008} for moment restriction models with missing data.
Asymptotically globally efficient estimators for ATE and ATT are studied by \citet{Hahn1998}, \citet{Hirano2003}, and \citet{Chen2008} among others, using nonparametric series/sieve estimation
on the propensity score, $\pi(X)= P(T=1|X)$, or the outcome regression function, $m_t(X) = E(Y|T=t,X)$, or both.
But various smoothness conditions are assumed for such methods and can sometimes be problematic with a high-dimensional covariate vector $X$ (\citealt{Robins1997}).

Alternatively, various methods are developed by using parametric working models on the propensity score $\pi(X)$ or the outcome regression function $m_t(X)$ or both, to
achieve desirable properties such as local efficiency, double robustness, and beyond.
This line of research has been well pursued for estimation of ATE (e.g., \citealt{Robins1994, Tan2006, Tan2010, Cao2009}).
See also \citet{KS2007} and its discussion.
For an estimator of ATE, double robustness means that the estimator remains consistent if either the propensity score model or the outcome regression model is correctly specified.
Local efficiency means that if both the propensity score model and the outcome regression model are correctly specified, then the estimator achieves
the semiparametric efficiency bound, which is the same whether the propensity score is known, paramtrically modeled, or completely unknown due to the ancillarity of the propensity score
for estimation of ATE (\citealt{Hahn1998}).
To our knowledge, there seems to be limited work explicitly dealing with locally efficient and doubly robust estimation of ATT (e.g., \citealt{Graham2015, Zhao2015}).

There are two possible reasons why ATT estimation has been studied much less extensively than ATE estimation.
On one hand, ATT can often be estimated by a simple modification or extension of estimators of ATE. \
On the other hand, semiparametric theory for estimation of ATT is complicated by the fact that the propensity score is no longer ancillary (\citealt{Hahn1998}).
The purpose of this article is twofold:
(i) we review semiparametric theory for ATT estimation and the use of efficient influence functions to derive augmented inverse probability weighted (AIPW) estimators of ATT,
and (ii)
we discuss the extension of related techniques for improved estimation of ATE (\citealt{Tan2006, Tan2010, Cao2009})
to develop calibrated estimators of ATT that achieve desirable properties beyond local efficiency and double robustness.
Demonstration of these ideas can also facilitate their applications to other missing-data problems,
for example, data combination discussed in \cite{Graham2015}.

There are several interesting phenomena clarified from our work, all different from familiar results for estimation of ATE. \
First, there are two AIPW estimators achieving local efficiency of different types.
If the propensity score and outcome regression models are correctly specified, the first estimator achieves the semiparametric efficiency bound, $V_{\mbox{\scriptsize NP}}$, calculated when the propensity score is unknown,
whereas the second estimator achieves the semiparametric efficiency bound, $V_{\mbox{\scriptsize SP}} \, (\le V_{\mbox{\scriptsize NP}})$, calculated under the parametric propensity score model used.
These two estimators are then referred to as locally nonparametric or, respectively, semiparametric efficient.

Second, the locally nonparametric efficient estimator AIPW of ATT is doubly robust, but the locally semiparametric efficient AIPW estimator is generally not.
Therefore, it is the efficient influence function calculated under the nonparametric model (i.e., when the propensity score as well as the outcome regression function is unknown) that leads to doubly robust estimation.
Incidentally, it can be shown that the doubly robust estimators of ATT in \citet{Graham2015} and \citet{Zhao2015} are also locally nonparametric efficient.

Third, due to the discrepancy between the locally nonparametric and semiparametric AIPW estimators, a direct application of the techniques in \citet{Tan2006, Tan2010} and \citet{Cao2009}
would fail to yield an improved estimator of ATT that is not only doubly robust and locally nonparametric efficient, but also intrinsically efficient
in achieving smaller variances than AIPW estimators when the propensity score model is correctly specified but the outcome regression model may be misspecified.
We show that such improved estimators can still be developed by introducing a simple idea, namely, working with
an augmented propensity score model which includes the fitted outcome regression functions as additional regressors.

To illustrate the advantage of the improved estimators, we present two simulation studies and an econometric application related to \citet{LaLonde1986}
and subsequent analyses (e.g., \citealt{Dehejia2002, Smith2005a}). In contrast with these previous works, we compare the performance of different methods by
examining not only the effect or bias estimates (where the experimental treatment or, respectively, control group is compared with a non-experimental comparison group),
but also how well the differences between the effect and bias estimates agree with the benchmark estimate (where
the experimental control and treatment groups are compared).
The latter comparisons are relevant even if the non-experimental group might inherently differ from the cohort on which the experiment was conducted.

\section{Setup} \label{setup}

To introduce the setup, suppose that a simple random sample of $n$ subjects is available from a population under study. The observed data consist
of independent and identically distributed observations $\{(Y_i, T_i, X_i): i=1,\ldots,n\}$ of $(Y,T,X)$, where $Y$ is an outcome variable,
$T$ is a dichotomous treatment variable ($T= 1$ if treated or $T=0$ otherwise),
and $X$ is a vector of measured covariates.
In the potential outcomes framework for causal inference (\citealt{Neyman1923, Rubin1974}), two potential outcomes
$(Y^0, Y^1)$ are defined to indicate what would be the response under treatment 0 or 1 respectively. By
consistency, the observed outcome $Y$ is assumed to be either $Y^0$ or $Y^1$, depending on whether $T= 0$ or $T= 1$.
Two causal parameters commonly of interest are the average treatment effect (ATE), defined as
$E(Y^1- Y^0) = \mu^1 - \mu^0$ with $\mu^t= E(Y^t)$, and the average treatment effect on the treated (ATT),
defined as $E(Y^1 - Y^0 |T=1) = \nu^1 - \nu^0$ with $\nu^t = E(Y^t | T=1)$.
In this article, we are concerned with estimation of the ATT.
See, for example, \citet{Imbens2004} for a review and \citet{Tan2006, Tan2010} for related works on estimation of ATE.

While the parameter $\nu^1$ is directly identifiable as $E(TY)/E(T)$, a fundamental difficulty in identification of $\nu^0$ is
that $Y^0$ is missing for treated subjects with $T=1$. Nevertheless,
it is known (e.g., \citealt{Imbens2004}) that the $\nu^0$ and hence ATT are identifiable from observed data under the two assumptions:
\begin{itemize} 
\item[(A1)] Unconfoundedness for controls: $T \perp Y^0 | X$, i.e., $T$ and $Y^0$ are conditionally independent given $X$;

\item[(A2)] Weak overlap: $0 \le P(T=1|X=x) < 1$ for all $x$.
\end{itemize}
%
%
%
Assumption (A2) allows that $P(T=1|X=x)$ is 0 for some values $x$, i.e.,
subjects with certain covariate values will always take treatment 0.

By the fact that $\nu^1=E(TY)/E(T)$, a consistent, nonparametric estimator of $\nu^1$ is $\hat\nu^1_{\text{\scriptsize NP}} = n_1^{-1} \sum_{i=1}^n T_iY_i$,
where $n_1=\sum_{i=1}^n T_i$ and $n_0=n-n_1$ are the sizes of treated and untreated groups respectively in the sample.
However, modeling (or dimension-reduction) assumptions, in addition to (A1)--(A2), are, in general, needed to
obtain consistent estimation of $\nu^0$ and ATT from finite samples with high-dimensional $X$.
There are broadly two modelling approaches as follows (e.g., \citealt{Tan2007}).

One approach is to build a (parametric) regression model for the outcome regression (OR) function, $m_t(X)=E(Y|T=t,X)$:
\begin{align}
E(Y|T=t,X)=m_t(X; \alpha_t)=\Psi\{\alpha_t^\T g_t(X)\}, \quad t=0 \mbox{ or } 1, \label{OR}
\end{align}
where $\Psi(\cdot)$ is an inverse link function, $g_t(X)$ is a vector of known functions of $X$ {\it including 1}, and $\alpha_t$ is a vector of unknown parameters.
For $t=0$ or 1, let $\hat\alpha_t$ be the maximum quasi-likelihood estimate of $\alpha_t$, and let $\hat m_t(X) = m_t(X; \hat\alpha_t)$.
If model (\ref{OR}) is correctly specified for $t=0$ or 1, then a consistent estimator for $\nu^t$ is
$\hat\nu^t_{\text{\scriptsize OR}} = n_1^{-1} \sum_{i=1}^n T_i \,\hat m_t(X_i)$. The ATT can be estimated by $\hat\nu^1_{\text{\scriptsize OR}} - \hat\nu^0_{\text{\scriptsize OR}} $.
In the special case where $\Psi(\cdot)$ is the identity link and parallel regression functions are assumed for the two treatment groups, i.e.,
$E(Y|T=t,X) = \alpha_{1,t} + \alpha^\T_{(1)} g_{(1)}(X)$ with $g_{(1)}(X)$ excluding 1, the ATT can be directly estimated as $\alpha_{1,1}- \alpha_{1,0}$.

An alternative approach is to build a (parametric) regression model for the propensity score (PS) (\citealt{Rosenbaum1983}),
$\pi(X) = P( T=1 | X) $:
\begin{align}
P(T=1|X)=\pi(X;\gamma)=\Pi\{\gamma^\T  f(X)\}, \label{PS}
\end{align}
where $\Pi(\cdot)$ is an inverse link function, $f(x)$ is a vector of known functions {\it including} 1, and $\gamma$ is a vector of unknown parameters.
The score function for $\gamma$ is
\begin{align*}
s_{\gamma}(T,X)
=  \left\{\frac{T}{ \pi(X; \gamma) } -  \frac{1-T}{1-\pi(X; \gamma)} \right\} \frac{\partial\pi(X;\gamma)}{\partial\gamma} .
\end{align*}
Typically, logistic regression is used: $\pi(X;\gamma)= [1+\exp\{-\gamma^\T  f(X)\}]^{-1}$, and
the score function is
$s_\gamma(T,X) = \{T-\pi(X;\gamma)\} f(X)$.
Let $\hat\gamma$ be the maximum likelihood estimator (MLE) of $\gamma$ and $\hat\pi(X)= \pi(X; \hat\gamma)$, satisfying the score equation $\tilde E\{ S_\gamma(T,X)\}=0$, which for logistic regression reduces to
\begin{align}
\tilde E \left[ \{T-\pi(X; \gamma)\} f(X) \right] = 0 , \label{score-eq}
\end{align}
where $\tilde E(\cdot)$ denotes a sample average, for example, $\tilde E(T) = n_1/n$. Then $\nu^0$ and ATT can be estimated by matching, stratification, or weighting on the fitted propensity
score $\hat\pi(X)$ (e.g., \citealt{Imbens2004}). We focus on inverse probability weighting (IPW), which is
central to rigorous theory of statistical estimation in missing-data problems (e.g., \citealt{Tsiatis2006}).
Two standard IPW estimators for $\nu^0$ are (e.g., \citealt{McCaffrey2004, Abadie2005})
\begin{align*}
 \hat{\nu}^0_{\text{\scriptsize IPW}}(\hat\pi) & =\tilde{E}\left\{\frac{(1-T)\hat{\pi}(X)Y}{1-\hat{\pi}(X)}\right\}\Big/ \tilde{E}(T), \\
 \hat{\nu}^0_{\text{\scriptsize IPW,ratio}}(\hat\pi) & =\tilde{E}\left\{\frac{(1-T)\hat{\pi}(X)Y}{1-\hat{\pi}(X)}\right\}\Big/ \tilde{E}\left\{\frac{(1-T)\hat{\pi}(X)}{1-\hat{\pi}(X)}\right\} .
 \end{align*}
The estimator of ATT based on $\hat{\nu}^0_{\text{\scriptsize IPW}}(\hat\pi)$ and $\hat\nu^1_{\text{\scriptsize NP}}$ is then
\begin{align*}
\hat\nu^1_{\text{\scriptsize NP}} - \hat{\nu}^0_{\text{\scriptsize IPW}}(\hat\pi) =\tilde E \left\{ \frac{T-\hat\pi(X)}{1-\hat\pi(X)} Y \right\} \Big/ \tilde E(T) .
\end{align*}
If model (\ref{PS}) is correctly specified, then the IPW estimators are consistent.
However, if model (\ref{PS}) is misspecified or even mildly so, these estimators can perform poorly,
especially due to the instability of inverse weighting to fitted propensity scores $\hat\pi(X_i)$ near 1 for some untreated subjects (e.g., \citealt{KS2007}).

\section{Semiparametric theory and AIPW estimation} \label{semiparametric}

For consistency, the estimator $\nu^0_{\text{\scriptsize OR}}$ requires a correctly specified OR model (\ref{OR}) for $t=0$, whereas
$\nu^0_{\text{\scriptsize IPW}}$ and $\nu^0_{\text{\scriptsize IPW,ratio}}$ require a correctly specified PS model (\ref{PS}).
Alternatively, it is desirable to develop estimators of $\nu^0$ and ATT using both OR model (\ref{OR}) and PS model (\ref{PS}) to gain efficiency and robustness,
similarly as in estimation of ATE.\
We review semiparametric theory and derive locally efficient and doubly robust estimators of $\nu^0$ and ATT in the form of augmented IPW (AIPW).
Understanding of these estimators facilitates our development of improved estimators in Section~\ref{improved}.

First, Proposition \ref{prop1} restates the efficient influence functions for estimation of $\nu^0$ under three different settings, based on \citet{Hahn1998} and \citet{Chen2008}.

\begin{pro}[\citealt{Hahn1998,Chen2008}] \label{prop1}
Let $q = E(T)$ and define
$$
\tau^0(\pi, h) = \frac{1-T}{1-\pi( X )}\pi( X )Y- \left\{ \frac{1-T}{1-\pi( X )}-1 \right\} h( X ) .
$$
The efficient influence function for estimation of $\nu^0$ is as follows, depending on assumptions on the propensity score.
\begin{enumerate}
\item[(i)]The efficient influence function is
$$
\varphi^0_{\text{\scriptsize NP}}(Y,T, X )=\left\{ \tau^0(\pi,m_0) - T \nu^0 \right\} \Big/q .
$$
\item[(ii)]If the propensity score $\pi( X )$ is known, then the efficient influence function is
\begin{align*}
\varphi^0_{\text{\scriptsize SP*}}(Y,T, X )
& = \left\{ \tau^0(\pi, \pi m_0) - \pi(X) \nu^0 \right\} \Big/ q\\
& =\varphi^0_{\text{\scriptsize NP}}(Y,T, X ) -\{T-\pi(X)\}\frac{m_0(X)-\nu^0}{q} .
\end{align*}
\item[(iii)]If the propensity score $\pi( X )$ is unknown but assumed to belong to a correctly specified parametric family $\pi(X;\gamma)$, then the efficient influence function is
$$
\varphi^0_{\text{\scriptsize SP}}(Y,T, X )=\varphi^0_{\text{\scriptsize SP*}}(Y,T, X )+\mbox{Proj}\left[ \{ T-\pi( X ) \}\frac{m_0( X )-\nu^0}{q}\Big| s_\gamma(T, X )\right] ,
$$
\end{enumerate}
where for two random vectors $Z_1$ and $Z_2$, $\mbox{Proj}(Z_2 | Z_1) = \cov(Z_2,Z_1) \var^{-1}(Z_1) Z_1$, i.e., the linear projection of $Z_2$ onto $Z_1$.
\end{pro}

As discussed in \citet{Hahn1998} and \citet{Chen2008},  the semiparamtric efficiency bounds satisfy the following order:
$V^0_{\text{\scriptsize NP}} \ge V^0_{\text{\scriptsize SP}} \ge V^0_{\text{\scriptsize SP*}} $,
with strict inequalities holding in general, where
$V^0_{\text{\scriptsize NP}}$, $V^0_{\text{\scriptsize SP}}$, and
$V^0_{\text{\scriptsize SP*}}$ are respectively the variances of $\varphi^0_{\text{\scriptsize NP}}$, $\varphi^0_{\text{\scriptsize SP}}$, and $\varphi^0_{\text{\scriptsize SP*}}$.
In fact, the efficient influence functions $\varphi^0_{\text{\scriptsize NP}}$, $\varphi^0_{\text{\scriptsize SP}}$, and $\varphi^0_{\text{\scriptsize SP*}}$ can all be expressed as the following functional
with suitable choices of $h(X)$:
\begin{align}
\varphi^0_h(Y,T, X) = \left\{ \tau^0(\pi, h) - T \nu^0 \right\} \Big/q .  \label{AIPW}
\end{align}
The minimum variance of $\varphi^0_h(Y,T, X)$ over possible choices of $h(X)$ is exactly $V^0_{\text{\scriptsize SP*}}$, corresponding to the choice $h(X) = \pi(X) m_0(X) + \{1-\pi(X)\} \nu^0$.

This ordering of efficiency bounds agrees with the usual comparison that the efficiency bound under a more restrictive model is no greater than under a less restrictive model.
But this relationship differs from the result that
the semiparametric efficiency bounds for estimation of $\mu^t=E(Y^t)$ are the same whether under the nonparametric model for $\pi(X)$, or under a parametric model for $\pi(X)$, or with
exact knowledge of $\pi(X)$.
Conceptually, these differences reflect the fact the propensity score is ancillary for estimation of ATE, but not ancillary for estimation of ATT (\citealt{Hahn1998}).

We now derive two estimators of $\nu^0$ that depend on both fitted outcome regression function $\hat m_0(X)$ and fitted propensity score $\hat \pi(X)$, by directly taking
the efficient influence functions in Proposition \ref{prop1} as estimating functions, with $\hat m_0(X)$ and $\hat \pi(X)$ in place of the unknown truth $m_0(X)$ and $\pi(X)$.
Proposition \ref{prop2} shows that only one estimator is doubly robust, whereas both estimators possess local efficiency but of different types according to
the semiparametric efficiency bounds achieved when the OR and PS models are correctly specified.
For clarity, the semiparametric efficiency bound $V^0_{\text{\scriptsize NP}}$ under the nonparametric model is hereafter referred to as the nonparametric efficiency bound.
See, for example, \citet{Newey1990}, \citet{Robins2001}, and \citet{Tsiatis2006} for general discussions on local efficiency and double robustness.

\vspace{-.05in}
\begin{pro} \label{prop2}
Under suitable regularity conditions (see Appendix~\ref{sec:tech-details} in the Supplementary Material), the following results hold.\vspace{-.1in}
\begin{enumerate}\addtolength{\itemsep}{-.1in}
\item[(i)] Define an estimator of $\nu^0$ as
\begin{align*}
\hat{\nu}^{0}_{\text{\scriptsize NP}} (\hat{\pi},\hat{m}_0)
&=\tilde{E} \left\{ \tau^0 (\hat{\pi},\hat{m}_0) \right\} \Big/\tilde{E}(T).
\end{align*}
Then $\hat{\nu}^{0}_{\text{\scriptsize NP}} (\hat{\pi},\hat{m}_0)$ is locally nonparametric efficient: it achieves the nonparametric efficiency bound $V^0_{\text{\scriptsize NP}}$
when both model (\ref{OR}) for $t=0$ and model (\ref{PS}) are correctly specified.
Moreover, $\hat{\nu}^{0}_{\text{\scriptsize NP}} (\hat{\pi},\hat{m}_0)$ is doubly robust: it remains consistent when either model (\ref{OR}) for $t=0$ or model (\ref{PS}) is
correctly specified.

\item[(ii)] Define an estimator of $\nu^0$ as
\begin{align*}
\hat{\nu}^{0}_{\text{\scriptsize SP}} (\hat{\pi},\hat{m}_0)  &=\tilde{E} \left\{ \tau^0 (\hat{\pi}, \hat\pi \hat{m}_0) \right\} \Big/\tilde{E}\{\hat\pi(X)\} .
\end{align*}
For logistic PS model (\ref{PS}), $\hat{\nu}^{0}_{\text{\scriptsize SP}} (\hat{\pi},\hat{m}_0)$ can be equivalently expressed as
\begin{align*}
\hat{\nu}^{0}_{\text{\scriptsize SP}} (\hat{\pi},\hat{m}_0)  & =\tilde{E} \left\{ \tau^0 (\hat{\pi}, \hat\pi \hat{m}_0) \right\} \Big/\tilde{E}(T) ,
\end{align*}
because $\tilde E(T) = \tilde E\{\hat\pi(X)\}$ by Eq.~(\ref{score-eq}) with $f(X)$ including 1.
Then $\hat{\nu}^{0}_{\text{\scriptsize SP}} (\hat{\pi},\hat{m}_0)$ is locally semiparametric efficient:  it achieves the semiparametric efficiency bound $V^0_{\text{\scriptsize SP}}$
when both model (\ref{OR}) for $t=0$ and model (\ref{PS}) are correctly specified. But $\hat{\nu}^{0}_{\text{\scriptsize SP}} (\hat{\pi},\hat{m}_0)$ is, generally, not doubly robust.
\end{enumerate}
\end{pro}

The estimators $\hat{\nu}^{0}_{\text{\scriptsize NP}} (\hat{\pi},\hat{m}_0)$  and, for a logistic PS model, $\hat{\nu}^{0}_{\text{\scriptsize SP}} (\hat{\pi},\hat{m}_0)$
belong to the following class of AIPW estimators, with the choice $h=\hat m_0$ or $h=\hat \pi \hat m_0$ respectively:
\begin{align*}
\hat{\nu}^{0}( \hat\pi, h) & = \tilde E \left\{ \tau^0 (\hat{\pi}, h ) \right\} \Big/\tilde{E}(T) \\
&= \tilde E \left[ \frac{1-T}{1-\hat\pi( X )} \hat\pi( X )Y- \left\{ \frac{1-T}{1-\hat\pi( X )}-1 \right\} h(X) \right] \Big/ \tilde E(T),
\end{align*}
which are defined by directly taking (\ref{AIPW}) as the estimating function with the fitted propensity score $\hat \pi(X)$ in place of the unknown truth $\pi(X)$.
Setting $h(X)\equiv 0$ leads to the simple estimator $\hat{\nu}^0_{\text{\scriptsize IPW}}$.
Although related estimators of $\nu^0$ may be implicit in previous works (e.g., \citealt{Stoczynski-Wooldridge2014}) and the idea of constructing estimators from
influence functions is generally known (e.g., \citealt{Tsiatis2006}), our application of this idea to derive the estimators
$\hat{\nu}^{0}_{\text{\scriptsize NP}} (\hat{\pi},\hat{m}_0)$  and $\hat{\nu}^{0}_{\text{\scriptsize SP}} (\hat{\pi},\hat{m}_0)$ for $\nu^0$
seems new and sheds light on subtle differences between the two estimators as discussed below.
Such differences lead to new challenges to be addressed in our development of improved estimators in Section~\ref{improved}; see the remarks after Proposition~\ref{prop-reg}.

By local semiparametric efficiency, the estimator $\hat{\nu}^{0}_{\text{\scriptsize SP}} (\hat{\pi},\hat{m}_0)$, but not
$\hat{\nu}^{0}_{\text{\scriptsize NP}} (\hat{\pi},\hat{m}_0)$, achieves the minimum asymptotic variance among all regular estimators
under PS model (\ref{PS}), including AIPW estimators
$\hat{\nu}^{0}_h( \hat\pi, \hat m_0)$ over possible choices of $h(X)$, when both model (\ref{OR}) for $t=0$  and model (\ref{PS}) are correctly specified.
However, $\hat{\nu}^{0}_{\text{\scriptsize SP}} (\hat{\pi},\hat{m}_0)$ is not doubly robust, and
$\hat{\nu}^{0}_{\text{\scriptsize NP}} (\hat{\pi},\hat{m}_0)$ is doubly robust.
This situation differs from the case where among the class of AIPW estimators of $\mu^0$, the estimator
\begin{align*}
\hat{\mu}^{0}_{\text{\scriptsize AIPW}} = \tilde E \left[ \frac{1-T}{1-\hat\pi( X )} Y- \left\{ \frac{1-T}{1-\hat\pi( X )}-1 \right\} \hat m_0(X) \right],
\end{align*}
is doubly robust, i.e., consistent when either OR model (\ref{OR}) for $t=0$ or PS model (\ref{PS}) is correctly specified,
and locally semiparamtric and nonparametric efficient, i.e., achieving the minimum asymptotic variance among all regular estimators under parametric PS model (\ref{PS}) and, respectively, under the nonparametric model
when model (\ref{OR}) for $t=0$ and model (\ref{PS}) are correctly specified.
As discussed after Proposition \ref{prop1}, the semiparametric efficient bound for estimation of $\mu^0$ under a parametric PS model coincides with that under the nonparametric model.

Next, we restate the efficient influence functions in Proposition \ref{prop3}
for estimation of $\nu^1$, based on \citet{Hahn1998} and \citet{Chen2008}. Similarly as for estimation of $\nu^0$, the efficiency bounds satisfy
$V^1_{\text{\scriptsize NP}} \ge V^1_{\text{\scriptsize SP}} \ge V^1_{\text{\scriptsize SP*}} $,
with strict inequalities in general.

\vspace{-.05in}
\begin{pro}[\citealt{Hahn1998,Chen2008}] \label{prop3}
The efficient influence function for estimation of $\nu^1$ is as follows, depending on assumptions on the propensity score.\vspace{-.1in}
\begin{enumerate}\addtolength{\itemsep}{-.1in}
\item[(i)]The efficient influence function is
$$
\varphi^1_{\text{\scriptsize NP}}(Y,T, X )= \left( TY-T\nu^1 \right) \Big/ q.
$$
\item[(ii)]If the propensity score $\pi( X )$ is known, then the efficient influence function is
\begin{align*}
\varphi^1_{\text{\scriptsize SP*}}(Y,T, X ) & = \left[ TY- \{T-\pi( X )\}m_1( X )- \pi(X) \nu^1 \right] \Big/q \\
& = \varphi^1_{\text{\scriptsize NP}}(Y,T, X ) -\{T-\pi(X)\}\frac{m_1(X)-\nu^1}{q} .
\end{align*}
\item[(iii)]If the propensity score $\pi( X )$ is unknown but assumed to belong to a correctly specified parametric family $\pi(X;\gamma)$, then the efficient influence function is
\begin{align*}
\varphi^1_{\text{\scriptsize SP}}(Y,T, X )&= \varphi^1_{\text{\scriptsize SP*}}(Y,T, X )+\mbox{Proj} \left[ \{T-\pi( X )\} \frac{m_1( X ) -\nu^1}{q} \Big| s_{\gamma}(T, X ) \right].
\end{align*}
\end{enumerate}
\end{pro}

The estimator $\hat\nu^{1}_{\text{\scriptsize NP}}=\tilde{E}(TY)\big/\tilde{E}(T)$ is always consistent and has the efficient influence function $\varphi^1_{\text{\scriptsize NP}}(Y,T, X )$.
Therefore, $\hat\nu^{1}_{\text{\scriptsize NP}}$ is fully robust to model misspecification, and globally nonparametric efficient.
Alternatively, taking $\varphi^1_{\text{\scriptsize SP*}}(Y,T, X )$ as an estimating function with $\hat m_1(X)$ and $\hat\pi(X)$ in place of $m_1(X)$ and $\pi(X)$
gives an estimator of $\nu^1$ that is locally semiparametric efficient, but not doubly robust.

\vspace{-.05in}
\begin{pro} \label{prop4}
Under suitable regularity conditions (see Appendix~\ref{sec:tech-details} in the Supplementary Material), the following results hold.\vspace{-.1in}
\begin{enumerate}\addtolength{\itemsep}{-.1in}
\item[(i)] The estimator $\hat\nu^{1}_{\text{\scriptsize NP}}=\tilde{E}(TY)\big/\tilde{E}(T)$ is consistent and achieves the nonparametric efficiency bound $V^1_{\text{\scriptsize NP}}$,
independently of model (\ref{OR}) for $t=1$ and model (\ref{PS}).

\item[(ii)] Define an estimator of $\nu^1$ as
\begin{align*}
\hat \nu^{1}_{\text{\scriptsize SP}}(\hat{\pi},\hat{m}_1)
=\tilde{E} \left[ TY-\{T-\hat{\pi}( X )\} \hat{m}_1( X ) \right] \big/ \tilde{E}\{ \hat\pi(X)\}.
\end{align*}
For logistic PS model (\ref{PS}), $\hat{\nu}^{1}_{\text{\scriptsize SP}} (\hat{\pi},\hat{m}_1)$ can be equivalently expressed as
\begin{align*}
\hat \nu^{1}_{\text{\scriptsize SP}}(\hat{\pi},\hat{m}_1)
=\tilde{E} \left[ TY-\{T-\hat{\pi}( X )\} \hat{m}_1( X ) \right] \big/ \tilde{E}(T).
\end{align*}
Then $\hat\nu^{1}_{\text{\scriptsize SP}}(\hat{\pi},\hat{m}_1)$ is locally semiparametric efficient: it attains the semiparametric efficiency bound $V^1_{\text{\scriptsize SP}}$
when both model (\ref{OR}) for $t=1$ and model (\ref{PS}) are correctly specified. But $\nu^{1}_{\text{\scriptsize SP}}(\hat{\pi},\hat{m}_1)$ is not doubly robust.
\end{enumerate}
\end{pro}

Finally, for estimation of ATT $=\nu^1-\nu^0$, the efficient influence function is the difference of the efficient influence functions for
estimation of $\nu^1$ and $\nu^0$ under each of the three settings in Propositions \ref{prop1} and \ref{prop3}.
Combining the estimators of $\nu^0$ and $\nu^1$ in Propositions \ref{prop2} and \ref{prop4} leads to the following results.

\vspace{-.05in}
\begin{cor} \label{cor5}
Under suitable regularity conditions (see Appendix~\ref{sec:tech-details} in the Supplementary Material), the following results hold. \vspace{-.1in}
\begin{enumerate}\addtolength{\itemsep}{-.1in}
\item[(i)] The estimator $\hat{\nu}^{1}_{\text{\scriptsize NP}} - \hat{\nu}^{0}_{\text{\scriptsize NP}} (\hat{\pi},\hat{m}_0)$ for ATT
is locally nonparametric efficient: it achieves the nonparametric efficiency bound, $\var\{\varphi^1_{\text{\scriptsize NP}}(Y,T, X ) - \varphi^0_{\text{\scriptsize NP}}(Y,T, X )\}$,
when both model (\ref{OR}) for $t=0$ and model (\ref{PS}) are correctly specified.
Moreover, this estimator is doubly robust: it remains consistent when either model (\ref{OR}) for $t=0$ or model (\ref{PS}) is
correctly specified.

\item[(ii)] The estimator $\hat{\nu}^{1}_{\text{\scriptsize SP}} (\hat{\pi},\hat{m}_0) - \hat{\nu}^{0}_{\text{\scriptsize SP}} (\hat{\pi},\hat{m}_0)$ for ATT
is locally semiparametric efficient:  it achieves the semiparametric efficiency bound, $\var\{\varphi^1_{\text{\scriptsize SP}}(Y,T, X ) - \varphi^0_{\text{\scriptsize SP}}(Y,T, X )\}$,
when both model (\ref{OR}) for $t=0,1$ and model (\ref{PS}) are correctly specified. But this estimator is, generally, not doubly robust.
\end{enumerate}
\end{cor}

\section{Improved estimation} \label{improved}

We develop estimators of $\nu^0$ that are not only locally nonparametric efficient and doubly robust, but
also intrinsically efficient: when the PS model (\ref{PS}) is correctly specified but the OR model (\ref{OR}) for $t=0$ may be misspecified,
these estimators achieve at least as small asymptotic variances
among a class of AIPW estimators, including $\hat{\nu}^{0}_{\text{\scriptsize NP}} (\hat{\pi},\hat{m}_0)$ but only with
$\hat\pi(X)$ replaced by the fitted value from a slightly augmented PS model as defined later in (\ref{augPS}).
The new estimators are then similar to $\hat{\nu}^{0}_{\text{\scriptsize NP}} (\hat{\pi},\hat{m}_0)$,
in being consistent when either the PS model or the OR model is correctly specified
and achieving the nonparametric efficiency bound $V^0_{\text{\scriptsize NP}}$ when both models are correctly specified,
but often achieve smaller variances over $\hat{\nu}^{0}_{\text{\scriptsize NP}} (\hat{\pi},\hat{m}_0)$ when
the PS model is correctly specified but the OR model is misspecified.

Similarly, we develop estimators of ATT that are not only locally nonparametric efficient and doubly robust,
but also often provide efficiency gains over
$\hat{\nu}^{1}_{\text{\scriptsize NP}} - \hat{\nu}^{0}_{\text{\scriptsize NP}} (\hat{\pi},\hat{m}_0)$
when the PS model is correctly specified but the OR model is misspecified.

Before proceeding, we point out that although, by symmetry, it also seems desirable to construct estimators of $\nu^0$ or ATT
that are not only locally nonparametric efficient and doubly robust, but also achieve efficiency gains approximately over
$\hat{\nu}^{0}_{\text{\scriptsize NP}} (\hat{\pi},\hat{m}_0)$ or $\hat{\nu}^{1}_{\text{\scriptsize NP}} - \hat{\nu}^{0}_{\text{\scriptsize NP}} (\hat{\pi},\hat{m}_0)$
when the OR model is correctly specified but the PS model is misspecified,
such estimators have not been obtained so far.

\subsection{Regression estimators} \label{reg-est}

We derive regression estimators for $\nu^0$ and ATT to achieve the desired properties,
similarly to regression estimators for ATE (\citealt{Tan2006}) but with an important new idea as follows.
For simplicity, assume in Sections \ref{reg-est}--\ref{lik-est} that PS model (\ref{PS}) is logistic regression. See Appendix~\ref{sec:tech-details}.6 for
an extension when PS model (\ref{PS}) is non-logistic regression.
Consider an augmented logistic PS model \vspace{-.05in}
\begin{align}
& P(T=1|X) = \pi_{\text{\scriptsize aug}}(X; \gamma,\delta, \hat\alpha) \nonumber \\
& = \mbox{expit} \left\{ \gamma^\T f(X) + \delta_0\, \hat m_0(X) + \delta_1\, \hat m_1(X) \right\}, \label{augPS}
\end{align}
where $\mbox{expit}(c)=\{1+\exp(-c)\}^{-1}$, $\hat\alpha=(\hat\alpha_0^\T,\hat\alpha_1^\T)^\T$ are estimates of
$\alpha=(\alpha_0^\T, \alpha_1^\T)^\T$ from OR model (\ref{OR}), and $\delta=(\delta_0, \delta_1)^\T$ are unknown coefficients for additional regressors $\hat m_0(X)$ and $\hat m_1(X)$.
Let $(\tilde\gamma, \tilde\delta)$ be the MLE of $(\gamma,\delta)$ and $\tilde\pi(X) = \pi_{\text{aug}}(X;\tilde\gamma,\tilde\delta, \hat\alpha)$.
An important consequence of including the additional regressors is that, by Eq.~(\ref{score-eq}), we have, in addition to $\tilde E[  \{T-\tilde\pi(X)\} f(X)]=0$, \vspace{-.05in}
\begin{align}
\tilde E\left [ \{T- \tilde\pi(X) \} \hat m_t(X) \right] = 0, \quad t=0,1. \label{augPS-eq}
\end{align}
For the augmented PS model, there may be linear redundancy in the variables, $\{f(X)$, $\hat m_0(X)$, $\hat m_1(X)\}$, in which case
the regressors need to be redefined accordingly.
In particular, consider the following condition:
\begin{itemize}
\item[(L)] $\hat m_t(X)$ is a linear combination of variables in $f(X)$ for $t=0,1$,
\end{itemize}
which is satisfied when all variables in $g_t(X)$ are included as components of $f(X)$, and
$\Psi(\cdot)$ is the identity link corresponding to linear regression in (\ref{OR}).
If Condition (L) holds, then the augmented model (\ref{augPS}) reduces to (\ref{PS})
and hence $\tilde \pi(X) = \hat \pi(X)$ subsequently.
Otherwise, $\tilde \pi(X)$ and $ \hat \pi(X)$ generally differ from each other.

With $\tilde\pi(X)$ the fitted value from the augmented PS model (\ref{augPS}),
we define the regression estimator of $\nu^t = E(Y^t|T=1)$ as
\begin{align*}
\tilde\nu^t_{\text{\scriptsize reg}}=\tilde{E}\left(\tilde{\eta}_t -\tilde{\beta }_t^\T  \tilde{\xi }_t \right) \big/ \tilde E(T), \quad t=0 \mbox{ or } 1,
\end{align*}
where $\tilde{\beta }_t= \tilde{E}^{-1}(\tilde{\xi }_t\tilde{\zeta }_t^\T ) \tilde{E}(\tilde{\xi }_t\tilde{\eta}_t)$ with
\begin{align*}
& \tilde{\eta}_1 =T Y, & \quad&  \tilde{\eta}_0 =\frac{1-T}{1-\tilde{\pi}( X )}\tilde{\pi}( X ) Y, \\
& \tilde{\xi }_1 =\left\{\frac{T}{\tilde{\pi}( X )}-1\right\}\frac{\tilde{h }( X )}{1-\tilde{\pi}( X )}, & \quad &
 \tilde{\xi }_0 \,(=-\tilde\xi_1)=\left\{\frac{1-T}{1-\tilde{\pi}( X )}-1\right\} \frac{\tilde{h }( X )}{\tilde{\pi}( X )}, \\
& \tilde{\zeta }_1 =\frac{T}{\tilde{\pi}( X )}\frac{\tilde{h }( X )}{1-\tilde{\pi}( X )}, & \quad &
 \tilde{\zeta }_0 =\frac{1-T}{1-\tilde{\pi}( X )}\frac{\tilde{h }( X )}{\tilde{\pi}( X )},
\end{align*}
and $\tilde{h }( X )=\{\tilde{h }_1^\T , (C\tilde{h }_2)^\T \}^\T(X) $ are defined with a constant matrix $C$ such that the variables in $\tilde h(X)$ are linearly independent, and
\begin{align*}
\tilde{h }_1( X ) &=\left[\{1-\tilde{\pi}( X )\}\tilde{v }_1^\T ( X ),\tilde{\pi}( X )\tilde{v }_0^\T ( X )\right]^\T , \\
\tilde{h }_2( X ) & = \tilde{\pi}( X )\{1-\tilde{\pi}( X )\} \left\{ f_{(1)}^\T(X), \hat{m}_0( X ) \right\}^\T, \\
 \tilde{v }_1( X ) & =\left\{ \tilde{\pi}( X ),\tilde{\pi}( X )\hat{m}_1( X )\right\}^\T , \quad
 \tilde{v }_0( X ) =\left\{ \tilde{\pi}( X ),\tilde{\pi}( X )\hat{m}_0( X )\right\}^\T .
\end{align*}
where $f_{(1)}(X)$ is the vector of nonconstant variables in $f(X)$, because $\tilde{\pi}( X )\{1-\tilde{\pi}( X )\} $ is already a component of $\{1-\tilde{\pi}( X )\}\tilde{v }_1^\T ( X )$ in $\tilde h_1(X)$.
For example, if Condition (L) holds for $t=0$ or 1, then $\tilde h(X)$ should be specified such that
one variable is removed from the vector
$\tilde{\pi}( X )\{1-\tilde{\pi}( X )\} f_{(1)}(X)$ in $\tilde h_2(X)$.

The variables in $\tilde h(X)$ are included for the following considerations. The variables $\tilde \pi(X) \hat m_0(X)$ and $\tilde \pi(X) \hat m_1(X)$ are included
in $\tilde v_0(X)$ and $\tilde v_1(X)$ respectively to achieve double robustness and local nonparametric efficiency, as later seen from Eq.~(\ref{DR-explain}).
Moreover, the variables in $\tilde h_2(X)$, in addition to $\{1-\tilde\pi(X)\} \tilde v_1(X)$, are included to
accommodate the variation of $(\tilde\gamma,\tilde\delta)$ for achieving intrinsic efficiency,
as later described in Proposition \ref{prop-reg}.
The corresponding variables in $\tilde\xi_0$ or $\tilde\xi_1$ are exactly the scores $\{T-\tilde\pi(X)\}\{f^\T(X), \hat m_0(X),\hat m_1(X)\}^\T$ for the augmented PS model (\ref{augPS}).
Finally, $\tilde \pi(X)$ is included in $\tilde v_0(X)$ and $\tilde v_1(X)$ to ensure efficiency gains over the ratio estimator $\hat{\nu}^{0}_{\text{\scriptsize IPW,ratio}}(\tilde\pi)$
under a correctly specified PS model, as discussed after Corollary \ref{cor-reg}.

The  name  ``regression  estimator"  is  adopted  from the literatures of survey sampling (\citealt{Cochran1977})
and  Monte  Carlo  integration (\citealt{Ham1964}),  and  should  be  distinguished
from  the estimator $\hat\nu^t_{\text{\scriptsize OR}}$ based on outcome regression in Section \ref{setup}.
The idea is to exploit the fact that if the PS model is correct, then $\tilde E(\tilde \eta_t)$
asymptotically  has  mean $E(T Y^t)$ (to  be  estimated)  and
$\tilde \xi_t$ mean 0 (known). That is,
$\tilde \xi_t$ serves as auxiliary variables (in
the terminology of survey sampling) or control variates
(in that of Monte Carlo integration).
The effect of variance reduction using regression estimators can be seen from in the following results.

\begin{pro} \label{prop-reg}
Under suitable regularity conditions (see Appendix~\ref{sec:tech-details} in the Supplementary Material), the estimator $\tilde\nu^t_{\text{\scriptsize reg}}$ for $\nu^t$
has the following properties for $t=0,1$.
\begin{enumerate}
\item[(i)] $\tilde\nu^t_{\text{\scriptsize reg}}$ is locally nonparametric efficient: it achieves the nonparametric efficiency bound $V^t_{\text{\scriptsize NP}}$
when both model (\ref{OR}) for the corresponding $t$ and model (\ref{PS}) are correctly specified.

\item[(ii)] $\tilde\nu^t_{\text{\scriptsize reg}}$ is doubly robust: it remains consistent when either model (\ref{OR}) for the corresponding $t$
or model (\ref{PS}) is correctly specified.

\item[(iii)] $\tilde\nu^t_{\text{\scriptsize reg}}$ is intrinsically efficient:
if  model (\ref{PS}) is correctly specified, then it
achieves the lowest asymptotic variance among the class of estimators
\begin{align}
\tilde{E}\left(\tilde{\eta}_t - b_t^\T \tilde{\xi }_t\right) \big/ \tilde E(T), \label{nu-class}
\end{align}
where $b_t $ is an arbitrary vector of constants.
\end{enumerate}
\end{pro}

\begin{cor} \label{cor-reg}
The estimator $\tilde\nu^1_{\text{\scriptsize reg}} - \tilde\nu^0_{\text{\scriptsize reg}}$ for ATT
has the following properties.
\begin{enumerate}
\item[(i)] $\tilde\nu^1_{\text{\scriptsize reg}} - \tilde\nu^0_{\text{\scriptsize reg}}$ is locally nonparametric efficient:
it achieves the nonparametric efficiency bound, $\var\{\varphi^1_{\text{\scriptsize NP}}(Y,T, X ) - \varphi^0_{\text{\scriptsize NP}}(Y,T, X )\}$,
when both model (\ref{OR}) for $t=0,1$ and model (\ref{PS}) are correctly specified.

\item[(ii)] $\tilde\nu^1_{\text{\scriptsize reg}} - \tilde\nu^0_{\text{\scriptsize reg}}$ is doubly robust:
it remains consistent when either model (\ref{OR}) for $t=0, 1$
or model (\ref{PS}) is correctly specified.

\item[(iii)] $\tilde\nu^1_{\text{\scriptsize reg}} - \tilde\nu^0_{\text{\scriptsize reg}}$ is intrinsically efficient:
if model (\ref{PS}) is correctly specified, then it
achieves the lowest asymptotic variance among the class of estimators
\begin{align*}
\tilde{E}\left(\tilde{\eta}_1 - \tilde \eta_0 - b_0^\T \tilde{\xi }_0 \right) \big/ \tilde E(T),
\end{align*}
where $b_0$ is an arbitrary vector of constants.
\end{enumerate}
\end{cor}

{\bf Double robustness}.\;
We point out that the use of augmented propensity scores $\tilde\pi(X)$ is crucial for $\tilde{\nu}^t_{\text{\scriptsize reg}}$ to be doubly robust, in particular,
to be consistent under a correctly specified OR model but a misspecified PS model.
[It is possible, for example, under Condition (L) that $\tilde\pi(X)$ reduces to $\hat \pi(X)$.]
If the OR model (\ref{OR}) for $t=0$ or 1 is correctly specified, then, as shown in the Appendix~\ref{sec:tech-details}.3,
the vector $\tilde\beta_t$ converges to a constant vector $\beta^*_t$ such that
\begin{align}
\tilde{\nu}^t_{\text{\scriptsize reg}} =\tilde{E}\left(\tilde{\eta}_t -{\beta_t^*}^\T  \tilde{\xi }_t \right) \big/ \tilde E(T) +o_p(n^{-1/2})
= \hat \nu^t_{\text{\scriptsize SP}}(\tilde \pi, \hat m_t) + o_p(n^{-1/2}), \label{DR-explain}
\end{align}
because $\tilde\pi(X)\hat m_0(X)$ is a linear combination of variables in $\tilde h(X)/\tilde{\pi}(X)$
and $\tilde\pi(X)\hat m_1(X)$ is a linear combination of those in $\tilde h(X)/\{1-\tilde{\pi}(X)\}$.
By Eq.~(\ref{augPS-eq}) for the augmented PS model, $\hat \nu^t_{\text{\scriptsize SP}}(\tilde \pi, \hat m_t)$ is identical to
$\hat \nu^0_{\text{\scriptsize NP}}(\tilde \pi, \hat m_0)$ for $t=0$, which is doubly robust similarly as $\hat \nu^0_{\text{\scriptsize NP}}(\hat\pi, \hat m_0)$ by Propsition~\ref{prop2},
or to $\hat \nu^1_{\text{\scriptsize NP}}$ for $t=1$, which is fully robust.
Therefore, $\tilde{\nu}^t_{\text{\scriptsize reg}}$ is consistent when the OR model (\ref{OR}) for the corresponding $t$ is correctly specified.
This result would not hold when $\tilde{\nu}^t_{\text{\scriptsize reg}}$  were defined with $\hat\pi(X)$ in place of $\tilde \pi(X)$.

{\bf Local efficiency}.\;
For $t=0$ or 1, the estimator $\tilde{\nu}^t_{\text{\scriptsize reg}}$ is locally nonparametric efficient, similarly as $\hat \nu^0_{\text{\scriptsize NP}}(\tilde \pi, \hat m_0)$ or $\hat \nu^1_{\text{\scriptsize NP}}$.
In addition, $\tilde{\nu}^t_{\text{\scriptsize reg}}$ is generally not locally semiparametric efficient with respect to PS model (\ref{PS}),
but locally semiparametric efficient with respect to PS model (\ref{augPS}):
$\tilde{\nu}^t_{\text{\scriptsize reg}}$ achieves the semiparametric efficiency bounded calculated under model (\ref{augPS}),
when both model (\ref{OR}) and model (\ref{PS}) are correctly specified. In fact, when model (\ref{PS}) holds,
the efficiency bound $V^t_{\text{\scriptsize SP}}$ under model (\ref{augPS}) coincides with the
nonparametric efficiency bound $V^t_{\text{\scriptsize NP}}$, because $\{T-\pi(X)\}\{m_t(X)-\nu^t\}$ is
a linear combination of the score function, which contains $\{T-\pi(X)\}\{1,m_0(X),m_1(X)\}^\T$ under model (\ref{augPS}) as shown in Appendix~\ref{sec:tech-details}.
On the other hand, $\tilde{\nu}^t_{\text{\scriptsize reg}}$ with $\tilde\pi(X)$ replaced by $\hat \pi(X)$ throughout would be locally semiparametric efficient
with respect to the original PS model (\ref{PS}), but generally
not doubly robust, similarly as $\hat \nu^t_{\text{\scriptsize SP}}(\tilde \pi, \hat m_t)$.

{\bf Intrinsic efficiency}.\;
A classical estimator of the optimal choice of $b_t $ in minimizing the asymptotic variance of (\ref{nu-class}) is
$\hat \beta_t = \tilde{E}(\tilde{\xi }_t\tilde{\xi }_t^\T )^{-1}\tilde{E}(\tilde{\xi }_t\tilde{\eta}_t)$, which differs from $\tilde\beta_t$ in a subtle manner.
It can be shown that the corresponding estimator,
$\hat{\nu}^t_{\text{\scriptsize reg}} = \tilde{E} (\tilde{\eta}_t -\hat{\beta }_t^\T  \tilde{\xi }_t )/\tilde E(T)$,  for $\nu^t$
is asymptotically equivalent to the first order to $\tilde{\nu}^t_{\text{\scriptsize reg}}$
when the PS model is correctly specified.
But $\hat{\nu}^t_{\text{\scriptsize reg}}$, unlike $\tilde{\nu}^t_{\text{\scriptsize reg}}$, is generally inconsistent for $\nu^t$,
even when the OR model is correctly specified and the PS model may be misspecified.
The particular form of $\tilde\beta_t$, although seems ad hoc in the above definition, can also be derived through empirical efficiency maximization (\citealt{Rubin2008, Tan2008})
and design-optimal regression estimation for survey calibration (\citealt{Tan2013}).
See further discussion related to calibration estimation after Proposition \ref{prop-lik}.

The advantage of achieving intrinsic efficiency is shown by the following comparison, where
$\hat{\nu}^{0}_{\text{\scriptsize IPW}}(\tilde\pi)$, $\hat{\nu}^{0}_{\text{\scriptsize IPW,ratio}}(\tilde\pi)$, and, as discussed below (\ref{DR-explain}),
$\hat{\nu}^{0}_{\text{\scriptsize NP}} (\tilde{\pi},\hat{m}_0)$, are obtained from
$\hat{\nu}^{0}_{\text{\scriptsize IPW}}(\hat\pi)$, $\hat{\nu}^{0}_{\text{\scriptsize IPW,ratio}}(\hat\pi)$, and
$\hat{\nu}^{0}_{\text{\scriptsize NP}} (\hat{\pi},\hat{m}_0)$ with $\hat\pi(X)$ replaced by $\tilde \pi(X)$.

\begin{cor} \label{cor-reg2}
Under the setting of Proposition~\ref{prop-reg}, if PS model (\ref{PS}) is correctly specified, then the estimator $\tilde\nu^0_{\text{\scriptsize reg}}$
is asymptotically at least as efficient as not only $\hat{\nu}^{0}_{\text{\scriptsize IPW}}(\tilde\pi)$ and  $\hat{\nu}^{0}_{\text{\scriptsize IPW,ratio}}(\tilde\pi)$ but also
$\hat{\nu}^{0}_{\text{\scriptsize NP}} (\tilde{\pi},\hat{m}_0)$,
and  the estimator $\tilde\nu^1_{\text{\scriptsize reg}}-\tilde\nu^0_{\text{\scriptsize reg}}$ for ATT is asymptotically at least as efficient as
$\hat{\nu}^{1}_{\text{\scriptsize NP}} -\hat{\nu}^{0}_{\text{\scriptsize IPW}}(\tilde\pi)$,
$\hat{\nu}^{1}_{\text{\scriptsize NP}} -\hat{\nu}^{0}_{\text{\scriptsize IPW,ratio}}(\tilde\pi)$, and
$\hat{\nu}^{1}_{\text{\scriptsize NP}} -\hat{\nu}^{0}_{\text{\scriptsize NP}} (\tilde{\pi},\hat{m}_0) $.
\end{cor}

A technical complication of using augmented propensity scores $\tilde\pi(X)$ is that
$\tilde\nu^0_{\text{\scriptsize reg}}$ may not, in general, be intrinsically efficient, when compared to the  class of estimators (\ref{nu-class}) with
$\tilde\pi(X)$ replaced by $\hat\pi(X)$ in $\tilde\eta_0$ and $\tilde\xi_0$.
[Nevertheless, such intrinsic efficiency holds in the special case with $\tilde\pi(X)=\hat\pi(X)$, where the OR model (\ref{OR}) for $t=0$ is linear regression with all variables in $g_0(X)$ also included in $f(X)$.]
Particularly, if the PS model (\ref{PS}) is correctly specified, then
$\tilde\nu^0_{\text{\scriptsize reg}}$ may not be as efficient as $\hat{\nu}^{0}_{\text{\scriptsize NP}} (\hat{\pi},\hat{m}_0)$ based on $\hat\pi(X)$
even though $\tilde\nu^0_{\text{\scriptsize reg}}$ is proven to be asymptotically at least as efficient as $\hat{\nu}^{0}_{\text{\scriptsize NP}} (\tilde{\pi},\hat{m}_0)$ based on $\tilde\pi(X)$
and, when the OR model (\ref{OR}) for $t=0$ is also correctly specified, asymptotically equivalent to $\hat{\nu}^{0}_{\text{\scriptsize NP}} (\hat{\pi},\hat{m}_0)$
and $\hat{\nu}^{0}_{\text{\scriptsize NP}} (\tilde{\pi},\hat{m}_0)$.
However, the increase in the asymptotic variance of
$\hat{\nu}^{0}_{\text{\scriptsize NP}} (\tilde{\pi},\hat{m}_0)$ over that of $\hat{\nu}^{0}_{\text{\scriptsize NP}} (\hat{\pi},\hat{m}_0)$
is usually small, caused by the use of a slightly augmented PS model (\ref{augPS}).
The estimator $\tilde\nu^0_{\text{\scriptsize reg}}$ may still often achieve efficiency gains over $\hat{\nu}^{0}_{\text{\scriptsize NP}} (\hat{\pi},\hat{m}_0)$
when the PS model is correctly specified but the OR model is misspecified, as shown in our simulation studies.

\subsection{Likelihood estimators} \label{lik-est}

A practical limitation of the regression estimators as well as AIPW estimators is that
they may lie outside either the sample or the population range of observed outcomes.
For example, $\tilde\nu^t_{\text{\scriptsize reg}}$ may take values outside the interval $(0,1)$ for binary outcomes.
Such behavior may occur due to the presence of fitted propensity scores $\tilde\pi(X_i)$ near 1 or, equivalently, large inverse probability weights $\{1-\tilde\pi(X_i)\}^{-1}$ among the untreated.
In this section, we derive likelihood estimators for $\nu^t$ that are not only doubly robust, locally nonparametric efficient, and intrinsically efficient similarly
to the regression estimators, but also sample-bounded in falling within the range of $\{Y_i: T_i=t, i=1,\ldots,n\}$.
These likelihood estimators are therefore much less sensitive to large weights than the regression and AIPW estimators.

There are two steps in constructing the desired likelihood estimators, similarly as for ATE estimation in \citet{Tan2010} but
using the fitted propensity scores $\tilde\pi(X)$ from augmented PS model (\ref{augPS}).
First, we derive intrinsically efficient, but non-doubly robust, likelihood estimators by the approach of empirical likelihood (\citealt{Owen2001})
taking $\tilde \eta_t-\nu^t T $ and $\tilde\xi_t$ as asymptotically unbiased estimating functions or, equivalently, the approach
of nonparametric likelihood (\citealt{Tan2006, Tan2010}).
Specifically, our approach is to maximize the log empirical likelihood, $\sum_{i=1}^n \log p_i$, subject to the constraints
\begin{align*}
\sum_{i=1}^n p_i \tilde \xi_{1,i} =0  \quad \text{and} \quad  \sum_{i=1}^n p_i (\tilde \eta_{t,i} - \nu^t T_i)=0 \text{ for } t=0,1,
\end{align*}
where $p_i$ is a nonnegative weight assigned to $(Y_i,T_i,X_i)$ for $i=1,\ldots,n$ with $\sum_{i=1}^n p_i=1$.
We show in the Appendix~\ref{sec:tech-details}.4 that the resulting estimates of $\nu^0$ and $\nu^1$ are
\begin{eqnarray*}
\hat{\nu}^0_{\text{\scriptsize lik}}&=&\tilde{E}\left\{\frac{(1-T)\tilde{\pi}(X )Y}{1-\omega(X ;\hat{\lambda })}\right\} \Big{/}  \tilde{E}\left\{\frac{(1-T)\tilde{\pi}(X )}{1-\omega(X ;\hat{\lambda })}\right\}, \\
\hat{\nu}^1_{\text{\scriptsize lik}}&=&\tilde{E}\left\{\frac{T\tilde{\pi}(X )Y}{\omega(X ;\hat{\lambda })}\right\} \Big{/} \tilde{E}\left\{\frac{T\tilde{\pi}(X )}{\omega(X ;\hat{\lambda })}\right\},
\end{eqnarray*}
where $\omega(X ;\lambda)=\tilde{\pi}(X )+\lambda^\T\tilde{h }(X )$ and $\hat{\lambda}$ is a maximizer of the function
\begin{align*}
\ell(\lambda)=\tilde{E}[ T\log{\omega(X ;\lambda)}+(1-T)\log\{ 1-\omega(X ;\lambda) \} ] ,
\end{align*}
subject to $\omega(X_i;\lambda)>0$ if $T_i=1$ and $\omega(X_i;\lambda)<1$ if $T_i=0$ for $i=1,\ldots,n$.
Setting the gradient of $\ell(\lambda)$ to zero shows that $\hat{\lambda}$ is a solution to
\begin{eqnarray}
\tilde{E}\left[ \frac{T-\omega(X ;\lambda)}{\omega(X ;\lambda) \{1-\omega(X ;\lambda)\}} \tilde{h }(X ) \right]=0 . \label{lik-score}
\end{eqnarray}
Because $\tilde \pi(X)$ is a linear combination of variables in $\tilde h(X)$, it follows from Eq.~(\ref{lik-score}) that
the two denominators, $\tilde{E} [(1-T)\tilde{\pi}(X )/\{1-\omega(X ;\hat{\lambda })\}]$ and $\tilde{E} [T\tilde{\pi}(X )/\omega(X ;\hat{\lambda })]$, in the definitions of
$ \hat{\nu}^0_{\text{\scriptsize lik}}$ and $ \hat{\nu}^1_{\text{\scriptsize lik}}$ are equal to each other.

The estimator
$\hat{\nu}^t_{\text{\scriptsize lik}}$ can be shown to be intrinsically efficient among the class of estimators (\ref{nu-class})
and locally nonparametric efficient, but generally not doubly robust.
We introduce the following modified likelihood estimators, to achieve double robustness but without affecting the first-order asymptotic behavior.

For $t=0$ or 1, partition $\tilde{h }$ as $\tilde h= \{\tilde{h }_{1t}^\T, \tilde{h }_{1(t)}^\T, (C\tilde{h }_2)^\T \}^\T$ for a constant matrix $C$ and accordingly
$\lambda$ as $\lambda =(\lambda _{1t}^\T,\lambda _{1(t)}^\T, \lambda _2^\T)^\T$,
where $\tilde h_{1t} = \tilde \pi \tilde v_0$ or $(1-\tilde\pi) \tilde v_1$ if $t=0$ or 1, and
$\tilde{h }_{1(t)}$ consists of the elements of $\tilde{h }_1$ excluding $\tilde{h }_{1t}$.
Moreover, write $R_t=0$ or 1, $\tilde \pi(t,X) = 1-\tilde\pi(X)$ or $\tilde\pi(X)$, and
$\omega (t,X; \lambda) = 1-\omega(X;\lambda)$ or $\omega(X;\lambda)$ respectively for $t=0$ or 1.
Define $\tilde \lambda^t = (\tilde \lambda_{1t}^\T, \hat\lambda_{1(t)}^\T, \hat\lambda_2^\T)^\T$, where $\hat\lambda_{1(t)}$ and $ \hat\lambda_2$ are obtained from $\hat\lambda$, and
$\tilde \lambda_{1t}$ is a maximizer of the function
$$
\kappa_t(\lambda _{1t})=\tilde{E}\left[ R_t\frac{\log\{\omega(t,X ;\lambda _{1t},\hat{\lambda }_{1(t)},\hat{\lambda }_2)\}-
\log\{\omega(t,X ;\hat{\lambda })\}}{1-\tilde{\pi}(t,X )}-\lambda _{1t}^\T v_t(X )\right],
$$
subject to $\omega(t,X _i;\lambda _{1t}, \hat\lambda _{1(t)},\hat{\lambda }_2)>0 $ if $T_i=t$ for $i=1,\ldots,n$.
Setting the gradient of $\kappa_t(\lambda _{1t})$ to 0 shows that $\tilde{\lambda }_{1t}$ is a solution to
\begin{eqnarray}
\tilde{E}\left[\left\{ \frac{R_t}{\omega(t,X ;\lambda _{1t},\hat{\lambda }_{1(t)},\hat{\lambda }_2)}-1\right\} \tilde{v}_t(X )\right]=0. \label{lik2-score}
\end{eqnarray}
For $t=0,1$, the resulting estimator of $\nu^t$ is
\begin{eqnarray*}
\tilde{\nu}^t_{\text{\scriptsize lik}}&=&\tilde{E}\left\{\frac{R_t\tilde{\pi}(X )Y}{\omega(t,X ;\tilde{\lambda }^t)}\right\} \Big/ \tilde{E}\left\{\frac{R_t\tilde{\pi}(X )}{\omega(t,X ;\tilde{\lambda }^t)}\right\}
=\tilde{E}\left\{\frac{R_t\tilde{\pi}(X )Y}{\omega(t,X ;\tilde{\lambda }^t)}\right\} \Big/ \tilde{E}(T),
\end{eqnarray*}
where the second equation holds due to Eq.~(\ref{lik2-score}) with $\tilde \pi(X)$ included in $\tilde v_0(X)$ and $\tilde v_1(X)$, and $\tilde E\{ T- \tilde \pi(X)\}=0$
by the score equation for model (\ref{augPS}).
The likelihood estimator $\tilde{\nu}^t_{\text{\scriptsize lik}}$ has several desirable properties as follows.

\begin{pro} \label{prop-lik}
Under suitable regularity conditions (see Appendix~\ref{sec:tech-details} in the Supplementary Material), the estimator $\tilde\nu^t_{\text{\scriptsize lik}}$ for $\nu^t$
has the following properties for $t=0,1$.
\begin{enumerate}
\item[(i)] $\tilde\nu^t_{\text{\scriptsize lik}}$ is sample-bounded: it lies within the range of $\{Y_i: T_i=t, i=1,\ldots,n\}$.

\item[(ii)] If model (\ref{PS}) is correctly specified, then
$\tilde\nu^t_{\text{\scriptsize lik}}$ is asymptotically equivalent, to the first order, to $\tilde\nu^t_{\text{\scriptsize reg}}$. Hence
$\tilde\nu^t_{\text{\scriptsize lik}}$ is intrinsically efficient among the class (\ref{nu-class}) and locally nonparametric efficient, similarly as $\tilde\nu^t_{\text{\scriptsize reg}}$ in Proposition \ref{prop-reg}.

\item[(iii)] $\tilde\nu^t_{\text{\scriptsize lik}}$ is doubly robust, similarly as $\tilde\nu^t_{\text{\scriptsize reg}}$ in Proposition \ref{prop-reg}.
\end{enumerate}
\end{pro}

The sample-boundedness of $\tilde\nu^t_{\text{\scriptsize lik}}$ holds because $\omega(t,X_i;\tilde{\lambda }^t)>0$ if $T_i=t$ for $i=1,\ldots,n$ and
$\tilde E \{ R_t \tilde\pi(X) / \omega(t,X_i;\tilde{\lambda }^t) \}= \tilde E \{\tilde \pi(X)\} = \tilde E(T)$ by Eq.~(\ref{lik2-score}).
The double robustness of  $\tilde\nu^t_{\text{\scriptsize lik}}$ follows mainly for two reasons:
$\tilde E \{ R_t \tilde\pi(X) \hat m_t(X) / \omega(t,X_i;\tilde{\lambda }^t) \} = \tilde E\{ \tilde\pi(X) \hat m_t(X) \}$  by Eq.~(\ref{lik2-score})
with $\tilde\pi(X) \hat m_t(X)$ included in $\tilde v_t(X)$,
and  $\tilde E\{ \tilde \pi(X) \hat m_t(X) \} $ $= \tilde E \{ T \hat m_t (X) \}$ by Eq.~(\ref{augPS-eq}) for the augmented PS model (\ref{augPS}).

Eq.~(\ref{lik2-score}), which underlies both sample-boundedness and double robustness as discussed above,
can be connected to calibration estimation using auxiliary information in survey sampling (\citealt{Deville1992, Tan2013}).
In fact, the inverse weighted average of $\tilde v_t(X) = \tilde\pi(X) \{1, \hat m_t(X)\}^\T$ is matched (or calibrated) with
the simple sample average of $\tilde v_t(X)$.
This is equivalent to saying that if $Y$ is replaced by $ \hat m_t(X)$, then the numerator in the definition of $\tilde\nu^t_{\text{\scriptsize lik}}$
yields exactly $\tilde E\{ \tilde\pi(X) \hat m_t(X) \}$.
A similar property holds for $\tilde\nu^t_{\text{\scriptsize reg}}$ :
if $Y$ is replaced by $ \hat m_t(X)$, then the numerator in the definition of $\tilde\nu^t_{\text{\scriptsize reg}}$
yields exactly $\tilde E\{ \tilde\pi(X) \hat m_t(X) \}$.
By this relationship, $\tilde\nu^t_{\text{\scriptsize reg}}$ and  $\tilde\nu^t_{\text{\scriptsize lik}}$
can be referred to as {\it calibrated} regression and likelihood estimators.

The implication of intrinsic efficiency for $\tilde\nu^t_{\text{\scriptsize lik}}$ is similar to that for $\tilde\nu^t_{\text{\scriptsize reg}}$ as discussed in Section \ref{reg-est}.
If the PS model (\ref{PS}) is correctly specified while the OR model (\ref{OR}) may be misspecified, then $\tilde\nu^0_{\text{\scriptsize lik}}$ is asymptotically at least as efficient as
$\hat{\nu}^{0}_{\text{\scriptsize NP}} (\tilde{\pi},\hat{m}_0)$, and
$\tilde\nu^1_{\text{\scriptsize lik}}-\tilde\nu^0_{\text{\scriptsize lik}}$ is asymptotically at least as efficient as
$\hat{\nu}^{1}_{\text{\scriptsize NP}} -\hat{\nu}^{0}_{\text{\scriptsize NP}} (\tilde{\pi},\hat{m}_0) $.

\section{Extensions and comparisons}

To possibly enhance numerical stability and finite-sample performance, we suggest the following versions of $\tilde\nu^t_{\text{\scriptsize reg}}$ and  $\tilde\nu^t_{\text{\scriptsize lik}}$
with simplifications of $\tilde \pi(X)$ and $\tilde h(X)$:
\begin{enumerate}
\item[(i)] Consider an augmented logistic PS model in place of (\ref{augPS}):
\begin{align}
& P(T=1|X) = \pi_{\text{\scriptsize aug2}}(X; \gamma_0,\delta, \hat\alpha, \hat\gamma) \nonumber \\
& = \mbox{expit} \left[ \mbox{logit}\{\hat\pi(X)\} + \gamma_0 + \delta_0\, \hat m_0(X) + \delta_1\, \hat m_1(X) \right], \label{augPS2}
\end{align}
where $\mbox{logit} (\hat\pi) = \log\{\hat\pi/(1-\hat\pi)\}$ is included as an offset, and
$\gamma_0$ and $\delta=(\delta_0,\delta_1)^\T$ are unknown coefficients.
Let $(\tilde\gamma_0, \tilde \delta)$ be the MLE of $(\gamma_0,\delta)$, and redefine $\tilde\pi(X) = \pi_{\text{\scriptsize aug2}}(X; \tilde\gamma_0, \tilde\delta, \hat\alpha, \hat \gamma) $.
In contrast with (\ref{augPS}), this model (\ref{augPS2}) is
meaningful even when the original model (\ref{PS}) is non-logistic regression or $\hat\gamma$ is obtained by non-maximum likelihood estimation, for example,
penalized estimation.

\item[(ii)] Redefine $\tilde h(X)= \tilde h_1(X)$, that is, with $\tilde h_2(X)$ removed. Then $\tilde\beta_t$ is defined by projection of $\tilde\eta_t$ on a lower-dimensional vector $\tilde\xi_t$,
and $\hat\lambda$ is defined by solving a lower-dimensional optimization problem. The dimension reduction may improve numerical stability and finite-sample performance of $\tilde\nu^t_{\text{\scriptsize reg}}$ and  $\tilde\nu^t_{\text{\scriptsize lik}}$.
\end{enumerate}
For concreteness, the resulting estimators
$\tilde\nu^t_{\text{\scriptsize reg}}$ and  $\tilde\nu^t_{\text{\scriptsize lik}}$ are denoted by $\tilde\nu^t_{\text{\scriptsize reg2}}$ and  $\tilde\nu^t_{\text{\scriptsize lik2}}$ respectively.
These simplified estimators can be shown to remain locally nonparametric efficient and doubly robust as in Propositions \ref{prop-reg} and \ref{prop-lik};
they are generally not intrinsically efficient, but are expected to asymptotically nearly as efficient as
$\tilde\nu^t_{\text{\scriptsize reg}}$ and  $\tilde\nu^t_{\text{\scriptsize lik}}$ when the PS model (\ref{PS}) is correctly specified.
Informally,
$\tilde\nu^t_{\text{\scriptsize reg2}}$ and $\tilde\nu^t_{\text{\scriptsize lik2}}$ would be intrinsically efficient if
$\hat\pi(X)=\pi(X;\hat\gamma)$ were replaced, in model (\ref{augPS2}) and the definition of $\tilde\pi(X)$, by $\pi(X;\gamma^*)$ with $\gamma^*$ the limit of $\hat \gamma$ in probability.

While $\tilde h_2(X)$ can be removed from $\tilde h(X)$ for dimension reduction, we point out that
$\tilde h_1(X)$ can be extended to include additional functions of $X$ for achieving calibration on those variables in addition to $\tilde v_t(X)$.
Specifically, let $c_t(X)$ be a vector of known but possibly data-dependent functions of $X$ {\it including} 1, for example, $g_t(X)$ in the OR model (\ref{OR}) for $t=0,1$.
Redefine the augmented PS model (\ref{augPS2}) as
\begin{align}
& P(T=1|X) = \pi_{\text{\scriptsize aug2}}(X; \gamma_0, \delta, \hat\gamma) \nonumber \\
& = \mbox{expit} \left[ \mbox{logit}\{\hat\pi(X)\} + \gamma_0 + \delta_0^\T c_{0(1)}(X) + \delta_1^\T c_{1(1)}(X) \right], \label{augPS3}
\end{align}
where $\gamma_0$ and $\delta = (\delta_0^\T, \delta_1^\T)$ are unknown coefficients, and $ c_{0(1)}(X)$ or $c_{1(1)}$ is the vector of nonconstant variables in $c_0(X)$ or $c_1(X)$ respectively.
Redefine $\tilde \pi(X) = \pi_{\text{\scriptsize aug2}}(X; \tilde\gamma_0, \tilde\delta, \hat \gamma) $ with $(\tilde\gamma_0,\tilde \delta)$ the MLE of $(\gamma_0,\delta)$ for model (\ref{augPS3}),
and redefine $\tilde h(X) = \tilde h_1(X)$ with $\tilde v_t(X) = \tilde \pi(X) c^\T_t (X)$ for $t=0,1$.
Then Eq.~(\ref{lik2-score}) in conjunction with the score equation for model (\ref{augPS2}) leads to calibration equations
\begin{align}
\tilde{E} \left\{ \frac{(1-T)\tilde\pi(X)}{1-\omega(X ;\tilde\lambda^0)} c_0(X) \right\}= \tilde E\{ \tilde\pi(X) c_0(X) \} = \tilde E\{ T c_0(X) \}, \label{g-cal0}\\
\tilde{E} \left\{ \frac{T\tilde\pi(X)}{\omega(X ;\tilde\lambda^1)} c_1(X) \right\} = \tilde E\{ \tilde\pi(X) c_1(X) \} = \tilde E\{ T c_1(X) \} . \label{g-cal1}
\end{align}
By the discussion after (\ref{DR-explain}), the resulting estimators  $\tilde\nu^t_{\text{\scriptsize reg2}}$ and  $\tilde\nu^t_{\text{\scriptsize lik2}}$
are doubly robust and locally nonparametric efficient under the following condition:
\begin{itemize}
\item[(R)] $\hat m_t(X)$ is a linear combination of $c_t(X)$ for $t=0,1$.
\end{itemize}
This condition is satisfied when all variables in $g_t(X)$ including 1 are contained in $c_t(X)$, and $\Psi(\cdot)$ is the identity link in the OR model (\ref{OR}).

In the rest of this section, we compare our calibrated methods and several related methods for estimating ATT,
including \citet{Qin2008}, \citet{Hainmueller2012}, \citet{Imai2014}, and \citet{Graham2015}.
The estimators of $\nu^0$ in \citet{Qin2008} and \citet{Graham2015} are in the form
\begin{align*}
\frac{1}{n_1} \sum_{i=1}^n \frac{(1-T_i) \hat \pi(X_i)}{w_i} Y_i ,
\end{align*}
where $\{w_i >0: T_i=0, i=1,\ldots,n\}$ are derived such that, similarly to (\ref{g-cal0})--(\ref{g-cal1}),
\begin{align*}
\sum_{i=1}^n \frac{(1-T_i) \hat \pi(X_i)}{w_i} c_0(X_i) =\sum_{i=1}^n \hat \pi(X_i) c_0(X_i) .
\end{align*}
\citet{Qin2008} studied asymptotic behavior of their estimator under a correctly specified PS model, but did not investigate local efficiency or double robustness
or address how $c_0(X)$ should be specified to gain efficiency or robustness over non-augmented IPW estimators.
For our current setting, \citet{Graham2015} showed that their estimator of $\nu^t$ is locally semiparametric efficient with respect to PS model (\ref{PS}) under Condition (R), and
doubly robust under the following condition:
\begin{enumerate}
\item[(R$^+$)]: Condition R holds, PS model (\ref{PS}) is logistic regression, and
each variable in $c_t(X)$ is a linear combination of $f(X)$ for $t=0,1$. \footnotemark[2]
\end{enumerate}
These results can be related to our results as follows.

\footnotetext[2]{This last condition should be added to condition (b) in Theorem 4 of \cite{Graham2015}.}

\begin{itemize}
\item[(i)] Similarly as discussed after Proposition \ref{prop-reg}, the semiparametric efficiency bound $V^t_{\text{\scriptsize SP}}$ with respect to model (\ref{PS}) coincides with the
nonparametric efficiency bound $V^t_{\text{\scriptsize NP}}$ when model (\ref{PS}) is logistic regression and $\{T-\pi(X)\}\{m_t(X)-\nu^t\}$ is
a linear combination of $\{T-\pi(X)\} f(X)$.
Therefore, under Condition R$^+$, the estimator of \citet{Graham2015} is doubly robust and locally both nonparametric and semiparamtric efficient
(see Proposition~\ref{prop2}).

\item[(ii)] If Condition (R$^+$) holds, then Condition (L) holds and hence $\tilde\pi(X)$ reduces to $\hat\pi(X)$.
In this case, our estimators $\tilde\nu^t_{\text{\scriptsize reg}}$ and $\tilde\nu^t_{\text{\scriptsize lik}}$, while using $\hat\pi(X)$ directly,
are not only doubly robust and locally nonparametric efficient, but also intrinsically efficient among the class of estimator (\ref{nu-class}) with $\tilde\pi(X)$ the same as $\hat\pi(X)$.
The estimator of \citet{Graham2015} can be shown to be asymptotically equivalent, to the first order, to some estimator in class (\ref{nu-class}) under a correctly specified PS model (\ref{PS}).
Therefore, under Condition (R$^+$), our estimators are proved to be asymptotically at least as efficient as the estimator of \citet{Graham2015}
when the PS model (\ref{PS}) is correctly specified but the OR model (\ref{OR}) is misspecified.

\item[(iii)] Our approach can be used to handle the general case where PS model (\ref{PS}) is non-logistic regression (see Appendix~\ref{sec:tech-details}.6), and construct
both AIPW estimators that are doubly robust and locally nonparametric efficient,
but also improved estimators that further achieve intrinsic efficiency.
\end{itemize}

If PS model (\ref{PS}) is logistic regression, then
the methods of \citet{Hainmueller2012} and \citet{Imai2014} seem to use the same estimator of $\nu^0$,
\begin{align*}
\hat \nu^0_{\text{\scriptsize HIR}} = & \frac{1}{n_1} \sum_{i=1}^n (1-T_i) r(X_i;\breve \gamma) Y_i
= \frac{\sum_{i=1}^n (1-T_i) r(X_i;\breve \gamma) Y_i } { \sum_{i=1}^n (1-T_i) r(X_i;\breve \gamma) },
\end{align*}
where $r(X;\gamma) = \pi(X;\gamma)/\{1-\pi(X;\gamma)\}=\exp\{\gamma^\T f(X)\}$ and $\breve\gamma$ is determined from the balancing equation similar to Eq.~(\ref{g-cal0})--(\ref{g-cal1}),
\begin{align}
\sum_{i=1}^n (1-T_i) r(X_i;\gamma) f(X_i) =\sum_{i=1}^n T_i f(X_i) .  \label{HIR-eq}
\end{align}
Eq.~(\ref{HIR-eq}) differs from balancing equations used for ATE estimation in \citet{Imai2014}.
The two expressions of $ \hat \nu^0_{\text{\scriptsize HIR}} $ follow from the fact that
$\sum_{i=1}^n (1-T_i) r(X_i;\gamma)=n_1$ by Eq.~(\ref{HIR-eq}) with $f(X)$ including a constant.
That is, $ \hat \nu^0_{\text{\scriptsize HIR}} $ can be seen as standard IPW estimators:
$\hat \nu^0_{\text{\scriptsize HIR}} = \hat \nu^0_{\text{\scriptsize IPW}}(\breve \pi) = \hat \nu^0_{\text{\scriptsize IPW,ratio}}(\breve \pi)$,
where $\breve \pi(X) = \pi(X;\breve \gamma)$ is substituted for $\hat \pi(X) = \pi(X;\hat\gamma)$ with the MLE $\hat\gamma$.
Under Condition (L), the estimator $\hat \nu^0_{\text{\scriptsize HIR}}$ can be shown to be doubly robust and locally nonparametric efficient (\citealt{Zhao2015}).
However, $\hat \nu^0_{\text{\scriptsize HIR}}$ is not intrinsically efficient and hence, similarly to the estimator of \citet{Graham2015}, not as efficient as our estimators
$\tilde\nu^0_{\text{\scriptsize reg}}$ and $\tilde\nu^0_{\text{\scriptsize lik}}$
when the PS model (\ref{PS}) is correctly specified but the OR model (\ref{OR}) is misspecified.

\section{Simulation studies} \label{simulation}

We conducted two simulation studies to compare the proposed and existing estimators. We present in Appendix~\ref{sec:add-simulation} the results under the simulation settings of \citet{KS2007}
and \citet{McCaffrey2007}. Here we present the results under the simulation settings of \citet{Qin2008} and \citet{Graham2015}.

The simulation setting of \citet{Qin2008} is originally designed in the context of difference-in-differences estimation, but can be equivalently recast for
estimation of ATT as shown in \citet{Graham2015}. Specifically, suppose that the covariate vector, $X=(X_1, X_2)$, is generated as \vspace{-.05in}
\begin{align*}
X_1 \sim N(0,1), \quad X_2|X_1 \sim N(1+0.6X_1,1) .
\end{align*}
The true propensity score is generated as a logistic regression function
\begin{align*}
\pi(X) =P(T=1|X )=\mbox{expit} (\gamma_0^* +\gamma_1^* X_1 +\gamma_2^* X_2),
\end{align*}
where $(\gamma_1^*,\gamma_2^*,\gamma_3^*) = (1.0,0.1,0.1)$, $(1.0,0.2,0.2)$, or $(1.0,0.5,0.5)$, corresponding to increasing selection bias into treatment.
The potential outcomes $(Y^1, Y^0)$ are generated (regardless of $T$ for exogenity) as
\begin{align*}
  Y^1|X,T \sim N \{m_1(X), X_2^2 \}, \quad Y^0|X,T \sim N \{m_0(X), X_2^2\},
\end{align*}
where $m_0(X)$ and $m_1(X)$ are set in two possible ways:
\begin{enumerate}
\item[(i)] LIN-OR: \;$m_1(X)=2+2X_1+2 X_2$, $m_0(X)= 2X_1  + 2 X_2$,
\item[(ii)] QUA-OR: \; $ m_1(X)=2+2X_1^2+3X_2^2 -X_2$, $m_0(X)= 2X_1^2 + 3X_2^2 -X_2 $.
\end{enumerate}
It is easily shown that the true value of ATT is always 2, because the regression functions $m_0(X)$ and $m_1(X)$ are parallel to each other.

For estimation of ATT, consider an outcome regression model (\ref{OR})
with the identity link $\Psi(\cdot)$ and the regressor vector
$g_0(X) = g_1(X) =  (1,X_1,X_2)^\T$ or $(1, X_1^2, X_2^2)^\T$, corresponding to a linear or quadratic OR model.
Under the LIN-OR setting, the linear or quadratic OR model is, respectively, correctly specified or misspecified.
Under the QUA-OR setting, both of the OR models are misspecified, but the quadratic OR model is misspecified to a lesser degree.
Similarly, consider a propensity score model (\ref{PS}) with the logistic link $\Pi(\cdot)$ and the regressor vector $f(X) = (1,X_1,X_2)^\T$ or $(1, X_1^2, X_2^2)^\T$,
corresponding to a logistic linear or quadratic PS model, which is, respectively, correctly specified or misspecified.

We implemented the following estimators of ATT: \vspace{-.05in}
\begin{itemize}\addtolength{\itemsep}{-.1in}
\item (OR) $\hat \mu^1_{\mbox{\scriptsize OR}} - \hat \mu^1_{\mbox{\scriptsize OR}}$;

\item (IPW) $\hat \mu^1_{\mbox{\scriptsize NP}} - \hat \mu^0_{\mbox{\scriptsize IPW}}(\hat\pi)$, (IPW.ratio) $\hat \mu^1_{\mbox{\scriptsize NP}} - \hat \mu^0_{\mbox{\scriptsize IPW,ratio}}(\hat\pi)$;

\item (AIPW) $\hat \mu^1_{\mbox{\scriptsize NP}} - \hat \mu^0_{\mbox{\scriptsize NP}}(\hat\pi, \hat m_0)$;

\item (LIK) $\tilde \mu^1_{\mbox{\scriptsize lik}} - \tilde \mu^0_{\mbox{\scriptsize lik}}$, (LIK2) $\tilde \mu^1_{\mbox{\scriptsize lik2}} - \tilde \mu^0_{\mbox{\scriptsize lik2}}$;

\item (HIR) $ \hat \mu^1_{\mbox{\scriptsize NP}} - \hat \mu^0_{\mbox{\scriptsize IPW}}(\breve \pi)$, (AIPW.HIR) $ \hat \mu^1_{\mbox{\scriptsize NP}} - \hat \mu^0_{\mbox{\scriptsize NP}}(\breve \pi, \hat m_0)$.
\end{itemize} \vspace{-.05in}

\begin{figure}
\begin{tabular}{c}
\includegraphics[width=6in, height=4in]{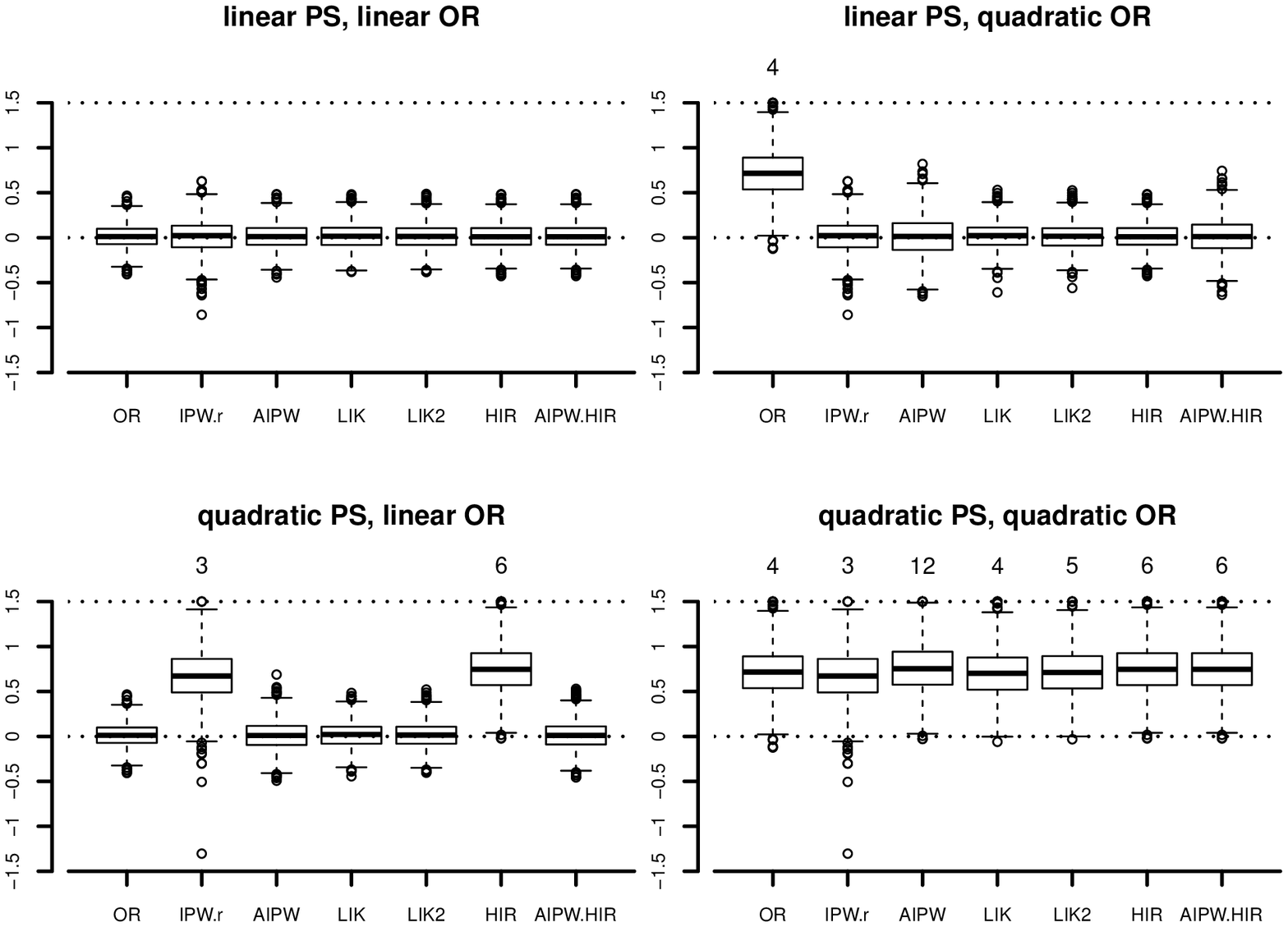}
\end{tabular} \vspace{-.4in}
\caption{\small Boxplots of estimates minus the truth under LIN-OR setting with $(\gamma_1^*,\gamma_2^*,\gamma_3^*)=(1.0,0.2,0.2)$.
All values are censored within the range of $y$-axis, and the
number of values that lie outside the range are indicated next to the lower and upper limits of $y$-axis. }
\label{plot:LIN-2}

\vspace{.4in}
\begin{tabular}{c}
\includegraphics[width=6in, height=4in]{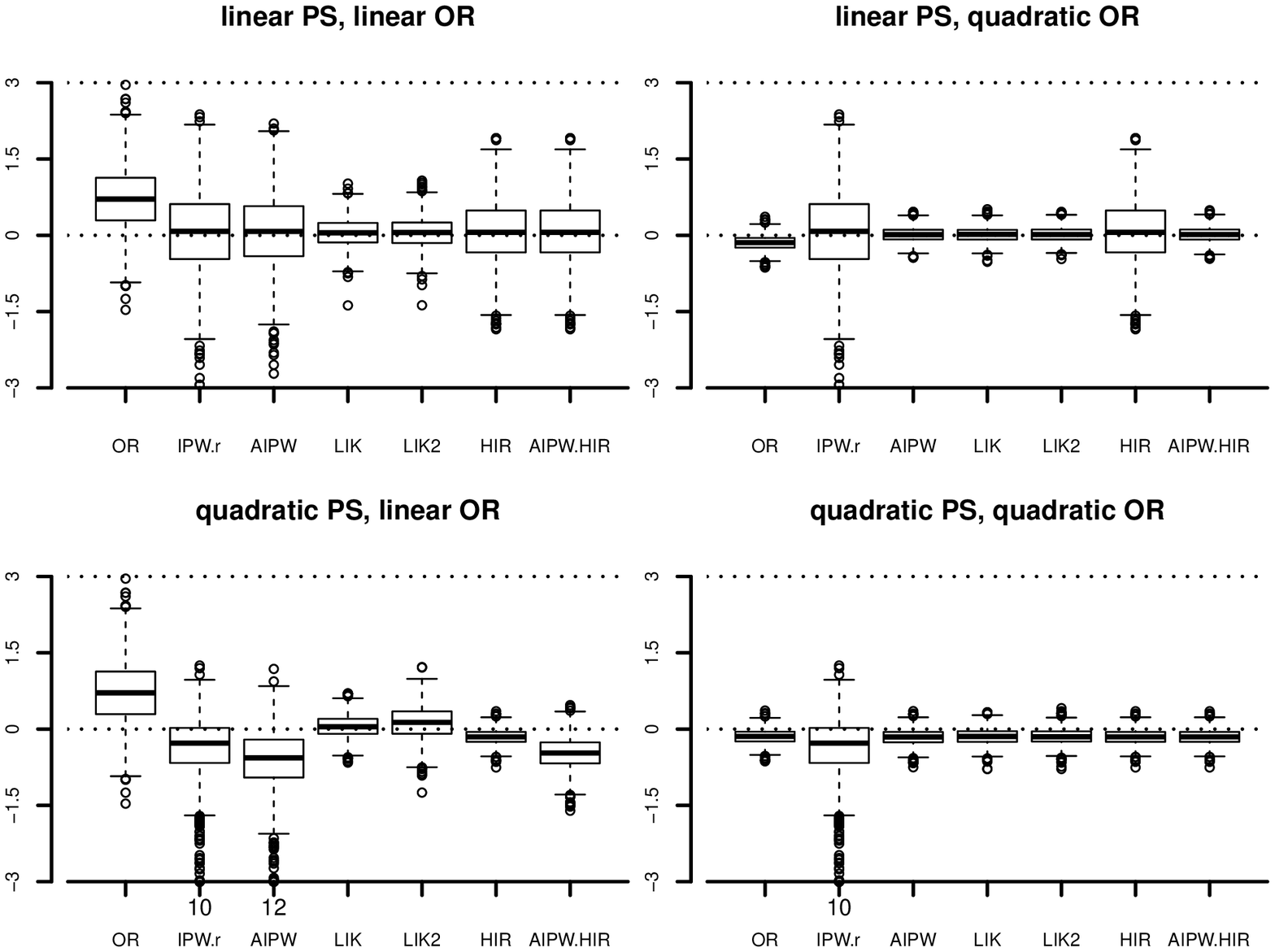}
\end{tabular} \vspace{-.3in}
\caption{\small Boxplots of estimates minus the truth under QUA-OR setting with $(\gamma_1^*,\gamma_2^*,\gamma_3^*)=(1.0,0.2,0.2)$}
\label{plot:QUA-2}
\end{figure}

Table~\ref{table:Qin_Zhang} and Figures \ref{plot:LIN-2}-\ref{plot:QUA-2} present the results for these estimators, from $1000$ Monte Carlo samples of size $n=1000$, under the PS setting with moderate selection bias,
$(\gamma_1^*,\gamma_2^*,\gamma_3^*) = (1.0,0.2,0.2)$.
In addition, results are reproduced under the same setting for two estimators in \citet{Qin2008} and \citet{Graham2015}.
See the Appendix~\ref{sec:add-simulation} for the results under the other two PS settings.

The following remarks can be drawn on the comparisons of various estimators. First, the OR estimator is approximately unbiased only when the OR model used is correctly specified (i.e., linear OR model under LIN-OR setting).

Second, the IPW and IPW.ratio estimators are approximately unbiased only when the PS model used is correctly specified (i.e., linear PS model), but they have large variances with noticeably outlying values.

\begin{table}[t]
\caption{ Qin--Zhang simulation results with $(\gamma_1^*,\gamma_2^*,\gamma_3^*)=(1.0,0.2,0.2)$} \vspace{-.1in}
\scriptsize
\label{table:Qin_Zhang}
\begin{center}
\def\arraystretch{0.8}%
\begin{tabular}{lccccccccc}
\hline\hline
\noalign{\medskip}
 Models   & OR       & IPW.r & AIPW     & LIK      & LIK2     & HIR      & AIPW.HIR       & EL  & AST     \\
\noalign{\medskip}\hline
\noalign{\medskip}
& \multicolumn{9}{c}{\bf Data generated under LIN-OR setting} \\
\noalign{\medskip}
linear PS,    & 0.0120   & 0.0147    & 0.0125   & 0.0118   & 0.0120   & 0.0123   & 0.0123   & 0.0031 & -0.0065    \\
linear OR     & (0.0175) & (0.0358)  & (0.0201) & (0.0209) & (0.0208) & (0.0200) & (0.0200) & (0.0275) & (0.0261)  \\
\noalign{\medskip}
linear PS,    & 0.7170   & 0.0147    & 0.0139   & 0.0168   & 0.0132   & 0.0123   & 0.0122   & -0.0009 & -0.0039    \\
quadratic OR & (0.0767) & (0.0358)  & (0.0500) & (0.0221) & (0.0225) & (0.0200) & (0.0431) & (0.0306) & (0.0371)  \\
\noalign{\medskip}
quadratic PS, & 0.0120   & 0.6655    & 0.0106   & 0.0125   & 0.0105   & 0.7501   & 0.0114   & $\cdots$         & $\cdots$         \\
linear OR     & (0.0175) & (0.0878)  & (0.0269) & (0.0212) & (0.0221) & (0.0756) & (0.0244) &  $\cdots$        & $\cdots$        \\
\noalign{\medskip}
quadratic PS, & 0.7170   & 0.6655    & 0.7644   & 0.7023   & 0.7120   & 0.7501   & 0.7501   &  $\cdots$        &  $\cdots$        \\
quadratic OR & (0.0767) & (0.0878)  & (0.0828) & (0.0746) & (0.0746) & (0.0756) & (0.0756) &  $\cdots$        &  $\cdots$        \\
\noalign{\medskip}\hline
\noalign{\medskip}
& \multicolumn{9}{c}{\bf Data generated under QUA-OR setting} \\
\noalign{\medskip}
linear PS,    & 0.7028   & 0.0414    & 0.0500   & 0.0471   & 0.0522   & 0.0553   & 0.0553   & 0.0477 & 0.0787      \\
linear OR    & (0.4176) & (0.6407)  & (0.5201) & (0.0796) & (0.0946) & (0.3683) & (0.3683) & (0.1227) & (0.3620)  \\
\noalign{\medskip}
linear PS,    & -0.1473  & 0.0414    & 0.0142   & 0.0120   & 0.0138   & 0.0553   & 0.0144 & 0.0028  & 0.0078      \\
quadratic OR  & (0.0238) & (0.6407)  & (0.0216) & (0.0224) & (0.0221) & (0.3683) & (0.0223)  & (0.0309) & (0.0218) \\
\noalign{\medskip}
quadratic PS, & 0.7028   & -0.4155   & -0.6468  & 0.0549   & 0.1256   & -0.1554  & -0.4657  & $\cdots$         & $\cdots$         \\
linear OR     & (0.4176) & (0.6381)  & (0.6286) & (0.0485) & (0.1044) & (0.0249) & (0.1021) & $\cdots$        & $\cdots$         \\
\noalign{\medskip}
quadratic PS, & -0.1473  & -0.4155   & -0.1599  & -0.1465  & -0.1493  & -0.1554  & -0.1554  &$\cdots$          &$\cdots$          \\
quadratic OR  & (0.0238) & (0.6381)  & (0.0263) & (0.0272) & (0.0258) & (0.0249) & (0.0249) &$\cdots$          &$\cdots$         \\
 \noalign{\medskip}
         \hline\hline
\end{tabular}
\end{center}
Note: In the upper rows are the Monte Carlo biases ($=$ means$-2$), and in the brackets are the corresponding Monte Carlo variances. EL: \citet{Qin2008} and AST: \citet{Graham2015}.
\end{table}

Third, the HIR estimator is approximately unbiased when the PS model is correctly specified, but becomes biased when the PS model is misspecified and even when the OR model is correctly specified (for example, quadratic PS model and linear OR model under LIN-OR setting). The HIR estimator is not doubly robust, because Condition L is not satisfied in this situation.

Fourth, the four estimators, AIPW, LIK, LIK2, and AIPW.HIR are doubly robust: they are approximately unbiased when either the PS model is correctly specified (i.e., linear PS model)
or the OR model is correctly specified (i.e., linear OR model under LIN-OR setting).
In accordance with local efficiency, these estimators have similar variances to each other when both the PS and OR models are correctly specified.
But LIK and LIK2 have smaller variances, sometimes substantially so, than AIPW and AIPW.HIR estimators when the PS model is correctly specified but the OR model is misspecified.
For example, for linear PS model and linear OR model under QUA-OR setting,
the variance of LIK is smaller than that of AIPW by a factor of $0.52/0.08 = 6.5$ and that of AIPW.HIR by a factor of $0.37/0.08 \approx 4.6$.
Such differences are supported by our theoretical results on intrinsic efficiency.

Fifth, in contrast with AIPW and AIPW.HIR, the LIK estimator appears to be approximately unbiased when the quadratic PS model and linear OR model are used under the QUA-OR setting (hence both PS and OR models are misspecified). This behavior is not indicated by general theory, but can be explained by the fact that  even though the
PS model (\ref{PS}) is misspecified, the augmented PS model (\ref{augPS}) happens to be correctly specified in this case:
$\{\hat m_0(X), \hat m_1(X)\}$ provide exactly the correct regressors $(X_1,X_2)$ up to linear transformation.

Finally, we compare our likelihood estimators with the estimators in \citet{Qin2008} and \citet{Graham2015}
when the PS model is correctly specified (i.e., linear PS model). Results for a misspecified PS model were not available in these previous simulation studies.
Similarly as in the comparisons with AIPW and AIPW.HIR, our likelihood estimators have smaller variances than those in Qin \& Zhang and Graham et al.\
when the PS model is correctly specified but the OR model is misspecified.
For example, for linear PS model and linear OR model under QUA-OR setting,
the variance of LIK is smaller than that of Qin \& Zhang by a factor of $0.12/0.08 = 1.5$ and that of Graham et al.\ by a factor of $0.36/0.08 = 4.5$.
Another interesting observation is that when the OR model is also correctly specified or approximately so,
our likelihood estimators and Graham et al.\ have similar variances, but smaller than that of Qin \& Zhang estimator,
indicating a lack of local efficiency for the latter estimator.
For example, the factor of efficiency gain is $.031/0.022 \approx 1.4$ for linear PS model and quadratic OR model under the QUA-OR setting.

\section{Analysis of LaLonde data} \label{LaLonde}

NSW (``National Supported Work Demonstration") is a randomized job training program implemented in 1970s to provide work experience for
individuals who had economic and social disadvantages. The randomized experiment provides
benchmark estimates of average treatment effects. To study econometric methods for program evaluation with
non-experimental data, \citet{LaLonde1986} constructed an
observational study by replacing the data from the experimental control group with survey data from
either Current Population Survey (CPS) or the Panel Study of Income Dynamics (PSID).
The question of interest is how well the experimental benchmark estimates of average treatment effects can be recovered by
econometric methods when applied to such composite observational studies.
\citet{LaLonde1986} showed that many commonly used methods failed to replicate the experimental results.

Analysis of LaLonde's composite data has since been extensively discussed in the evaluation and causal inference literature. \citet{Dehejia1999, Dehejia2002} obtained effect estimates that have low biases from
the experimental benchmark, while applying propensity score matching methods to a particular subsample of LaLonde's original data.
\citet{Smith2005a} raised the criticism that the propensity score matching estimates are highly sensitive to
both the analysis sample used and the specification of propensity score models. They calculated direct estimates of the bias by
applying matching to the experimental control group and a non-experimental comparison group (either CPS or PSID),
whereas LaLonde and Dehejia \& Wahba calculated the bias by applying matching to the experimental treatment group and a non-experimental comparison group
and then comparing the resulting estimate to the experimental benchmark.
See \citet{Diamond2013}, \citet{Hainmueller2012}, and \citet{Imai2014}, among others, for more recent analyses.

We investigate the performances of the proposed and existing estimators for analyzing LaLonde's original composite data.
Specifically, we apply various estimators of ATT as listed in Section \ref{simulation} in the following analyses:
\begin{itemize}
\item Analysis (i): NSW experimental treatment group is combined with either CPS or PSID non-experimental comparison group for effect estimation or, equivalently, for bias
estimation by subtracting the experimental benchmark from all effect estimates;

\item Analysis (ii): NSW experimental control group is combined with either CPS or PSID non-experimental comparison group for bias estimation.
\end{itemize}
For each application, we consider two possible PS models and two possible OR models, as specified in Table \ref{model}.
The quadratic PS model differs, only by a few terms, from the PS model obtained in an iterative
model-building approach by \citet{Dehejia2002} for analyzing NSW$+$CPS or NSW$+$PSID composite data.

For propensity score estimation, we use either the experimental treatment group in (i)
or experimental control group in (ii) as treated observations ($T=1$) and the non-experimental comparison group as
untreated observations ($T=0$).
This strategy is in line with \citet{LaLonde1986} and \citet{Dehejia1999, Dehejia2002}, but differs from
\citet{Smith2005a} and \citet{Imai2014}. In the latter articles, both the experimental treatment and control groups
are used as treated observations ($T=1$) when estimating propensity scores, but then
either the experimental treatment or control group is used in, respectively, effect or bias estimation.
This scheme does not mimic the practical situation of econometric analysis where a single dataset is used, and
may not even be desirable as discussed in \citet{Dehejia2005b}.

Before turning to our results, we provide some remarks to explain how the relative performances of estimators will be assessed from such
empirical results.
First, as discussed in \citet{Dehejia2005a} in response to \citet{Smith2005a}, applications of propensity score
methods should involve searching for a propensity score model that leads to balance of covariates between treatment groups.
The approach suggested in \citet{Rosenbaum1984} and \citet{Dehejia1999, Dehejia2002} is conceptually useful but
leaves open the issue of how PS models can actually be built to achieve covariate balance.
Alternatively, simple PS models such as in Table \ref{model} may often be used in applied research.
Second, \citet{Smith2005b} presented additional analyses in response to \citet{Dehejia2005a} to argue that the low-bias matching estimates in \citet{Dehejia1999, Dehejia2002}
are sensitive not only in regard to the sample and propensity score specification as shown in \cite{Smith2005a},
but also, among other factors, to whether the propensity score and subsequently the bias are estimated using the experimental treatment or control group, as
in Analyses (i) and (ii) described above.
Third, a criterion typically used in previous analyses of LaLonde data is that
the bias estimates should be close to 0 for a good method.
But the true bias can be 0 only when the exogeneity assumption (A1) holds on the composite sample, i.e.,
potential outcomes are influenced by the measured covariates in the experimental sample in the same way as in the comparison sample (CPS or PSID).
Nevertheless, the difference between the two bias estimates from Analyses (i) and (ii), as examined in \citet{Smith2005b}, can be shown to be 0 (up to random variation) even when
the exogeneity assumption (A1) fails on the composite sample. See Appendix~\ref{sec:tech-details}.7 for details.
By all the preceding considerations, we will assess the relative performances of estimators
mainly in terms of how close the two bias estimates from Analyses (i) and (ii) are to each other,
depending on PS and OR models used.

\begin{table}
\caption{PS and OR models for LaLonde data} \label{model}
\scriptsize
\begin{center}
\begin{tabular*}{.9\textwidth}{ll} \hline\hline
      Name    & Regressors $f(X)$ in PS model or $g(X)$ in OR model  \\\noalign{\smallskip}\hline\noalign{\medskip}
      Linear   & $(1, age, school, black, hisp, married, nodegr, re74, re75, u74, u75)$  \\
      Quadratic & $(1, age, school, black, hisp, married, nodegr, re74, re75, u74, u75, age^2, school^2, re74^2, re75^2)$ \\ \hline
\end{tabular*} \\[1ex]
\parbox{.9\textwidth}{Note: The variables are defined as in Table 2 of \citet{Dehejia2002}.
The PS model is $T|X \sim f(X)$ with logistic link. The OR model is $Y | (T=t,X) \sim t+ g(X)$ with identity link.}
\end{center}

\caption{Bootstrap results from Analyses (i) and (ii) on NSW$+$PSID composite data}
\scriptsize
\label{table:PSID}
\begin{center}
\def\arraystretch{0.8}%
\begin{tabular}{lcccccccc}
\hline\hline
\noalign{\medskip}
          &           & OR    & IPW.ratio & AIPW  & LIK2  & HIR   & AIPW.HIR \\
          \noalign{\medskip}\hline\noalign{\medskip}
Linear PS, Linear OR & Treatment Effect     & -1690 & 901       & 1109  & 555   & 475   & 475      \\
             &            & (650) & (781)     & (852) & (616) & (598) & (598)    \\
           \noalign{\medskip}
           & Evaluation Bias       & -2941 & -6        & 337   & -211  & -118  & -118     \\
           &            & (636) & (764)     & (815) & (523) & (496) & (496)    \\
          \noalign{\medskip}
           & Difference & 1251  & 907       & 772   & 765   & 594   & 594      \\
           &            & (590) & (669)     & (757) & (563) & (549) & (549)   \\
\noalign{\medskip}\hline\noalign{\medskip}
Linear PS, Quadratic OR & Treatment Effect     & -1577 & 901       & 613   & 378   & 475   & 441      \\
                &            & (803) & (781)     & (729) & (613) & (598) & (601)    \\
            \noalign{\medskip}
           & Evaluation Bias       & -2674 & -6        & -7    & -365  & -118  & -177     \\
           &            & (807) & (764)     & (653) & (529) & (496) & (501)    \\
          \noalign{\medskip}
           & Difference & 1096  & 907       & 620   & 743   & 594   & 618      \\
           &            & (610) & (669)     & (641) & (560) & (549) & (553) \\
\noalign{\medskip}\hline \noalign{\medskip}
Quadratic PS, Linear OR  & Treatment Effect     & -1690 & 901   & 1216  & 573   & 393   & 477   \\
                        &            & (650) & (799) & (896) & (623) & (606) & (601) \\
             \noalign{\medskip}
              & Evaluation Bias       & -2941 & -9    & 451   & -236  & -254  & -142  \\
              &            & (636) & (791) & (862) & (537) & (505) & (498) \\
             \noalign{\medskip}
              & Difference & 1251  & 910   & 765   & 809   & 647   & 618   \\
              &            & (590) & (685) & (804) & (560) & (556) & (547) \\
\noalign{\medskip}\hline \noalign{\medskip}
Quadratic PS, Quadratic OR  & Treatment Effect     & -1577 & 901   & 571   & 332   & 393   & 393   \\
                     &            & (803) & (799) & (742) & (618) & (606) & (606) \\
            \noalign{\medskip}
              &  Evaluation Bias       & -2674 & -9    & -27   & -429  & -254  & -254  \\
              &            & (807) & (791) & (666) & (531) & (505) & (505) \\
            \noalign{\medskip}
              & Difference & 1096  & 910   & 598   & 761   & 647   & 647   \\
              &            & (610) & (685) & (651) & (560) & (556) & (556)\\
\noalign{\medskip}
         \hline
\end{tabular}
\end{center}
Note: In the upper rows are the bootstrap means, and in the brackets are the corresponding bootstrap standard errors.
Treatment Effect is obtained from Analysis (i), and Evaluation Bias from Analysis (ii). The difference is to be
compared with the experimental benchmark \$886 with standard error \$488.
To tackle numerical non-convergence when computing estimates during bootstrapping,
the following procedure is used. We performed Principle Component Analysis to the regressors from the composite data, NSW (treatment$+$control) $+$ PSID, and dropped
principle components whose sample variances are less than $(0.3)^2$ of the component with the largest sample variance.
Then we resampled the entire composite dataset and conducted Analyses (i) and (ii) on each bootstrap sample.
\end{table}

\begin{figure}
\begin{tabular}{c}
\includegraphics[width=6in, height=4in]{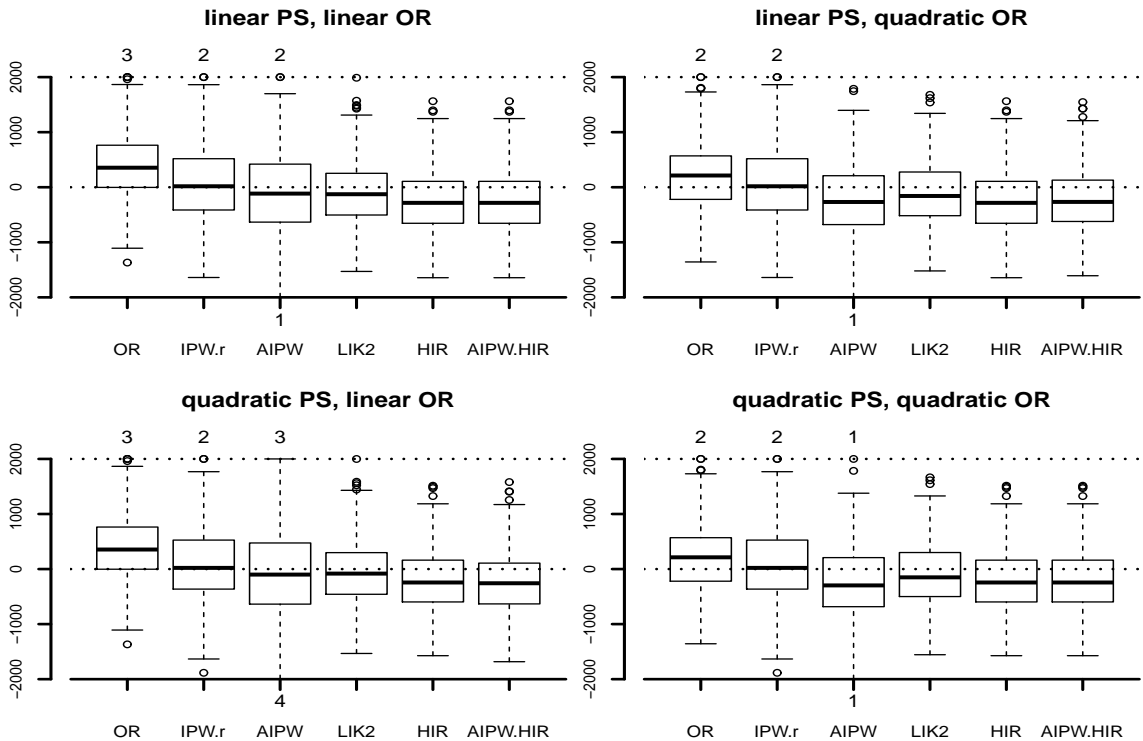} \vspace{-.1in}
\end{tabular}
\caption{\small Bootstrap boxplots of differences of bias estimates from Analyses (i) and (ii) on NSW$+$PSID composite data.
All values are censored within the range of $y$-axis, with
number of values laying outside indicated next to the lower and upper limits of $y$-axis.}
\label{plot:PSID}
\end{figure}

Table~\ref{table:PSID} and Figure~\ref{plot:PSID} present the results from Analyses (i) and (ii) for various estimators as listed in Section \ref{simulation},
based on 500 bootstrap samples of the NSW$+$PSID composite data.
See the Appendix~\ref{sec:add-analysis} for the results on the NSW$+$CPS composite data, where the relative performances of
estimators are more similar to each other than on the NSW$+$PSID composite data.

Among all estimators studied, the IPW.ratio estimator yields point estimates of effect closest to the experimental benchmark \$886 and
estimates of bias closest to 0 from Analyses (i) and (ii), using either the linear or quadratic PS model.
But the bootstrap variances for IPW.ratio are among the highest for all estimators studied.
Although such point estimates of effect are much closer to the experimental benchmark than various previously obtained estimates on LaLonde NSW$+$PSID data (e.g., \citealt{Diamond2013, Imai2014}), these results may not present real evidence for any advantage of IPW.ratio for reasons discussed above.

In terms of how close the difference between effect and bias estimates is to the experimental benchmark (i.e., how close the two bias estimates are close to each other) from
Analyses (i) and (ii), the estimators IPW.ratio, AIPW, and LIK2 yield the most accurate point estimates among all estimators studied, regardless of PS and OR models used.
But the bootstrap variances for LIK2 are much smaller than those of IPW.rato and AIPW.
As explained above, these results present strong evidence for the advantage of the proposed estimator LIK2.

\vspace{-.1in}
\section{Conclusion} \label{conclusion}

We study the problem of estimating ATTs from observational data and make the following contributions.
In spite of non-ancillarity of the propensity score, we show how efficient influence functions from semiparametric theory can be harnessed to derive
AIPW estimators that are locally efficient and doubly robust.
Furthermore, we develop calibrated regression and likelihood estimators that achieve desirable properties in efficiency and boundedness beyond
local efficiency and double robustness.
From two simulation studies and reanalysis of \citet{LaLonde1986} data,
the proposed methods perform overall the best compared with various existing methods.

The ideas developed in this article can be extended in various directions.
For example, it is interesting to consider marginal and nested structural models for ATTs in subpopulations, i.e.,
$E(Y^1- Y^0 | T=1, V)$ with some selected covariates $V$, and develop calibrated regression and likelihood estimators.
Moreover, as seen from \citet{Graham2015},
estimation of ATT can be put in a broader class of data combination problems.
The methods developed here can be extended in that direction.

\setlength{\bibsep}{4pt}

\bibliographystyle{myapa}
    \bibliography{thesis}

\begin{thebibliography}{}

\bibitem[\protect\astroncite{Kang \& Schafer}{2007}]{KS2007}
Kang, J. D.~Y. and Schafer, J.~L. (2007).
\newblock Demystifying double robustness: a comparison of alternative
  strategies for estimating a population mean from incomplete data (with
  discussions).
\newblock {\em Statistical Science}, 22:523--539.

\bibitem[\protect\astroncite{LaLonde}{1986}]{LaLonde1986}
LaLonde, R.~J. (1986).
\newblock Evaluating the econometric evaluations of training programs with
  experimental data.
\newblock {\em The American Economic Review}, 76:604--620.

\bibitem[\protect\astroncite{McCaffrey et~al.}{2007}]{McCaffrey2007}
McCaffrey, D.~F., Ridgeway, G., and Morral, A.~R. (2007).
\newblock Comment: Demystifying double robustness: A comparison of alternative
  strategies for estimating a population mean from incomplete data.
\newblock {\em Statistical Science}, 22:540--543.

\bibitem[\protect\astroncite{Qin \& Lawless}{1994}]{Qinlawless1994}
Qin, J. and Lawless, J. (1994).
\newblock Empirical likelihood and general estimating equations.
\newblock {\em The Annals of Statistics}, 22:300--325.

\bibitem[\protect\astroncite{Qin \& Zhang}{2008}]{Qin2008}
Qin, J. and Zhang, B. (2008).
\newblock Empirical-likelihood-based difference-in-differences estimators.
\newblock {\em Journal of the Royal Statistical Society: Series B (Statistical
  Methodology)}, 70:329--349.

\bibitem[\protect\astroncite{Robins}{1999}]{Robins1999}
Robins, J.~M. (1999).
\newblock Association, causation, and marginal structural models.
\newblock {\em Synthese}, 121:151--179.

\bibitem[\protect\astroncite{Robins et~al.}{1994}]{Robins1994}
Robins, J.~M., Rotnitzky, A., and Zhao, L.~P. (1994).
\newblock Estimation of regression coefficients when some regressors are not
  always observed.
\newblock {\em Journal of the American Statistical Association}, 89:846--866.

\bibitem[\protect\astroncite{Tan}{2006}]{Tan2006}
Tan, Z. (2006).
\newblock A distributonal approach for causal inference using propensity score.
\newblock {\em Journal of the American Statistical Association},
  101:1619--1637.

\bibitem[\protect\astroncite{Tan}{2010}]{Tan2010}
Tan, Z. (2010).
\newblock Bounded, efficient and doubly robust estimation with inverse
  weighting.
\newblock {\em Biometrika}, 97:661--682.

\bibitem[\protect\astroncite{White}{1982}]{White1982}
White, H. (1982).
\newblock Maximum likelihood estimation of misspecified models.
\newblock {\em Econometrica}, 50:1--25.

\end{thebibliography}


\begin{thebibliography}{}

\bibitem[\protect\astroncite{Abadie}{2005}]{Abadie2005}
Abadie, A. (2005).
\newblock Semiparametric difference-in-differences estimators.
\newblock {\em Review of Economic Studies}, 72:1--19.

\bibitem[\protect\astroncite{Cao et~al.}{2009}]{Cao2009}
Cao, W., Tsiatis, A.~A., and Davidian, M. (2009).
\newblock Improving efficiency and robustness of the doubly robust estimator
  for a population mean with incomplete data.
\newblock {\em Biometrika}, 96:723--734.

\bibitem[\protect\astroncite{Chen et~al.}{2008}]{Chen2008}
Chen, X., Hong, H., and Tarozzi, A. (2008).
\newblock Semiparametric efficiency in {GMM} models with auxiliary data.
\newblock {\em The Annals of Statistics}, 36:808--843.

\bibitem[\protect\astroncite{Cochran}{1977}]{Cochran1977}
Cochran, W.~G. (1977).
\newblock {\em Sampling Techniques}.
\newblock John Wiley \& Sons, 3 edition.

\bibitem[\protect\astroncite{Dehejia}{2005a}]{Dehejia2005a}
Dehejia, R. (2005a).
\newblock {\GG{20051}} {P}ractical propensity score matching: {A} reply to
  {Smith} and {Todd}.
\newblock {\em Journal of Econometrics}, 125:355--364.

\bibitem[\protect\astroncite{Dehejia}{2005b}]{Dehejia2005b}
Dehejia, R. (2005b).
\newblock {\GG{20052}} {D}oes matching overcome {LaLonde's} critique of
  nonexperimental estimators? {A} postscript.
\newblock Manuscript.

\bibitem[\protect\astroncite{Dehejia \& Wahba}{1999}]{Dehejia1999}
Dehejia, R. and Wahba, S. (1999).
\newblock Causal effects in nonexperimental studies: Reevaluating the
  evaluation of training programs.
\newblock {\em Journal of the American Statistical Association}, 94:1053--1062.

\bibitem[\protect\astroncite{Dehejia \& Wahba}{2002}]{Dehejia2002}
Dehejia, R. and Wahba, S. (2002).
\newblock Propensity score-matching methods for nonexperimental causal studies.
\newblock {\em The Review of Economics and Statistics}, 84:151--161.

\bibitem[\protect\astroncite{Deville \& Sarndal}{1992}]{Deville1992}
Deville, J.-C. and Sarndal, C.-E. (1992).
\newblock Calibration estimators in survey sampling.
\newblock {\em Journal of the American Statistical Association}, 87:376--382.

\bibitem[\protect\astroncite{Diamond \& Sekhon}{2013}]{Diamond2013}
Diamond, A. and Sekhon, J.~S. (2013).
\newblock Genetic matching for estimating causal effects: A general
  multivariate matching method for achieving balance in observational studies.
\newblock {\em The Review of Economics and Statistics}, 95:932--945.

\bibitem[\protect\astroncite{Graham et~al.}{2016}]{Graham2015}
Graham, B.~S., de~Xavier~Pinto, C.~C., and Egel, D. (2016).
\newblock Efficient estimation of data combination models by the method of
  auxiliary-to-study tilting ({AST}).
\newblock {\em Journal of Business and Economic Statistics}, 34:288--301.

\bibitem[\protect\astroncite{Hahn}{1998}]{Hahn1998}
Hahn, J. (1998).
\newblock On the role of the propensity score in efficient semiparametric
  estimation of average treatment effects.
\newblock {\em Econometrica}, 66:315--331.

\bibitem[\protect\astroncite{Hainmueller}{2012}]{Hainmueller2012}
Hainmueller, J. (2012).
\newblock Entropy balancing for causal effects: A multivariate reweighting
  method to produce balanced samples in observational studies.
\newblock {\em Political Analysis}, 20:25--46.

\bibitem[\protect\astroncite{Hammersley \& Handscomb}{1964}]{Ham1964}
Hammersley, J.~M. and Handscomb, D.~C. (1964).
\newblock {\em Monte Carlo Methods}.
\newblock Methuen.

\bibitem[\protect\astroncite{Heckman et~al.}{1997}]{Heckman1997}
Heckman, J.~J., LaLonde, R.~J., and Smith, J.~A. (1997).
\newblock Matching as an econometric evaluation estimator: Evidence from
  evaluating a job training program.
\newblock {\em Review of Economic Studies}, 64:605--654.

\bibitem[\protect\astroncite{Heckman \& Robb}{1985}]{Heckman1985}
Heckman, J.~J. and Robb, R. (1985).
\newblock Alternative methods for evaluating the impact of interventions.
\newblock In Heckman, J.~J. and Singer, B., editors, {\em Longitudinal Analysis
  of Labor Market Data}, pages 156--246. Cambridge University Press, New York.

\bibitem[\protect\astroncite{Hirano et~al.}{2003}]{Hirano2003}
Hirano, K., Imbens, G.~W., and Ridder, G. (2003).
\newblock Efficient estimation of average treatment effects using the estimated
  propensity score.
\newblock {\em Econometrica}, 71:1161--1189.

\bibitem[\protect\astroncite{Imai \& Ratkovic}{2014}]{Imai2014}
Imai, K. and Ratkovic, M. (2014).
\newblock Covariate balancing propensity score.
\newblock {\em Journal of the Royal Statistical Society}, 76:243--263.

\bibitem[\protect\astroncite{Imbens}{2004}]{Imbens2004}
Imbens, G.~W. (2004).
\newblock Nonparametric estimation of average treatment effects under
  exogeneity: A review.
\newblock {\em Review of Economics and Statistics}, 86:4--29.

\bibitem[\protect\astroncite{Kang \& Schafer}{2007}]{KS2007}
Kang, J. D.~Y. and Schafer, J.~L. (2007).
\newblock Demystifying double robustness: a comparison of alternative
  strategies for estimating a population mean from incomplete data (with
  discussions).
\newblock {\em Statistical Science}, 22:523--539.

\bibitem[\protect\astroncite{LaLonde}{1986}]{LaLonde1986}
LaLonde, R.~J. (1986).
\newblock Evaluating the econometric evaluations of training programs with
  experimental data.
\newblock {\em The American Economic Review}, 76:604--620.

\bibitem[\protect\astroncite{McCaffrey et~al.}{2004}]{McCaffrey2004}
McCaffrey, D.~F., Ridgeway, G., and Morral, A.~R. (2004).
\newblock Propensity score estimation with boosted regression for evaluating
  causal effects in observational studies.
\newblock {\em Psychological Methods}, 9:403--425.

\bibitem[\protect\astroncite{McCaffrey et~al.}{2007}]{McCaffrey2007}
McCaffrey, D.~F., Ridgeway, G., and Morral, A.~R. (2007).
\newblock Comment: Demystifying double robustness: A comparison of alternative
  strategies for estimating a population mean from incomplete data.
\newblock {\em Statistical Science}, 22:540--543.

\bibitem[\protect\astroncite{Newey}{1990}]{Newey1990}
Newey, W.~K. (1990).
\newblock Semiparametric efficiency bounds.
\newblock {\em Journal of Applied Econometrics}, 5:99--135.

\bibitem[\protect\astroncite{Neyman}{1923}]{Neyman1923}
Neyman, J. (1923).
\newblock On the application of probability theory to agricultural experiments:
  Essay on principles, {Section} 9.
\newblock {\em Statistical Science}, 5:465--472.

\bibitem[\protect\astroncite{Owen}{2001}]{Owen2001}
Owen, A.~B. (2001).
\newblock {\em Empirical Likelihood}.
\newblock Chapman \& Hall/CRC.

\bibitem[\protect\astroncite{Qin \& Zhang}{2008}]{Qin2008}
Qin, J. and Zhang, B. (2008).
\newblock Empirical-likelihood-based difference-in-differences estimators.
\newblock {\em Journal of the Royal Statistical Society: Series B (Statistical
  Methodology)}, 70:329--349.

\bibitem[\protect\astroncite{Robins \& Ritov}{1997}]{Robins1997}
Robins, J.~M. and Ritov, Y. (1997).
\newblock Toward a curse of dimensionality appropriate ({CODA}) asymptotic
  theory for semi-parametric models.
\newblock {\em Statistics in Medicine}, 16:285--319.

\bibitem[\protect\astroncite{Robins \& Rotnitzky}{2001}]{Robins2001}
Robins, J.~M. and Rotnitzky, A. (2001).
\newblock Comment on the {Bickel} and {Kwon} article, `{I}nference for
  semiparametric models: Some questions and an answer'.
\newblock {\em Statistica Sinica}, 11:920--936.

\bibitem[\protect\astroncite{Robins et~al.}{1994}]{Robins1994}
Robins, J.~M., Rotnitzky, A., and Zhao, L.~P. (1994).
\newblock Estimation of regression coefficients when some regressors are not
  always observed.
\newblock {\em Journal of the American Statistical Association}, 89:846--866.

\bibitem[\protect\astroncite{Rosenbaum \& Rubin}{1983}]{Rosenbaum1983}
Rosenbaum, P.~R. and Rubin, D.~B. (1983).
\newblock The central role of the propensity score in observational studies for
  causal effects.
\newblock {\em Biometrika}, 70:41--55.

\bibitem[\protect\astroncite{Rosenbaum \& Rubin}{1984}]{Rosenbaum1984}
Rosenbaum, P.~R. and Rubin, D.~B. (1984).
\newblock Reducing bias in observational studies using subclassification on the
  propensity score.
\newblock {\em Journal of the American Statistical Association}, 79:516--524.

\bibitem[\protect\astroncite{Rubin}{1974}]{Rubin1974}
Rubin, D.~B. (1974).
\newblock Estimating causal effects of treatments in randomized and
  nonrandomized studies.
\newblock {\em Journal of Educational Statistics}, 66:688--701.

\bibitem[\protect\astroncite{Rubin \& van~der Laan}{2008}]{Rubin2008}
Rubin, D.~B. and van~der Laan, M.~J. (2008).
\newblock Empirical efficiency maximization: Improved locally efficient
  covariate adjustment in randomized experiments and survival analysis.
\newblock {\em International Journal of Biostatistics}, 4:1557--4679.

\bibitem[\protect\astroncite{Smith \& Todd}{2005a}]{Smith2005a}
Smith, J. and Todd, P. (2005a).
\newblock Does matching overcome lalonde's critique of nonexperimental
  estimators?
\newblock {\em Journal of Econometrics}, 125:305--353.

\bibitem[\protect\astroncite{Smith \& Todd}{2005b}]{Smith2005b}
Smith, J. and Todd, P. (2005b).
\newblock Rejoinder.
\newblock {\em Journal of Econometrics}, 125:365--375.

\bibitem[\protect\astroncite{Stoczynski \&
  Wooldridge}{2014}]{Stoczynski-Wooldridge2014}
Stoczynski, T. and Wooldridge, J. (2014).
\newblock A general double robustness result for estimating average treatment
  effects.
\newblock {\em IZA Discussion Paper}.
\newblock No. 8084.

\bibitem[\protect\astroncite{Tan}{2006}]{Tan2006}
Tan, Z. (2006).
\newblock A distributonal approach for causal inference using propensity score.
\newblock {\em Journal of the American Statistical Association},
  101:1619--1637.

\bibitem[\protect\astroncite{Tan}{2007}]{Tan2007}
Tan, Z. (2007).
\newblock Comment: Understanding or, ps and dr.
\newblock {\em Statistical Science}, 22:560--568.

\bibitem[\protect\astroncite{Tan}{2008}]{Tan2008}
Tan, Z. (2008).
\newblock Improved local efficiency and double robustness, comment on
  `empirical efficiency maximization: Improved locally efficient covariate
  adjustment in randomized experiments and survival analysis' by {Rubin} and
  {van der Laan}.
\newblock {\em International Journal of Biostatistics}, 4:Article 10.

\bibitem[\protect\astroncite{Tan}{2010}]{Tan2010}
Tan, Z. (2010).
\newblock Bounded, efficient and doubly robust estimation with inverse
  weighting.
\newblock {\em Biometrika}, 97:661--682.

\bibitem[\protect\astroncite{Tan}{2013}]{Tan2013}
Tan, Z. (2013).
\newblock Simple design-efficient calibration estimators for rejective and
  high-entropy sampling.
\newblock {\em Biometrika}, 100:399--415.

\bibitem[\protect\astroncite{Tsiatis}{2006}]{Tsiatis2006}
Tsiatis, A.~A. (2006).
\newblock {\em Semiparametric Theory and Missing Data}.
\newblock Springer Series in Statistics. Springer.

\bibitem[\protect\astroncite{Zhao \& Percival}{2017}]{Zhao2015}
Zhao, Q. and Percival, D. (2017).
\newblock Entropy balancing is doubly robust.
\newblock {\em Journal of Causal Inference}, 5:20160010.

\end{thebibliography}


\clearpage
\setcounter{page}{1}
\setcounter{section}{0}

\renewcommand{\theequation}{S\arabic{equation}}
\renewcommand{\thesection}{\Roman{section}}
\renewcommand\thefigure{S\arabic{figure}}
\renewcommand\thetable{S\arabic{table}}
\setcounter{figure}{0}
\setcounter{table}{0}

\centerline{\bf\large Supplementary Material}
\centerline{\bf for ``Improved Estimation of Average Treatment Effects on the Treated:}
\centerline{\bf Local Efficiency, Double Robustness, and Beyond" by Shu \& Tan}
\vspace{.2in}

The Supplementary Material contains Appendices~\ref{sec:tech-details}--\ref{sec:add-analysis}.

\vspace{-.1in}
\section{Technical details} \label{sec:tech-details}

\subsection{Preparation}

Throughout, we make the following assumptions regarding the estimators $\hat \alpha_t$ for OR model (\ref{OR}), $\hat\gamma$ for PS model (\ref{PS}),
and $(\tilde\gamma,\tilde\delta)$ for augmented PS model (\ref{augPS}), allowing for possible model misspecification (e.g., \citealtappend{White1982}).

\begin{enumerate}
\item[(C1)] Assume that
$\hat\alpha_t$ converges to a constant $\alpha_t^*$ such that $\hat\alpha_t - \alpha_t^* = O_p(n^{-1/2})$ for $t=0,1$.
Write $m^*_t(X) = m_t(X; \alpha^*_t)$.
If model (\ref{OR}) is correctly specified, then $m_t^*(X) = m_t(X)$. In general, $m_t^*(X)$ and $m_t(X)$ may differ from each other.

\item[(C2)] Assume that $\hat\gamma$ converges to a constant $\gamma^*$ such that
\begin{align*}
\hat \gamma - \gamma^* = V^{-1} \, \tilde E \left\{ s_{\gamma^*}(T,X) \right\} + o_p(n^{-1/2}),
\end{align*}
where $E\{s_{\gamma^*}(T,X)\}=0$, and the matrix $V= - E\{ \partial s_\gamma(T,X)/\partial \gamma^\T\} | _{\gamma=\gamma^*}$ is nonsingular.
Write $\pi^*(X) = \pi(X; \gamma^*)$.
If model (\ref{PS}) is correctly specified, then $\pi^*(X) = \pi(X)$
and $V = \var \{ s_{\gamma^*}(T,X)  \}$. In general, $\pi^*(X)$ and
$\pi(X)$ may differ from each other.

\item[(C3)] For augmented PS model (\ref{augPS}), define
$$
s^\dag(T,X; \gamma,\delta, \alpha)= \{T- \pi_{\text{\scriptsize aug}}(X; \gamma, \delta, \alpha)\} \{ f^\T(X), m_0(X; \alpha_0), m_1(X; \alpha_1)\}^\T.
$$
Assume that $(\tilde\gamma, \tilde\delta)$ converges to a constant $(\gamma^\dag, \delta^*)$ such that
\begin{align*}
\left( \begin{array} {c}
 \tilde \gamma - \gamma^\dag \\
 \tilde\delta - \delta^*
 \end{array} \right) = {V^\dag}^{-1} \, \tilde E \left\{ s^\dag(T,X; \gamma^\dag, \delta^*, \hat\alpha) \right\} + o_p(n^{-1/2}),
\end{align*}
where
$E\{s^\dag(T,X; \gamma^\dag, \delta^*, \alpha^*)\}=0$, and the matrix
$V^\dag= - E\{ \partial s^\dag(T,X; \gamma, \delta, \alpha^*)/ $ $\partial (\gamma^\T, \delta^\T)\} |_{(\gamma,\delta)=(\gamma^\dag,\delta^*)}$ is nonsingular.
Write $\pi^\dag(X) = \pi_{\text{\scriptsize aug}}(X; \gamma^\dag, \delta^*, \alpha^*)$.
If model (\ref{PS}) is correctly specified, then $(\gamma^\dag,\delta^*)=(\gamma^*,0)$, $\pi^\dag(X) = \pi(X)$,
$V ^\dag= \var \{ s^\dag(T,X; \gamma^*, 0, $ $\alpha^* )  \}$,
and the asymptotic expansion for $(\tilde\gamma,\tilde\delta)$ reduces to
\begin{align*}
\left( \begin{array} {c}
 \tilde \gamma - \gamma^* \\
 \tilde\delta
 \end{array} \right) = {V^\dag}^{-1} \, \tilde E \left\{ s^\dag_{(\gamma^*,0)}(T,X) \right\} + o_p(n^{-1/2}),
\end{align*}
where $s^\dag_{(\gamma^*, 0)}(T,X) = s^\dag(T,X; \gamma^*, 0, \alpha^*)$.
\end{enumerate}

In addition, we assume that the following regularity conditions hold (e.g., \citealtappend{Robins1994}, Appendix B).
\begin{enumerate}
\item[(C4)] $E\{ (Y^t)^2 \} < \infty$ and $E\{ {m_t^*}^2(X) \}<\infty$ for $t=0,1$.

\item[(C5)] There exists $\epsilon>0$ such that $0<\pi^*(x) \le 1-\epsilon$ and $0<\pi^\dag(x) \le 1-\epsilon$ for all $x$.

\item[(C6)] There exists a neighborhood $N_{1,t}$ of $\alpha_t^*$ such that $E\{ \sup_{\alpha_t \in N_{1,t} } \| \partial m_t(X;\alpha_t)/ \partial \alpha_t \|^2 \} $ $  < \infty$ for $t=0,1$,
where $\| A\| = (\sum_{ij} A_{ij}^2)^{1/2}$ for any matrix with element $A_{ij}$.

\item[(C7)]  There exists a neighborhood $N_2$ of $\gamma^*$ such that $E\{ \sup_{\gamma \in N_2 } \| \partial \pi(X;\gamma)/ \partial \gamma \|^2\}  < \infty$
and $E\{ \sup_{\gamma \in N_2 } \| \partial^2 \pi(X;\gamma)/ \partial \gamma\partial\gamma^\T \|^2 \}  < \infty$.

\item[(C8)]  There exists a neighborhood $N_3$ of $(\gamma^*,\delta^*,\alpha^*)$ such that $E\{ \sup_{ \theta \in N_3 } \| \partial \pi_{\text{\scriptsize aug}}(X;\theta)/ $ $ \partial\theta \|^2 \}  < \infty$
and $E\{ \sup_{\theta \in N_3 } \| \partial^2 \pi_{\text{\scriptsize aug}}(X;\theta)/ \partial \theta\partial\theta^\T \|^2 \}  < \infty$, with $\theta=(\gamma^\T,\delta^\T, \alpha^\T)^\T$.
\end{enumerate}

We provide the following lemma on asymptotic expansions of AIPW estimators.

\begin{lem} \label{lem}
Assume that $E\{h^2(X)\} < \infty$. If the PS model (\ref{PS}) is correctly specified, then the following results hold.
\begin{enumerate}
\item[(i)]
$\hat \nu^0 (\hat \pi, h) $ admits the asymptotic expansion,
\begin{align*}
\hat \nu^0 (\hat \pi, h) -\nu^0 & = q^{-1} \tilde E \Big( \phi_h^0(Y,T,X) - T \nu^0 - \mbox{Proj}\{\phi_h^0(Y,T,X) | s_{\gamma^*}(T,X)\}  \\
& \quad\quad + \mbox{Proj} \left[ \{T-\pi(X)\} m_0(X) | s_{\gamma^*}(T,X) \right] \Big) + o_p(n^{-1/2}),
\end{align*}
where $\phi_h^0(Y,T,X) = [(1-T)/\{1-\pi(X)\}] \pi(X) Y - [(1-T)/\{1-\pi(X)\}-1]h(X)$.

\item[(ii)]
Define $\hat \nu^1(\hat\pi, h) = \tilde E [TY - \{T-\hat\pi(X)\} h(X) ] / \tilde E(T)$.
Then $\hat \nu^1 (\hat \pi, h) $ admits the asymptotic expansion,
\begin{align*}
& \hat \nu^1 (\hat \pi, h) - \nu^1 \\
= & q^{-1} \tilde E \Big( \phi_h^1(Y,T,X) - T \nu^1
 + \mbox{Proj} \left[ \{T-\pi(X)\} h(X) | s_{\gamma^*}(T,X) \right] \Big) + o_p(n^{-1/2}), \\
= & q^{-1} \tilde E \Big( \phi_h^1(Y,T,X) - T \nu^1 - \mbox{Proj}\{\phi_h^1(Y,T,X) | s_{\gamma^*}(T,X)\} \\
& \quad\quad + \mbox{Proj} \left[ \{T-\pi(X)\} m_1(X) | s_{\gamma^*}(T,X) \right] \Big) + o_p(n^{-1/2}),
\end{align*}
where $\phi_h^1(Y,T,X) = T Y - \{T-\pi(X)\}h(X)$.
\end{enumerate}
\end{lem}

\noindent{\bf Proof of Lemma \ref{lem}.}
By direct calculation and Slutsky theorem, we have
\begin{align*}
\hat \nu^0 (\hat \pi, h) - \nu^0
& = q^{-1} \tilde E \left[ \frac{1-T}{1-\hat\pi(X)}\hat\pi(X)Y - \left\{ \frac{1-T}{1-\hat\pi(X)}-1\right\} h(X) - T \nu^0 \right] + o_p(n^{-1/2}).
\end{align*}
By a Taylor expansion for $\hat\gamma$ about $\gamma^*$ and direct calculation, we have
\begin{align*}
& \tilde E  \left[ \frac{1-T}{1-\hat\pi(X)} \pi(X) Y - \left\{ \frac{1-T}{1-\hat\pi(X)}-1\right\} h(X) \right] \\
= & \tilde E  \left[   \frac{1-T}{1- \pi(X)} \pi(X) Y - \left\{ \frac{1-T}{1-\pi(X)}-1\right\} h(X) \right] \\
& \quad\quad + E \left[\frac{1-T}{\{1-\pi(X)\}^2} \frac{\partial \pi(X;\gamma^*)}{\partial\gamma} \{\pi(X)Y- h(X)\} \right] (\hat\gamma - \gamma^*) + o_p(n^{-1/2}) \\
= & \tilde E \Big( \phi_h^0(Y,T,X) - \mbox{Proj}\{\phi_h^0(Y,T,X) | s_{\gamma^*}(T,X)\} \Big) + o_p(n^{-1/2}).
\end{align*}
By similar arguments, we have
\begin{align*}
& \tilde E  \left[ \frac{1-T}{1-\hat\pi(X)} \{\hat\pi(X)-\pi(X) \} Y  \right] \\
= & E \left[ \frac{1-T}{1-\pi(X)} \frac{\partial \pi(X;\gamma^*)}{\partial\gamma} Y \right] (\hat\gamma-\gamma^*) + o_p(n^{-1/2}) \\
= & \tilde E \Big( \mbox{Proj} \left[ \{T-\pi(X)\} m_0(X) | s_{\gamma^*}(T,X) \right] \Big) + o_p(n^{-1/2}).
\end{align*}
Combining the preceding three expansions gives the desired expansion for  $\hat \nu^0 (\hat \pi, h) $.
Similarly, the expansion for $\hat \nu^1 (\hat \pi, h) $ can be shown. $\Box$

\subsection{Proofs of Propositions \ref{prop2} \& \ref{prop4}}

First, we show the local nonparametric efficiency of $\hat \nu^0_{\text{\scriptsize NP}} (\hat \pi, \hat m_0) $.
If both model (\ref{OR}) for $t=0$ and model (\ref{PS}) are correctly specified, then by Slutsky theorem,
\begin{align*}
 \hat \nu^0_{\text{\scriptsize NP}} (\hat \pi, \hat m_0) &=
 \tilde E \left[ \frac{1-T}{1-\hat\pi(X)}\hat\pi(X)Y - \left\{ \frac{1-T}{1-\hat\pi(X)}-1\right\} m_0(X) \right] \Big/ \tilde E(T) + o_p(n^{-1/2}).
\end{align*}
The leading term can be reexpressed as
\begin{align*}
& \tilde E \left[\frac{1-T}{1-\hat\pi(X)}\hat\pi(X)\{Y- m_0(X)\} + T m_0(X) \right] \Big/ \tilde E(T)
\end{align*}
and, by Slutsky theorem, approximated by
\begin{align*}
\tilde E \left[\frac{1-T}{1- \pi(X)} \pi(X)\{Y- m_0(X)\} + T m_0(X) \right] \Big/ \tilde E(T) + o_p(n^{-1/2}),
\end{align*}
which gives the desired result.
Alternatively, the result follows from Lemma \ref{lem}(i) with $h(X)= m_0(X)$
and the fact that
$\phi_{m_0}^0(Y,T,X)  = \phi_{\pi m_0}^0(Y,T,X)  + \{T-\pi(X)\} m_0(X)$ and hence
$\mbox{Proj}\{\phi_{m_0}^0(Y,T,X) | s_{\gamma^*}(T,X)\} = \mbox{Proj}[\{ T-\pi(X)\}m_0(X) | s_{\gamma^*}(T,X)\}$.

Second, we show the double robustness of $\hat \nu^0_{\text{\scriptsize NP}} (\hat \pi, \hat m_0) $.
If PS model (\ref{PS}) is correctly specified, then $\tilde E ( [(1-T)/\{1-\hat \pi(X)\}-1] \hat m_0(X) ) = \tilde E ( [(1-T)/\{1-\pi(X)\}-1] m_0^*(X) ) +O_p(n^{-1/2}) = O_p(n^{-1/2})$ and hence
\begin{align*}
 \hat \nu^0_{\text{\scriptsize NP}} (\hat \pi, \hat m_0) =
 \tilde E \left\{\frac{1-T}{1-\hat\pi(X)}\hat\pi(X)Y \right\} \Big/ \tilde E(T) + O_p(n^{-1/2}) = \nu^0 + O_p(n^{-1/2}).
\end{align*}
On the other hand,   $\hat \nu^0_{\text{\scriptsize NP}} (\hat \pi, \hat m_0) $ can be reexpressed as
\begin{align*}
\hat \nu^0_{\text{\scriptsize NP}} (\hat \pi, \hat m_0) = \tilde E \left[\frac{1-T}{1-\hat\pi(X)}\hat\pi(X)\{Y-\hat m_0(X)\} + T \hat m_0(X) \right] \Big/ \tilde E(T) .
\end{align*}
If OR model (\ref{OR}) for $t=0$ is correctly specified, then $\tilde E ( [(1-T)/\{1-\hat\pi(X)\}] \hat\pi(X) \{Y-\hat m_0(X)\} )=
\tilde E ( [(1-T)/\{1-\pi^*(X)\}] \pi^*(X) \{Y-m_0(X)\} )+O_p(n^{-1/2}) = O_p(n^{-1/2})$
and hence
\begin{align*}
\hat \nu^0_{\text{\scriptsize NP}} (\hat \pi, \hat m_0) = \tilde E \left\{ T \hat m_0(X) \right\} \Big/ \tilde E(T) = \nu^0 + O_p(n^{-1/2}).
\end{align*}

Third, we show the local semiparametric efficiency of $\hat \nu^0_{\text{\scriptsize SP}} (\hat \pi, \hat m_0) $.
If both model (\ref{OR}) for $t=0$ and model (\ref{PS}) are correctly specified, then
\begin{align*}
& \hat \nu^0_{\text{\scriptsize SP}} (\hat \pi, \hat m_0) - \nu^0 \\
= & q^{-1} \tilde E \left[ \frac{1-T}{1-\hat\pi(X)}\hat\pi(X)Y - \left\{ \frac{1-T}{1-\hat\pi(X)}-1\right\} \hat\pi(X) m_0(X)  - \hat\pi(X) \nu^0 \right] + o_p(n^{-1/2}) \\
= & q^{-1} \tilde E \left[ \frac{1-T}{1-\hat\pi(X)}\hat\pi(X)(Y-\nu^0) - \left\{ \frac{1-T}{1-\hat\pi(X)}-1\right\} \pi(X) \{m_0(X)-\nu^0\}  \right]  + o_p(n^{-1/2}).
\end{align*}
by direct calculation and Slutsky theorem. Applying, to the above, Lemma \ref{lem}(i) with $Y$ replaced by $Y-\nu^0$ and $h(X)=\pi(X) \{ m_0(X) -\nu^0\}$ yields
\begin{align*}
\hat \nu^0_{\text{\scriptsize SP}} (\hat \pi, \hat m_0) - \nu^0
& = q^{-1} \tilde E \Big( \phi_h^0(Y-\nu^0,T,X) - \mbox{Proj}\{\phi_h^0(Y-\nu^0,T,X) | s_{\gamma^*}(T,X)\}  \\
& \quad\quad + \mbox{Proj} \left[ \{T-\pi(X)\}\{ m_0(X)-\nu^0\} | s_{\gamma^*}(T,X) \right] \Big) + o_p(n^{-1/2}),
\end{align*}
The desired results follows because
$\phi_h^0(Y-\nu^0,T,X) = \tau^0(\pi, \pi m_0) - \pi(X) \nu^0$ by direct calculation,
and the variable $\phi_h^0(Y-\nu^0,T,X)$ is uncorrelated with the score $s_{\gamma^*}(T,X)$ and hence
$\mbox{Proj}\{\phi_h^0(Y-\nu^0,T,X) | s_{\gamma^*}(T,X)\}=0$.

Finally, we show the local semiparametric efficiency of $\hat \nu^1_{\text{\scriptsize SP}} (\hat \pi, \hat m_1) $.
If both model (\ref{OR}) for $t=1$ and model (\ref{PS}) are correctly specified, then
\begin{align*}
& \hat \nu^1_{\text{\scriptsize SP}} (\hat \pi, \hat m_1) - \nu^1
= q^{-1} \tilde E \left[ T Y  -\{T-\hat\pi(X)\} m_1(X) - \hat\pi(X) \nu^1 \right] + o_p(n^{-1/2}) \\
&= q^{-1} \tilde E \left[ T(Y-\nu^1) -\{T-\hat\pi(X)\} \{m_1(X)-\nu^1\}  \right]  + o_p(n^{-1/2}),
\end{align*}
by direct calculation and Slutsky theorem. Applying, to the above, Lemma \ref{lem}(ii) with $Y$ replaced by $Y-\nu^1$ and $h(X)= m_1(X) -\nu^1$ yields
\begin{align*}
\hat \nu^1_{\text{\scriptsize SP}} (\hat \pi, \hat m_1) - \nu^1
& = q^{-1} \tilde E \Big( \phi_h^1(Y-\nu^1,T,X) \\
& \quad\quad + \mbox{Proj} \left[ \{T-\pi(X)\}\{ m_1(X)-\nu^1\} | s_{\gamma^*}(T,X) \right] \Big) + o_p(n^{-1/2}).
\end{align*}
The desired result follows because
$\phi_h^1(Y-\nu^1,T,X) =TY- \{T-\pi(X)\} m_1(X) - \pi(X)\nu^1$ by direct calculation. $\Box$

\subsection{Proof of Proposition \ref{prop-reg}}

First, it is straightforward to show that $\tilde\beta_t = \beta_t^* + o_p(1)$, where
$\beta^*_t = E^{-1} ( \xi^*_t {\zeta_t^*}^\T) E( \xi^*_t \eta^*_t)$
and $\eta^*_t$, $\xi^*_t$, $\zeta^*_t$, and $h^*(X)$ are defined as $\tilde \eta_t$, $\tilde\xi_t$, $\tilde\zeta_t$, and $\tilde h(X)$ respectively
but with $\pi^\dag(X)$ and $m^*_t(X)$ in place of $\tilde \pi(X)$ and $\hat m_t(X)$ throughout.

Second, we show the local nonparametric efficiency and double robustness of $\tilde\nu^t_{\text{\scriptsize reg}}$.
By the discussion in Section \ref{reg-est}, it suffices to show that
if the OR model (\ref{OR}) for $t=0$ or 1 is correctly specified, then asymptotic expansion (\ref{DR-explain}) holds for the corresponding $t$.
By construction, $\tilde \pi(X) \tilde m_0(X)$ is a linear combination of $\tilde h(X)/\tilde \pi(X)$, that is,
$ \tilde \pi(X) \tilde m_0(X) = c_0^\T \tilde h(X)/ \tilde \pi(X)$ for some constant vector $c_0$.
Then $\pi^\dag(X) m^*_0(X) = c_0^\T h^*(X)/ \pi^\dag(X)$ also holds for the same vector $c_0$. If model (\ref{OR}) for $t=0$ holds, then $m^*(x) = m_0(X)$
and hence $\pi^\dag(X) m_0(X) = c_0^\T h^*(X)/ \pi^\dag(X)$. By direct calculation, we have
\begin{align*}
\beta^*_0 = E^{-1} \left\{ \xi^*_0 \frac{1-T}{1-\pi^\dag(X)} \frac{{h^*}^\T(X)}{\pi^\dag(X)} \right\} E\left\{ \xi^*_0 \frac{1-T}{1-\pi^\dag(X)} \pi^\dag(X) m_0(X) \right\} = c_0.
\end{align*}
and hence asymptotic expansion (\ref{DR-explain}) holds for $t=0$. Similarly, because $\tilde \pi(X) \tilde m_1(X)$ is a linear combination of $\tilde h(X)/\{1-\tilde \pi(X)\}$,
it can be shown that
if the OR model (\ref{OR}) for $t=1$ is correctly specified, then expansion (\ref{DR-explain}) holds for $t=1$.

Third, we show the intrinsic efficiency of $\tilde\nu^0_{\text{\scriptsize reg}}$ among the class of estimators (\ref{nu-class}) for $t=0$, denoted by $\tilde \nu^0(b_0)$.
By direct calculation and Slutsky theorem, we have
\begin{align*}
\tilde\nu^0 (b_0) - \nu^0 & = q^{-1} \tilde E ( \tilde\eta_0 - b_0^\T \tilde \xi_0 - \nu^0 T ) + o_p(n^{-1/2}) \\
& = q^{-1} \tilde E \left[ \tilde \eta_0 - b_0^\T \left\{ \frac{1-T}{1-\tilde\pi(X)}-1 \right\} \frac{h^*(X)}{\pi (X)} - \nu^0 T\right] + o_p(n^{-1/2}) .
\end{align*}
If PS model (\ref{PS}) is correctly specified, then applying, to the above, Lemma \ref{lem}(i) with $\hat\pi(X)$ replaced by $\tilde \pi(X)$ and
$h(X)=b_0^\T h^*(X)/\pi (X)$ yields
\begin{align*}
\tilde\nu^0 (b_0) - \nu^0  & = q^{-1} \tilde E \Big( \eta^*_0 - b_0^\T \xi^*_0 - \pi(X) \nu^0 - \mbox{Proj}\{ \eta^*_0 - b_0^\T \xi^*_0  | s^\dag_{(\gamma^*,0)}(T,X)\}  \\
& \quad\quad + \mbox{Proj} \left[ \{T-\pi(X)\} \{m_0(X) -\nu^0\} | s^\dag_{(\gamma^*,0)}(T,X) \right] \Big) + o_p(n^{-1/2}) .
\end{align*}
where $\phi_h^0(Y,T,X)  =\eta^*_0 - b_0^\T \xi^*_0 $ and
$T\nu^0$ is decomposed as $\pi(X) \nu^0 + \{T-\pi(X)\}\nu^0 =
\pi(X)\nu^0 + \mbox{Proj} [ \{T-\pi(X)\}  \nu^0| s^\dag_{(\gamma^*,0)}(T,X) ]$ because $ T-\pi(X)$ is contained in $s^\dag_{(\gamma^*,0)}(T,X)$.
The first term inside $\tilde E()$ above, $\eta^*_0 - \pi(X) \nu^0 - b_0^\T \xi^*_0 - \mbox{Proj}\{ \eta^*_0 - b_0^\T \xi^*_0  | s^\dag_{(\gamma^*,0)}(T,X)\} $, is uncorrelated with
the second term, $\mbox{Proj} [ \{T-\pi(X)\} \{m_0(X) -\nu^0\} | s^\dag_{(\gamma^*,0)}(T,X) ] $, which is independent of $b_0$.
Moreover, the first term can be expressed as $\eta^*_0 - \pi(X) \nu^0 -a_0^\T \xi^*_0 $ for some constant vector $a_0$,
because, by construction, each variable in $s^\dag_{(\gamma^*,0)}(T,X)$ is a linear combination of varibles in $\xi^*_0$.
By combining these two facts, we see that the asymptotic variance of $\tilde\nu^0 (b_0)$ is achieved when $a_0$ is equal to
\begin{align*}
 \var^{-1}(\xi^*_0) \cov  \left\{ \xi^*_0, \, \eta^*_0 - \pi(X) \nu^0\right\} = E^{-1} ( \xi^*_0 {\zeta^*_0}^\T)  E(\xi^*_0 \eta^*_0) = \beta_0^* .
\end{align*}
But to make $a_0$ equal to $\beta_0^*$, it suffices to set $b_0 = \beta_0^*$, because $\eta^*_0 - {\beta^*_0}^\T \xi^*_0$ is uncorrelated with $s^\dag_{(\gamma^*,0)}(T,X)$
and hence $\mbox{Proj}\{ \eta^*_0 - {\beta^*_0}^\T \xi^*_0  | s^\dag_{(\gamma^*,0)}(T,X)\}  = 0$.
If PS model (\ref{PS}) is correctly specified, then $\tilde\nu^0_{\text{\scriptsize reg}} = \tilde \nu^0( \beta^*_0) + o_p(n^{-1/2})$.
Therefore, $\tilde\nu^0_{\text{\scriptsize reg}} $ is intrinsically efficient among the class of estimators $\tilde \nu^0(b_0)$.

Finally, we show the intrinsic efficiency of $\tilde\nu^1_{\text{\scriptsize reg}}$ among the class of estimators (\ref{nu-class}) for $t=1$, denoted by $\tilde \nu^1(b_1)$.
By direct calculation and Slutsky theorem, we have
\begin{align*}
\tilde\nu^1 (b_1) - \nu^1 & = q^{-1} \tilde E ( \tilde\eta_1 - b_1^\T \tilde \xi_1 - \nu^1 T ) + o_p(n^{-1/2}) \\
& = q^{-1} \tilde E \left[ \tilde \eta_1 - b_1^\T \{ T- \tilde\pi(X)\} \frac{h^*(X)}{\pi (X) \{1-\pi (X)\}} - \nu^1 T\right] + o_p(n^{-1/2}) .
\end{align*}
If PS model (\ref{PS}) is correctly specified, then applying, to the above, Lemma \ref{lem}(i) with $\hat\pi(X)$ replaced by $\tilde \pi(X)$ and
$h(X)=b_1^\T h^*(X)/[\pi (X)\{1-\pi (X)\}]$ yields
\begin{align*}
\tilde\nu^1 (b_1) - \nu^1  & = q^{-1} \tilde E \Big( \eta^*_1 - b_1^\T \xi^*_1 - \pi(X) \nu^1 - \mbox{Proj}\{ \eta^*_1 - b_1^\T \xi^*_1  | s^\dag_{(\gamma^*,0)}(T,X)\}  \\
& \quad\quad + \mbox{Proj} \left[ \{T-\pi(X)\} \{m_1(X) -\nu^1\} | s^\dag_{(\gamma^*,0)}(T,X) \right] \Big) + o_p(n^{-1/2}),
\end{align*}
where $\phi_h^1(Y,T,X)  =\eta^*_1 - b_1^\T \xi^*_1 $. The intrinsic efficiency of $\tilde\nu^0_{\text{\scriptsize reg}}$
can be similarly obtained as above for the intrinsic efficiency of $\tilde\nu^1_{\text{\scriptsize reg}}$ .

\subsection{Derivation of empirical likelihood estimates}

The empirical likelihood estimate of $\nu^t$ is $\hat\nu^t_{\text{\scriptsize lik}} =\sum_{i=1}^n \hat p_i \tilde \eta_{t,i} /\sum_{i=1}^n \hat p_i T_i$, where
$(\hat p_1, \ldots, \hat p_n)$ are obtained from the constrained maximization problem:
\begin{align*}
\max_{p_1 \ge 0, \ldots, p_n \ge 0} \quad & \sum_{i=1}^n \log p_i \\
\mbox{subject to} \quad &  \sum_{i=1}^n p_i=1 \mbox{ and } \sum_{i=1}^n p_i \tilde\xi_{1,i}= 0 .
\end{align*}
By standard calculation (\citealtappend{Qinlawless1994}), we have
\begin{align*}
\hat p_i = \frac{n^{-1}}{1 + \hat\lambda^\T \tilde \xi_{1,i}},
\end{align*}
where $\hat\lambda$ is a maximizer of the function
\begin{align*}
\ell_{\text{\scriptsize EL}}(\lambda) = \frac{1}{n} \sum_{i=1}^n \log \left( 1 + \lambda^\T \tilde \xi_{1,i} \right).
\end{align*}
Write $\tilde\pi_i = \tilde \pi(X_i)$, $\tilde h_i = \tilde h(X_i)$, and $\omega_i = \omega(X_i;\lambda)$ for $i=1,\ldots,n$. By direct calculation, $\ell_{\text{\scriptsize EL}}(\lambda)$ can be reexpressed as
\begin{align*}
\ell_{\text{\scriptsize EL}}(\lambda) & = \frac{1}{n}\sum_{i=1}^n \log \left\{  1+ \lambda^\T \frac{T_i-\tilde{\pi}_i}{\tilde{\pi_i}(1-\tilde{\pi}_i)}\tilde{h}_i  \right\} \\
& = \frac{1}{n}\sum_{i=1}^n \left\{ T_i \log\left(  1+ \lambda^\T \frac{\tilde{h}_i }{\tilde{\pi}_i} \right) + (1- T_i ) \log \left(  1- \lambda^\T \frac{\tilde{h}_i }{1-\tilde{\pi}_i} \right) \right\} \\
& = \frac{1}{n}\sum_{i=1}^n \left\{ T_i\log\omega_i+ (1-T_i)\log(1-\omega_i) \right\} -\frac{1}{n}\sum_{i=1}^n \left\{ T_i\log\tilde{\pi}_i + (1-T_i)\log(1-\tilde{\pi}_i) \right\},
\end{align*}
which equals $\ell(\lambda)$ up to an additive constant.
Therefore, $\hat\lambda$ is a maximizer of $\ell(\lambda)$. The desired expressions for $\hat\nu^1_{\text{\scriptsize lik}}$ and $\hat\nu^0_{\text{\scriptsize lik}}$ hold because, by direct calculation,
\begin{align*}
& \sum_{i=1}^n \hat p_i \tilde \eta_{1,i} = \frac{1}{n} \sum_{i=1}^n \frac{\tilde\eta_{1,i}}{1+ \hat\lambda^\T \tilde\xi_{1,i}}
=  \frac{1}{n} \sum_{i=1}^n \frac{T_i Y_i}{ 1 + \hat\lambda^\T \frac{\tilde h_i}{\tilde \pi_i}} =  \frac{1}{n} \sum_{i=1}^n \frac{T_i \tilde \pi_i Y_i}{\hat \omega_i}, \\
&  \sum_{i=1}^n \hat p_i \tilde \eta_{0,i} = \frac{1}{n} \sum_{i=1}^n \frac{\tilde\eta_{0,i}}{1+ \hat\lambda^\T \tilde\xi_{1,i}}
=  \frac{1}{n} \sum_{i=1}^n \frac{(1-T_i) \frac{\tilde\pi_i}{1-\tilde\pi_i} Y_i}{ 1 - \hat\lambda^\T \frac{\tilde h_i}{1-\tilde \pi_i}} =  \frac{1}{n} \sum_{i=1}^n \frac{(1-T_i)\tilde \pi_i Y_i}{1-\hat\omega_i},
\end{align*}
where $\hat\omega_i = \omega(X_i; \hat\lambda)$ for $i=1,\ldots,n$. $\Box$

\subsection{Proof of Corollary~\ref{cor-reg2}}

The simple estimator $\hat{\nu}^{0}_{\text{\scriptsize IPW}}(\tilde\pi)$ based on $\tilde\pi(X)$ falls in the class (\ref{nu-class}) for $t=0$, with $b_0=0$.
The ratio estimator $\hat{\nu}^{0}_{\text{\scriptsize IPW,ratio}}(\tilde\pi)$ does not directly fall in the class (\ref{nu-class}), but
can be shown to be asymptotically equivalent to the first order, under a correctly specified PS model, to
$ \tilde E( \hat \eta_0 - [(1-T)/\{1-\tilde \pi(X)\}-1]\nu^0 )/ \tilde E(T)$, which falls in class (\ref{nu-class}) for $t=0$ becauase
$1$ is a linear combination of the variables, $\tilde\pi(X)$ and $1-\tilde\pi(X)$, in $\tilde h(X)/\tilde{\pi}(X)$.
The estimator $\hat{\nu}^{0}_{\text{\scriptsize NP}} (\tilde{\pi},\hat{m}_0)$ falls in the class (\ref{nu-class})  for $t=0$
because $\hat m_0(X)$ is a linear combination of the variables, $\tilde \pi(X)\hat m_0(X)$ and $\{1-\tilde \pi(X)\} \hat m_0(X)$, included in $\tilde h(X)/\tilde{\pi}(X)$.
The comparison then follows from Proposition~\ref{prop-reg}.

The estimator $\hat{\nu}^{1}_{\text{\scriptsize NP}}=\tilde E(\hat \eta_1)$ falls in the class (\ref{nu-class}) for $t=1$, with $b_1=0$.
By Corollary~\ref{cor-reg}, the estimator $\tilde\nu^1_{\text{\scriptsize reg}}-\tilde\nu^0_{\text{\scriptsize reg}}$ for ATT is asymptotically at least as efficient as
$\hat{\nu}^{1}_{\text{\scriptsize NP}} -\hat{\nu}^{0}_{\text{\scriptsize NP}} (\tilde{\pi},\hat{m}_0) $ when the PS model is correctly specified. $\Box$

\subsection{Proof of Proposition \ref{prop-lik}}

We need only to show that if model (\ref{PS}) is correctly specified, then
$\tilde\nu^t_{\text{\scriptsize lik}}$ is asymptotically equivalent, to the first order, to $\tilde\nu^t_{\text{\scriptsize reg}}$ for $t=0,1$.
By direct calculation and Slutsky theorem, we have
\begin{align*}
\tilde{\nu}^0_{\text{\scriptsize lik}} - \nu^0 & = q^{-1} \tilde{E}\left[ \frac{(1-T)\tilde{\pi}(X )Y}{1-\omega(X ;\tilde\lambda^0)} - T\nu^0\right] + o_p(n^{-1/2}).
\end{align*}
If model (\ref{PS}) is correctly specified, then
\begin{align*}
\tilde{E}\left[ \frac{(1-T)\tilde{\pi}(X )Y}{1-\omega(X ;\tilde\lambda^0)} \right] & =\tilde{E}\left[ \frac{(1-T)\tilde{\pi}(X )Y}{1-\omega(X ;\hat\lambda)} \right] + o_p(n^{-1/2}),
\end{align*}
by a Taylor expansion for $\tilde \lambda^0$ about $\hat\lambda$ and the fact that $\tilde E ( [ (1-T)/\{1-\omega(X;\hat\lambda)\}-1 ] \tilde \pi(X) ) = o_p(n^{-1/2})$, similarly
as in the asymptotic expansion of the calibrated likelihood estimator in \citetappend{Tan2010}.
Moreover, if model (\ref{PS}) is correctly specified, then $\hat\lambda$ converges to $0$ in probability and
\begin{align*}
\tilde{E}\left[ \frac{(1-T)\tilde{\pi}(X )Y}{1-\omega(X ;\hat\lambda)} \right]& =\tilde{E}\left(\tilde{\eta}_0 - {\beta^*_0}^\T \tilde{\xi }_0 \right) + o_p(n^{-1/2}).
\end{align*}
by a Taylor expansion for $\hat\lambda$ about $0$, similarly as in the asymptotic expansion of the non-calibrated likelihood estimator in \citetappend{Tan2010}.
The desired result for $\tilde\nu^0_{\text{\scriptsize lik}}$ then follows from the preceding expansions.
Similarly, the result for $\tilde\nu^1_{\text{\scriptsize lik}}$ can be shown. $\Box$

\subsection{Extension with non-logistic PS model}

We discuss an extension of the regression and likelihood estimators $\tilde\nu^t_{\text{\scriptsize reg}}$ and $\tilde\nu^t_{\text{\scriptsize lik}}$
when the PS model (\ref{PS}) is non-logistic regression. Consider an augmented PS model
\begin{align*}
& P(T=1|X) = \pi_{\text{\scriptsize aug}}(X; \gamma,\gamma_0, \delta, \hat\alpha) \nonumber \\
& =\Pi \left\{ \gamma^\T f(X) + \gamma_0 \hat \rho^{-1}(X) + \delta_0\, \hat \rho^{-1}(X) \hat m_0(X) + \delta_1\, \hat \rho^{-1}(X) \hat m_1(X) \right\},
\end{align*}
where $\hat\rho(X) = \rho(X;\hat\gamma)$, $\rho(X;\gamma) = \Pi'\{\gamma^\T f(X)\}/ [\pi(X;\gamma)\{1-\pi(X;\gamma)\}]$,
and $\Pi'()$ is the derivative of $\Pi()$.
For logistic regression, $\rho(X;\gamma)$ reduces to a constant 1.
Let $(\tilde\gamma, \tilde\gamma_0, \tilde\delta)$ be the estimates of $(\gamma,\gamma_0,\delta)$ solving the estimating equations
\begin{align*}
\tilde E \left[ \{T-\pi_{\text{\scriptsize aug}}(X; \gamma,\gamma_0, \delta, \hat\alpha)\} \{ \hat\rho(X) f^\T(X), 1, \hat m_0(X), \hat m_1(X)\}^\T  \right] = 0 .
\end{align*}
Let $\tilde \pi(X) = \pi_{\text{\scriptsize aug}}(X; \tilde \gamma, \tilde\gamma_0, \tilde \delta, \hat\alpha)$,
and define the estimators $\tilde\nu^t_{\text{\scriptsize reg}}$ and $\tilde\nu^t_{\text{\scriptsize lik}}$ same as before, except that $\tilde h(X)$ is defined with
$$
\tilde h_2(X) = \tilde \pi(X) \{1-\tilde \pi(X)\} \{ \hat\rho(X) f^\T(X), \hat m_0 (X)\}^\T .
$$
Then Propositions \ref{prop-reg} and \ref{prop-lik} can be shown to hold as before.

Particularly, to establish intrinsic efficiency, it can be shown that if PS model (\ref{PS}) is correctly specified, then
the estimates $(\tilde\gamma, \tilde\gamma_0, \tilde\delta)$ are asymptotically equivalent to the first order to
the MLE of $(\gamma,\gamma_0,\delta)$ from the following ``model,"
\begin{align*}
& P(T=1|X) = \pi^*_{\text{\scriptsize aug}}(X; \gamma,\gamma_0, \delta, \hat\alpha) \nonumber \\
& =\Pi \left\{ \gamma^\T f(X) + \gamma_0 \rho^{*-1}(X) + \delta_0\, \rho^{*-1}(X) m^*_0(X) + \delta_1\, \rho^{*-1}(X) m_1^*(X) \right\},
\end{align*}
where $\rho^*(X) = \rho(X;\gamma^*)$. That is, the random variation in $\hat\rho(X)$, $\hat m_0(X)$,
and $\hat m_1(X)$ does not affect the asymptotic behavior of $(\tilde\gamma, \tilde\gamma_0, \tilde\delta)$ to the first order.
The proofs of Propositions \ref{prop-reg} and \ref{prop-lik} can be completed similarly as before.

\subsection{Violation of the exogeneity assumption}

We present large-sample limits for estimators of ATT when the exogeneity assumption (A1) may be violated, i.e., $T$ and $Y^0$ may not be conditionally independent given $X$.
Similar results are known for estimators of ATE under possible violation of exogeneity assumptions (e.g., \citealtappend{Robins1999}; \citealtappend{Tan2006}).
We mainly use these results to justify how various estimators are compared in our analysis of LaLonde data in Section \ref{LaLonde},
although the results can be broadly used.

Suppose that the exogeneity ssumption (A1) may be violated. The following results can be shown by similar calculations as under Assumption (A1).
\begin{enumerate}
\item[(i)] If the the OR model (\ref{OR}) is correctly specified for $t=0$, then $\hat\nu^0_{\mbox{\scriptsize OR}}$, $\hat \nu^0_{\mbox{\scriptsize NP}}(\hat\pi, \hat m_0)$,
$\tilde \nu^0_{\mbox{\scriptsize reg}}$, and $\tilde \nu^0_{\mbox{\scriptsize lik}}$ converge in probability as $n\to\infty$ to
$E \{ T m_0(X)\} / E(T)$, which reduces to $E(Y^0 | T=1)$ when Assumption (A1) holds but not generally so.
Moreover, if the the OR model (\ref{OR}) is correctly specified for $t=1$, then
$\tilde \nu^1_{\mbox{\scriptsize reg}}$ and $\tilde \nu^1_{\mbox{\scriptsize lik}}$ converge in probability as $n\to\infty$ to
$E(Y| T=1)$.

\item[(ii)] If the PS model (\ref{PS}) is correctly specified, then $\hat \nu^0_{\mbox{\scriptsize IPW}}(\hat\pi)$, $\hat \nu^0_{\mbox{\scriptsize NP}}(\hat\pi, \hat m_0)$,
$\tilde \nu^0_{\mbox{\scriptsize reg}}$, and $\tilde \nu^0_{\mbox{\scriptsize lik}}$ converge in probability as $n\to\infty$ to
$E \{ T m_0(X)\} / E(T) $.
\end{enumerate}

In the context of LaLonde analysis,
let $T$ be the indicator for the NSW cohort, i.e., $T=1$ for the NSW treatment group in Analysis (i) or NSW control group in Analysis (ii) and $T=0$ for the comparison group,
and let $D$ be the indicator for job training, i.e., $D=1$ for the NSW treatment group and $D=0$ for the NSW control group and the comparison group.
Define $Y^{11}$ as the potential outcome that would be observed if an individual was selected into NSW cohort and assigned to treatment,
$Y^{01}$ as the potential outcome that would be observed if an individual was selected into NSW cohort and assigned to control,
and $Y^{00}$ as the potential outcome that would be observed if an individual was selected into the comparison cohort and hence no job training.
It is not necessary that $Y^{01} \equiv Y^{00}$, which would rule out any placebo effect such that
earnings could be affected by merely participating in the NSW experiment.
The exogeneity assumption (A1), $T \perp Y^{00}  | X$, means that
the NSW and comparison cohorts would have similar distributions of
of earnings, at each covariate level $x$, if both placed in the comparison cohort and not assigned to job training.
This assumption is implicitly made in all previous studies starting from \citeappend{LaLonde1986},
but can potentially be violated.

Because the NSW treatment and control groups are randomized, the difference
\begin{align*}
E(Y^{11} |T=1) - E(Y^{01}|T=1)
\end{align*}
is the experimental benchmark.
For Analysis (i) with NSW treatment group combined with a comparison group,
a valid ATT estimator should be close to $E( Y^{11} |T=1) - E\{ T m_0(X) \}/E(T)$,
and the corresponding bias be close to
\begin{align*}
& E( Y^{11} |T=1) - E\{ T m_0(X) \}/E(T) - \{ E(Y^{11} |T=1) - E(Y^{01}|T=1)\} \\
& = E( Y^{01} |T=1) - E\{ T m_0(X) \}/E(T) ,
\end{align*}
where $m_0(X) = E(Y^{00} | T=0,X)$.
For Analysis (ii) with NSW control group combined with a comparison group,
a valid ATT estimator should be close to
\begin{align*}
E( Y^{01} |T=1) - E\{ T m_0(X) \}/E(T).
\end{align*}
Therefore, the two bias estimates separately from Analyses (i) and (ii) should be close to each other for a good method,
even when the exogeneity assumption (A1) is violated.
This relationship forms the basis in our assessment of relative performances of various estimators of ATT in Section \ref{LaLonde}.

\newpage

\section{Additional simulation results} \label{sec:add-simulation}

\subsection{Qin--Zhang simulation}

\begin{figure}[H]

\begin{tabular}{c}
\includegraphics[width=6in, height=4in]{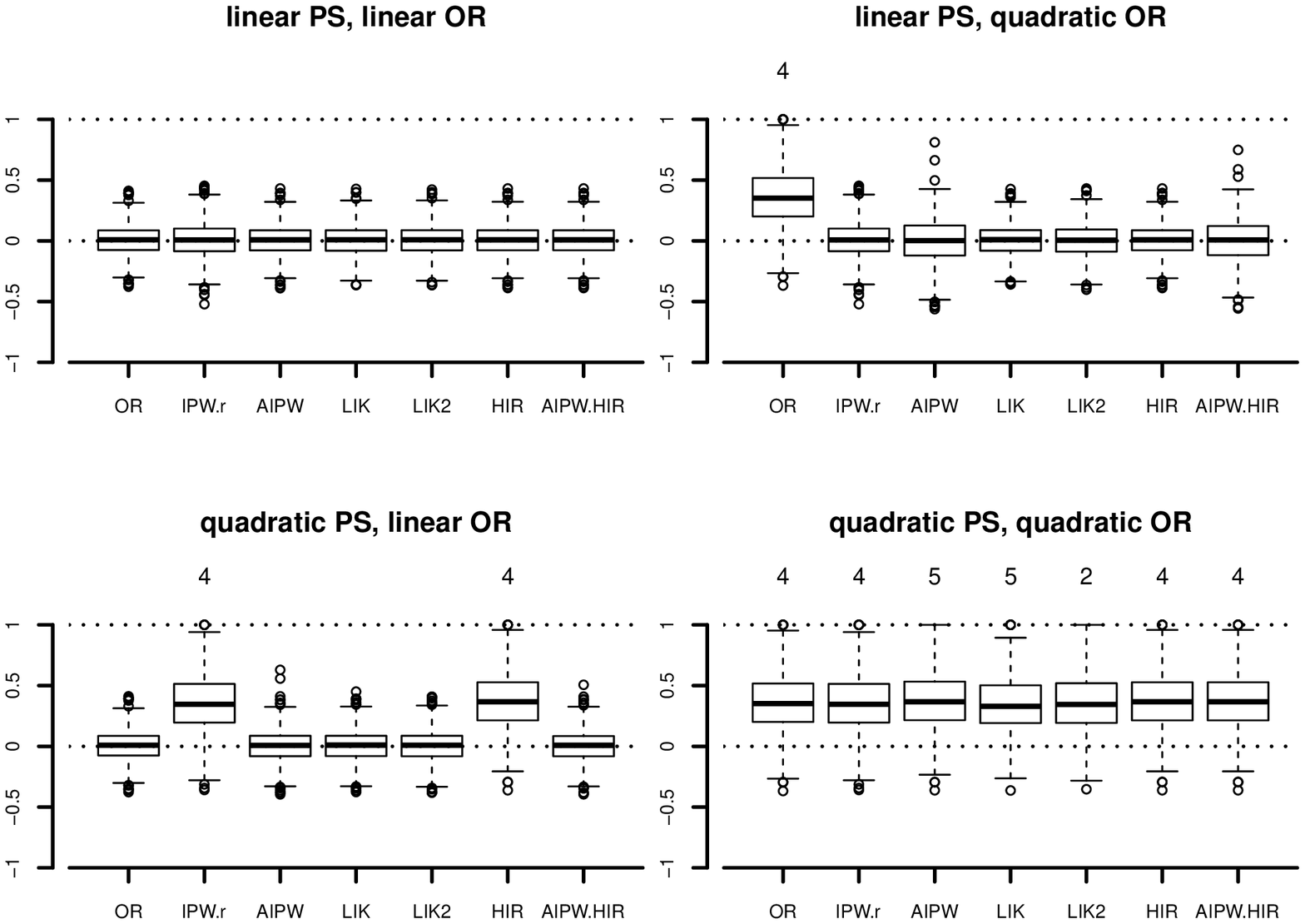}
\end{tabular} \vspace{-.4in}
\caption{\small Boxplots of estimates minus the truth under LIN-OR setting with $(\gamma_1^*,\gamma_2^*,\gamma_3^*)=(1.0,0.1,0.1)$. }
\label{plot:LIN-1}

\vspace{.4in}
\begin{tabular}{c}
\includegraphics[width=6in, height=4in]{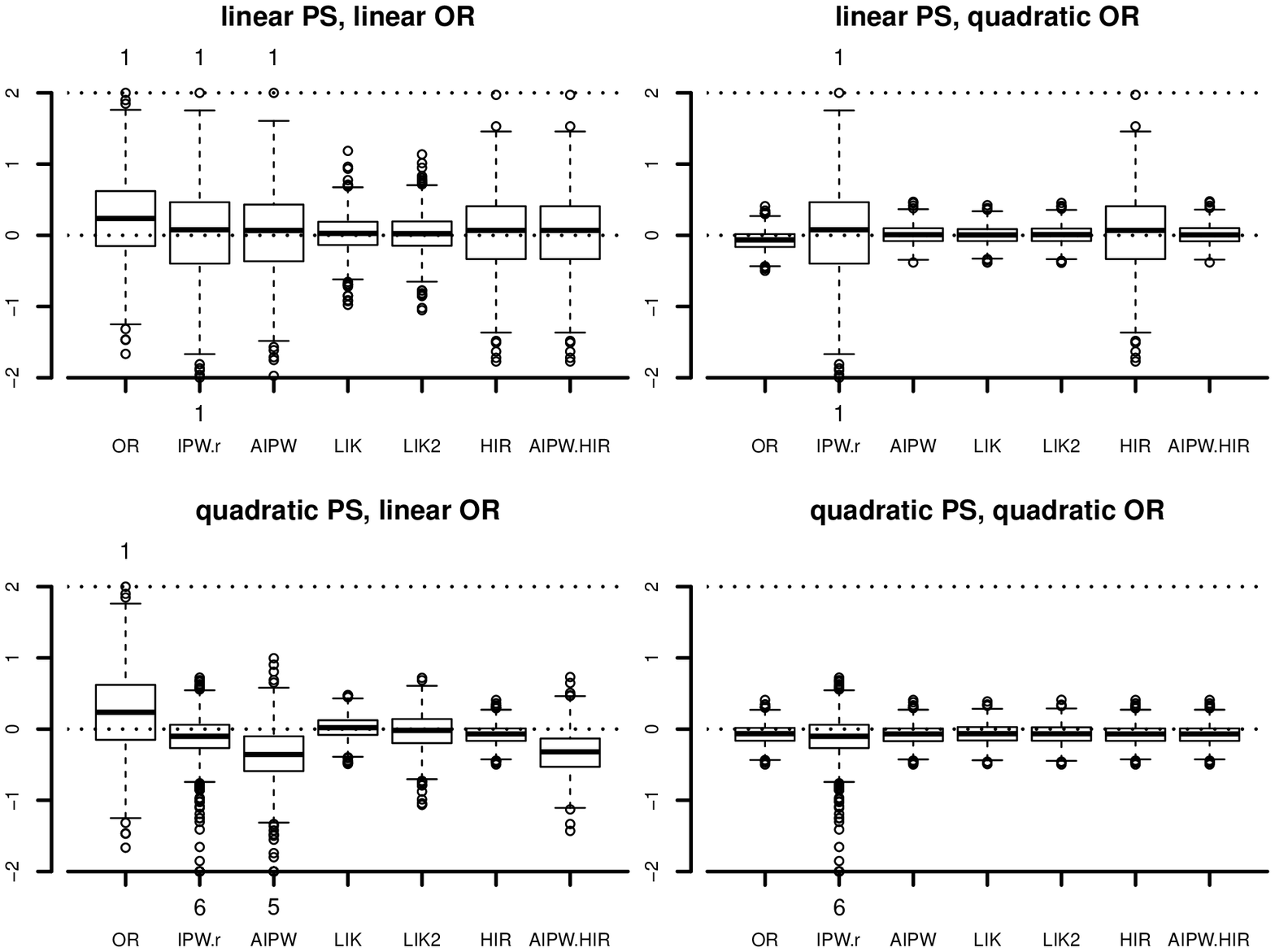}
\end{tabular}
\caption{\small Boxplots of estimates minus the truth under QUA-OR setting with $(\gamma_1^*,\gamma_2^*,\gamma_3^*)=(1.0,0.1,0.1)$.}
\label{plot:QUA-1}
\vspace{-.1in}
\end{figure}

\begin{figure}[H]
\begin{tabular}{c}
\includegraphics[width=6in, height=4in]{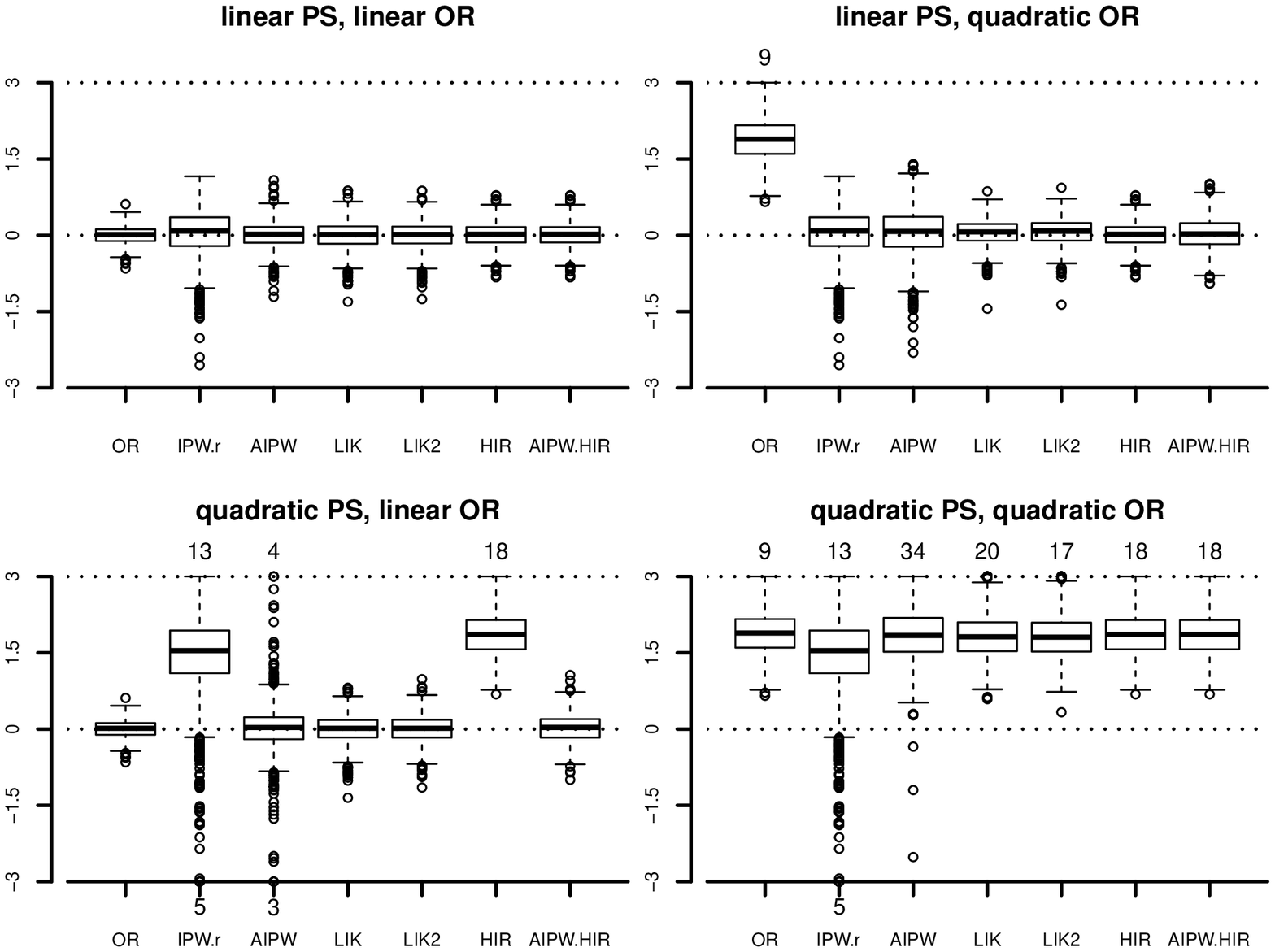}
\end{tabular}
\caption{\small Boxplots of estimates minus the truth under LIN-OR setting with $(\gamma_1^*,\gamma_2^*,\gamma_3^*)=(1.0,0.5,0.5)$.}
\label{plot:LIN-5}
\vspace{-.1in}
\end{figure}

\begin{figure}[H]
\begin{tabular}{c}
\includegraphics[width=6in, height=4in]{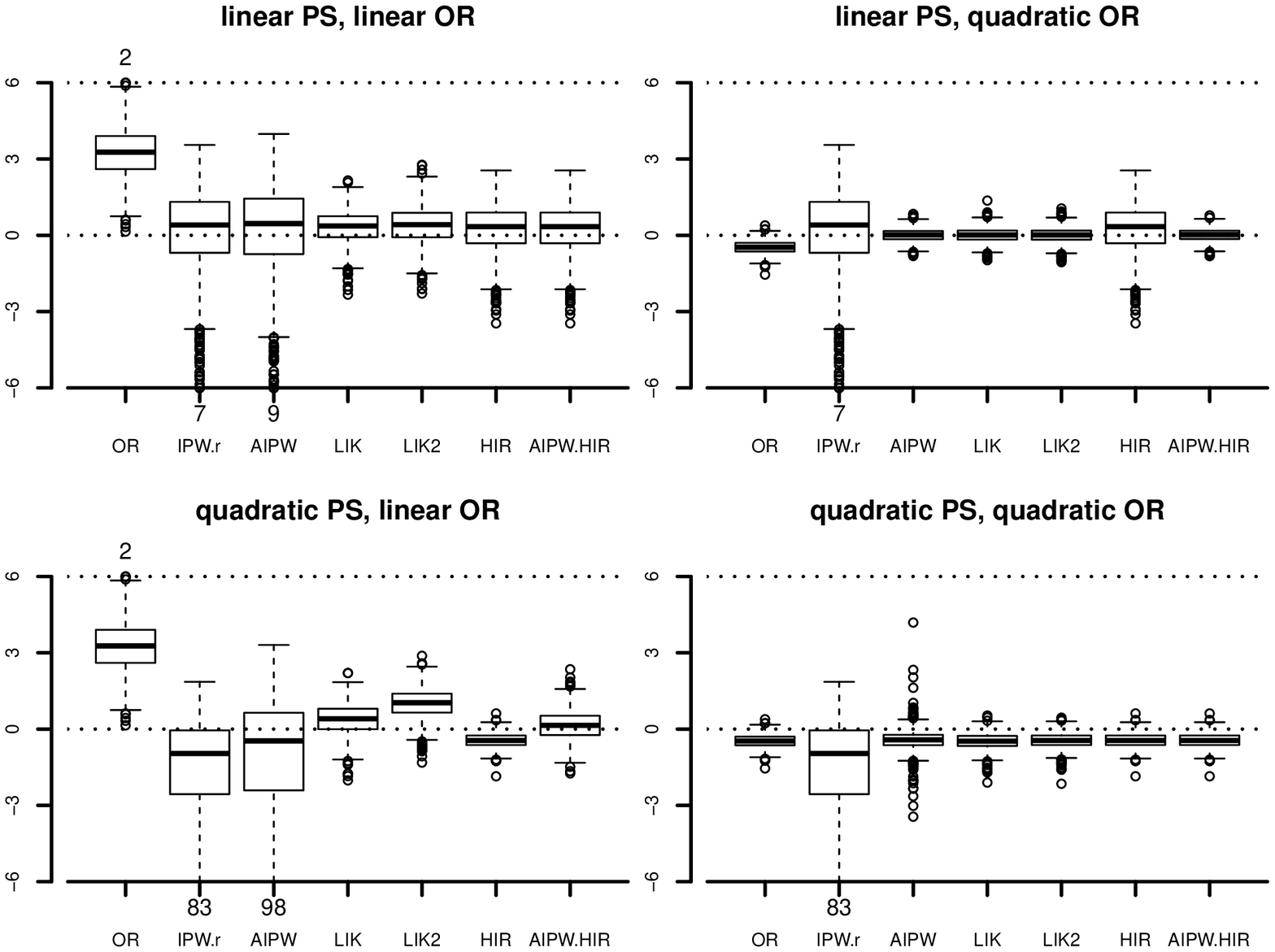}
\end{tabular}
\caption{\small Boxplots of estimates minus the truth under QUA-OR setting with $(\gamma_1^*,\gamma_2^*,\gamma_3^*)=(1.0,0.5,0.5)$.}
\label{plot:QUA-5}
\vspace{-.1in}
\end{figure}

Table~\ref{table:Qinzhang_2} and Figures \ref{plot:LIN-1}-\ref{plot:QUA-1} present the results from $1000$ Monte Carlo samples of size $n=1000$, under the PS setting with small selection bias,
$(\gamma_1^*,\gamma_2^*,\gamma_3^*) = (1.0,0.1,0.1)$.
Table~\ref{table:Qinzhang_3} and Figures \ref{plot:LIN-5}-\ref{plot:QUA-5} present the results from $1000$ Monte Carlo samples of size $n=1000$, under the PS setting with large selection bias,
$(\gamma_1^*,\gamma_2^*,\gamma_3^*) = (1.0,0.5,0.5)$.

The relative performances of the estimators under study are similar to those under the PS setting with large selection bias,
$(\gamma_1^*,\gamma_2^*,\gamma_3^*) = (1.0,0.2,0.2)$. In particular, efficiency gains of the calibrated likelihood estimators
over the doubly robust estimators, AIPW and AIPW.HIR, remain considerable across these settings, when the PS model is
correctly specified but the OR model is misspecified.

\begin{table}[H]
\caption{Qin--Zhang simulation results with $(\gamma_1^*,\gamma_2^*,\gamma_3^*)=(1.0,0.1,0.1)$}
\scriptsize
\label{table:Qinzhang_2}
\begin{center}
\def\arraystretch{0.8}%
\begin{tabular}{lccccccccc}
\hline\hline
\noalign{\medskip}
Models  & OR & IPW.r & AIPW & LIK & LIK2 & HIR & AIPW.HIR & EL & AST    \\
\noalign{\smallskip}\hline \noalign{\smallskip}
& \multicolumn{9}{c}{\bf Data generated under LIN-OR setting} \\
\noalign{\medskip}
linear PS,    & 0.0070   & 0.0076    & 0.0069    & 0.0062   & 0.0066   & 0.0069   & 0.0069  & 0.0038  & -0.0004    \\
linear OR    & (0.0147) & (0.0200)  & (0.0153)  & (0.0157) & (0.0156) & (0.0153) & (0.0153) & (0.0204) & (0.0154)  \\
\noalign{\medskip}
linear PS,    & 0.3551   & 0.0076    & 0.0032    & 0.0072   & 0.0040   & 0.0069   & 0.0032  & 0.0040   & -0.0083    \\
quadratic OR  & (0.0562) & (0.0200)  & (0.0320)  & (0.0163) & (0.0185) & (0.0153) & (0.0312) & (0.0241) & (0.0285)  \\
\noalign{\medskip}
quadratic PS, & 0.0070   & 0.3488    & 0.0062    & 0.0072   & 0.0070   & 0.3687   & 0.0063   & $\cdots$         & $\cdots$          \\
linear OR     & (0.0147) & (0.0553)  & (0.0176)  & (0.0160) & (0.0167) & (0.0557) & (0.0171) & $\cdots$        & $\cdots$         \\
\noalign{\medskip}
quadratic PS, & 0.3551   & 0.3488    & 0.3721    & 0.3428   & 0.3516   & 0.3687   & 0.3687   & $\cdots$         &$\cdots$          \\
quadratic OR  & (0.0562) & (0.0553)  & (0.0576)  & (0.0544) & (0.0541) & (0.0557) & (0.0557) &$\cdots$          &$\cdots$          \\
\noalign{\smallskip}\hline\noalign{\smallskip}
& \multicolumn{9}{c}{\bf Data generated under QUA-OR setting} \\
\noalign{\medskip}
linear PS,    & 0.2235   & 0.0275    & 0.0291    & 0.0233   & 0.0249   & 0.0302   & 0.0302   & 0.0347  & 0.0009      \\
linear OR     & (0.3152) & (0.4034)  & (0.3335)  & (0.0647) & (0.0730) & (0.2999) & (0.2999) & (0.1561)  & (0.3050)  \\
\noalign{\medskip}
linear PS,    & -0.0690  & 0.0275    & 0.0094    & 0.0071   & 0.0084   & 0.0302   & 0.0094  & 0.0029   & -0.0011     \\
quadratic OR  & (0.0190) & (0.4034)  & (0.0173)  & (0.0162) & (0.0170) & (0.2999) & (0.0178) & (0.0226)  & (0.0168)  \\
\noalign{\medskip}
quadratic PS, & 0.2235   & -0.1398   & -0.3619   & 0.0214   & -0.0387  & -0.0731  & -0.3250  & $\cdots$         &$\cdots$          \\
linear OR     & (0.3152) & (0.1555)  & (0.1949)  & (0.0241) & (0.0635) & (0.0191) & (0.0906) & $\cdots$         & $\cdots$         \\
\noalign{\medskip}
quadratic PS, & -0.0690  & -0.1398   & -0.0742   & -0.0672  & -0.0698  & -0.0731  & -0.0731  &$\cdots$          &$\cdots$          \\
quadratic OR  & (0.0190) & (0.1555)  & (0.0193)  & (0.0198) & (0.0195) & (0.0191) & (0.0191) &$\cdots$          &$\cdots$          \\
 \noalign{\medskip}
         \hline
\end{tabular}
\end{center}
\end{table}

\begin{table}[H]
\caption{Qin--Zhang simulation results with $(\gamma_1^*,\gamma_2^*,\gamma_3^*)=(1.0,0.5,0.5)$}
\scriptsize
\label{table:Qinzhang_3}
\begin{center}
\def\arraystretch{0.8}%
\begin{tabular}{lccccccccc}
\hline\hline
\noalign{\medskip}
Models & OR & IPW.r & AIPW & LIK & LIK2 & HIR & AIPW.HIR & EL & AST   \\
\noalign{\smallskip}\hline \noalign{\smallskip}
& \multicolumn{9}{c}{\bf Data generated under LIN-OR setting} \\
\noalign{\medskip}
linear PS,    & 0.0089   & 0.0323    & 0.0107    & 0.0009   & 0.0030   & 0.0109   & 0.0109   & 0.0051   & 0.0024   \\
linear OR    & (0.0280) & (0.2078)  & (0.0608)  & (0.0733) & (0.0698) & (0.0547) & (0.0547) & (0.0900) & (0.0537)  \\
\noalign{\medskip}
linear PS,    & 1.8926   & 0.0323    & 0.0471    & 0.0527   & 0.0665   & 0.0109   & 0.0294    & -0.0089  & 0.0244   \\
quadratic OR  & (0.1748) & (0.2078)  & (0.2414)  & (0.0642) & (0.0663) & (0.0547) & (0.0998)  & (0.1103)  & (0.1015)  \\
\noalign{\medskip}
quadratic PS, & 0.0089   & 1.3964    & 0.0262    & 0.0026   & 0.0059   & 1.8722   & 0.0169   &$\cdots$          &$\cdots$          \\
linear OR     & (0.0280) & (0.9500)  & (0.3731)  & (0.0739) & (0.0770) & (0.1931) & (0.0731) &$\cdots$          &$\cdots$          \\
\noalign{\medskip}
quadratic PS, & 1.8926   & 1.3964    & 1.8918    & 1.8529   & 1.8459   & 1.8722   & 1.8722   &$\cdots$          &$\cdots$          \\
quadratic OR & (0.1748) & (0.9500)  & (0.4227)  & (0.2195) & (0.2220) & (0.1931) & (0.1931)  &$\cdots$          &$\cdots$          \\
\noalign{\smallskip}\hline \noalign{\smallskip}
& \multicolumn{9}{c}{\bf Data generated under QUA-OR setting} \\
\noalign{\medskip}
linear PS,    & 3.2822   & 0.1296    & 0.1560    & 0.3212   & 0.3819   & 0.2428   & 0.2428  & 0.1969    & 0.1943     \\
linear OR     & (0.9469) & (3.0404)  & (3.7185)  & (0.4148) & (0.5712) & (0.9017) & (0.9017) & (0.2647)  & (0.7010)  \\
\noalign{\medskip}
linear PS,    & -0.4663  & 0.1296    & 0.0077    & 0.0091   & 0.0061   & 0.2428   & 0.0156  & 0.0075   & 0.0095      \\
quadratic OR  & (0.0593) & (3.0404)  & (0.0657)  & (0.0796) & (0.0798) & (0.9017) & (0.0603) & (0.1026)   & (0.0549) \\
\noalign{\medskip}
quadratic PS, & 3.2822   & -1.9909   & -1.9277   & 0.3801   & 0.9864   & -0.4403  & 0.1483   &$\cdots$          &$\cdots$          \\
linear OR     & (0.9469) & (13.0100) & (34.4996) & (0.3366) & (0.3682) & (0.0742) & (0.3204) &$\cdots$          &$\cdots$          \\
\noalign{\medskip}
quadratic PS, & -0.4663  & -1.9909   & -0.4319   & -0.4754  & -0.4449  & -0.4403  & -0.4403  &$\cdots$          &$\cdots$          \\
quadratic OR & (0.0593) & (13.0100) & (0.1954)  & (0.0918) & (0.0858) & (0.0742) & (0.0742) &$\cdots$          &$\cdots$        \\
 \noalign{\medskip}
         \hline\hline
\end{tabular}
\end{center}
\end{table}

\subsection{Kang--Schafer simulation}

In addition to the simulation study with the design of \citetappend{Qin2008}, we also conducted a simulation study with the design of \citetappend{KS2007} and a modified design defined in \citetappend{McCaffrey2007}.

In \citetappend{KS2007}, the data are generated as $z=(z_1,z_2,z_3,z_4)^\T$,
  $y=210+27.4z_1+13.7z_2+13.7z_3+13.7z_4+\epsilon$, and $T=1\{U\leqslant\text{expit}(-z_1+0.5z_2-0.25z_3-0.1z_4)\}$,
   where $(z_1,z_2,z_3,z_4,\epsilon,U)$ are mutually independent, $(z_1,z_2,z_3,z_4,\epsilon)$ are marginally normally
   distributed with mean 0  and variance 1 and $U$ is uniformly distributed on $(0,1)$. Let $x=(x_1,x_2,x_3,x_4)^\T$,
   $x_1=\text{exp}(0.5z_1)$, $x_2=z_2/\{1+\text{exp}(z_1)\}+10$, $x_3=(0.04z_1z_3+0.6)^3$, and $x_4=(z_2+z_4+20)^2$.

Two OR models (\ref{OR}) are specified with the identity link $\Psi(\cdot)$ and the regressor vector $g_0(z)=g_1(z)=(1,z_1,z_2,z_3,z_4)^\T$ or $(1,x_1,x_2,x_3,x_4)^\T$, corresponding to a correctly specified or misspecified OR model (denoted by OR z or OR x). Similarly, two PS models (\ref{PS}) are specified with the logistic link $\Pi(\cdot)$ and the regressor vector $f(z)=(1,z_1,z_2,z_3,z_4)^\T$ or $(1,x_1,x_2,x_3,x_4)^\T$, corresponding to a correctly specified or misspecified PS model (denoted by PS z or PS x).

The modified design in \citetappend{McCaffrey2007} is defined the same as above, except that an interaction term is added when generating the response, $y=210+27.4z_1+13.7z_2+13.7z_3+13.7z_4+20z_1z_2+\epsilon$.
Three possible OR models (\ref{OR}) are specified with the identity link $\Psi(\cdot)$ and the regressor vector $g_0(z)=g_1(z)=(1,z_1,z_2,z_3,z_4,z_1z_2)^\T$, $(1,z_1,z_2,z_3,z_4)^\T$ or $(1,x_1,x_2,x_3,x_4)^\T$, corresponding to a correctly specified, slightly misspecified, or misspecified OR model (denoted by OR z2, OR z, or OR x). Two possible PS models (\ref{PS}) are specified the same as above.

\begin{figure}[H]
\begin{tabular}{c}
\includegraphics[width=6in, height=4in]{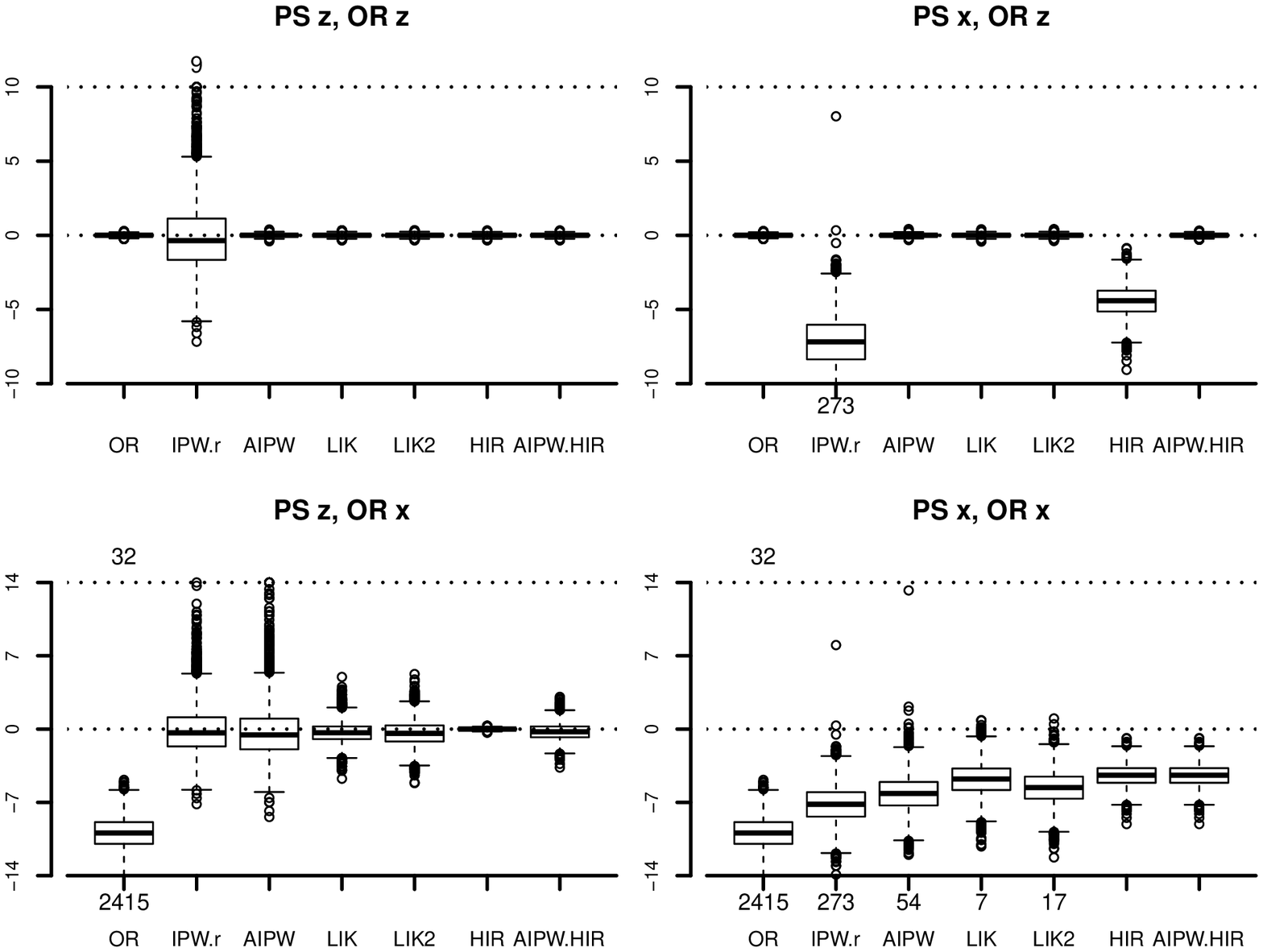}
\end{tabular}
\caption{\small Boxplots of estimates under the Kang--Schafer design.
The values are censored within the range of the $y$-axis, and the
number of values that lie outside the range are indicated next to the lower and upper limits of the $y$-axis.}
\label{plot:KS_no_inter}
\vspace{-.1in}
\end{figure}

\begin{figure}[H]
\begin{tabular}{c}
\includegraphics[width=6in, height=6in]{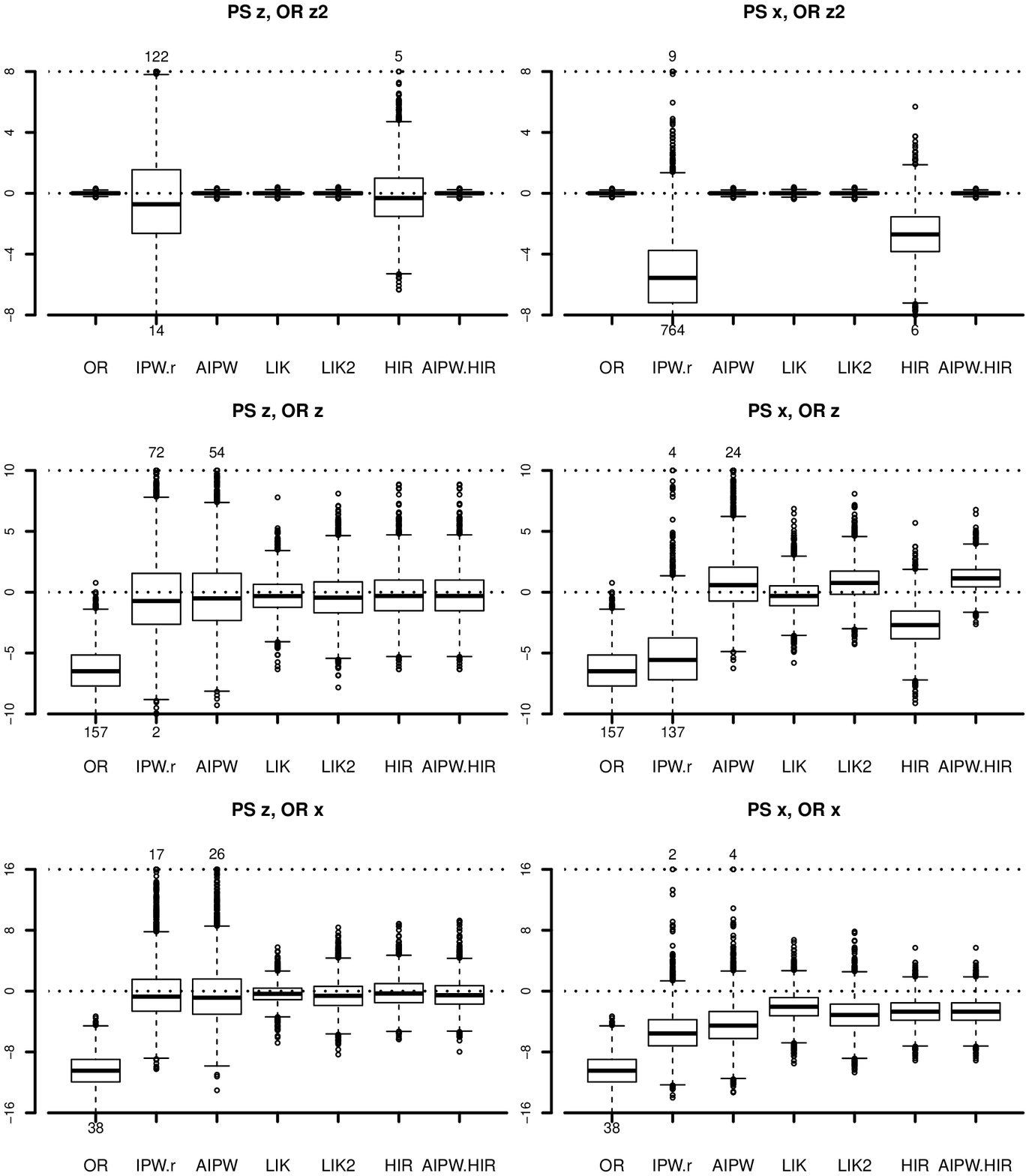}
\end{tabular}
\caption{\small Boxplots of estimates under the McCaffrey-et-al design (with interaction).}
\label{plot:KS_inter}
\vspace{-.1in}
\end{figure}

For these two designs, Table~\ref{KS} and Figure~\ref{plot:KS_no_inter}-\ref{plot:KS_inter} present the results for various estimators from 5000 Monte Carlo sample with size $n=1000$.
The true value of ATT is easily shown to be always 0.

The relative performances of the estimators under study are overall similar to those found in the Qin--Zhang simulation study.
A seemingly unexpected phenomenon, in view of intrinsic efficiency of LIK, is that the HIR and AIPW.HIR estimators have smaller variances than
LIK and LIK2 estimators in the Kang--Schafer design when PS z and OR x models (which are correctly specified and misspecified respectively) are used.
But this difference can be explained as follows. In this case, because the true $m_0(X)$ is a linear combination of $f(X)$ used, the HIR estimator
can be shown to achieve the nonparametric efficiency bound by similar arguments as in the proof of local nonparametric efficiency of $\hat \nu^0_{\mbox{\scriptsize NP}}(\hat\pi, \hat m_0)$.
This can also be seen numerically from Monte Carlo standard errors.
The estimator AIPW.HIR (which is doubly robust) has a moderately inflated from that of HIR (which is non-doubly robust) and hence smaller than those of LIK and LIK2.
This phenomenon depends on the particular way in which the Kang--Schafer design is defined; it does not occur in the McCaffrey-et-al design when PS z and OR z or OR x models are used.

\vspace{.2in}
\begin{table}[H]
\caption{Kang--Schafer and McCaffrey-et-al simulation results}
\scriptsize
\label{KS}
\begin{center}
\def\arraystretch{0.8}%
\begin{tabular}{lccccccc}
\hline\hline
\noalign{\medskip}
Models   & OR        & IPW.ratio & AIPW      & LIK       & LIK2      & HIR       & AIPW.HIR  \\
   \noalign{\medskip}\hline \noalign{\medskip}
& \multicolumn{7}{c}{\bf Kang--Schafer design} \\
    \noalign{\medskip}
PS z, OR z & -0.00021  & -0.15529  & 0.00009   & 0.00038   & 0.00038   & -0.00001  & -0.00001  \\
           & (0.07881) & (2.29565) & (0.08899) & (0.09119) & (0.09014) & (0.08815) & (0.08815) \\
            \noalign{\medskip}
PS x, OR z & -0.00021  & -7.19115  & -0.00027  & 0.00019   & -0.00015  & -4.43418  & -0.00035  \\
           & (0.07881) & (1.76502) & (0.08423) & (0.09345) & (0.09431) & (1.03883) & (0.08538) \\
            \noalign{\medskip}
PS z OR x  & -9.94070  & -0.15529  & -0.28659  & -0.34182  & -0.41053  & -0.00001  & -0.25263  \\
           & (1.53143) & (2.29565) & (2.51075) & (0.98906) & (1.19155) & (0.08815) & (0.78813) \\
            \noalign{\medskip}
PS x, OR x & -9.94070  & -7.19115  & -6.15538  & -4.80166  & -5.60558  & -4.43418  & -4.43418  \\
           & (1.53143) & (1.76502) & (1.70318) & (1.54315) & (1.56532) & (1.03883) & (1.03883)\\
            \noalign{\medskip}\hline \noalign{\medskip}
& \multicolumn{7}{c}{\bf McCaffrey-et-al design (with interaction)} \\
   \noalign{\medskip}
PS z, OR z2 & -0.00001  & -0.25727  & 0.00018   & 0.00031   & 0.00030   & -0.26256  & 1e-6   \\
                       & (0.08057) & (3.74352) & (0.08915) & (0.09289) & (0.09261) & (1.90160) & (0.08844) \\
                       \noalign{\medskip}
PS x, OR z2 & -0.00001  & -5.36684  & -0.00024  & -0.00048  & -0.00090  & -2.68423  & -0.00039  \\
                       & (0.08057) & (2.74036) & (0.08382) & (0.09746) & (0.09833) & (1.69244) & (0.08490) \\
                       \noalign{\medskip}
PS z, OR z           & -6.45619  & -0.25727  & -0.16704  & -0.29360  & -0.41220  & -0.26256  & -0.26256  \\
                       & (1.92221) & (3.74352) & (3.44095) & (1.43785) & (1.94467) & (1.90160) & (1.90160) \\
                       \noalign{\medskip}
PS x, OR z           & -6.45619  & -5.36684  & 0.79425   & -0.25981  & 0.81471   & -2.68423  & 1.16681   \\
                       & (1.92221) & (2.74036) & (2.42957) & (1.31109) & (1.48044) & (1.69244) & (1.06564) \\
                       \noalign{\medskip}
PS z, OR x           & -10.47822 & -0.25727  & -0.38755  & -0.35741  & -0.59751  & -0.26256  & -0.47727  \\
                       & (2.17768) & (3.74352) & (4.11415) & (1.22501) & (1.92895) & (1.90160) & (1.89506) \\
                       \noalign{\medskip}
PS x, OR x           & -10.47822 & -5.36684  & -4.36236  & -2.06547  & -3.12139  & -2.68423  & -2.68423  \\
                       & (2.17768) & (2.74036) & (2.86418) & (1.84705) & (2.21782) & (1.69244) & (1.69244)\\
 \noalign{\medskip}
         \hline\hline
\end{tabular}
\end{center}
Note: In the upper rows are the Monte Carlo means, and in the brackets are the corresponding Monte Carlo variances.
\end{table}

\newpage

\section{Additional results from LaLonde analysis} \label{sec:add-analysis}

\begin{figure}[H]
\begin{tabular}{c}
\includegraphics[width=6in, height=4in]{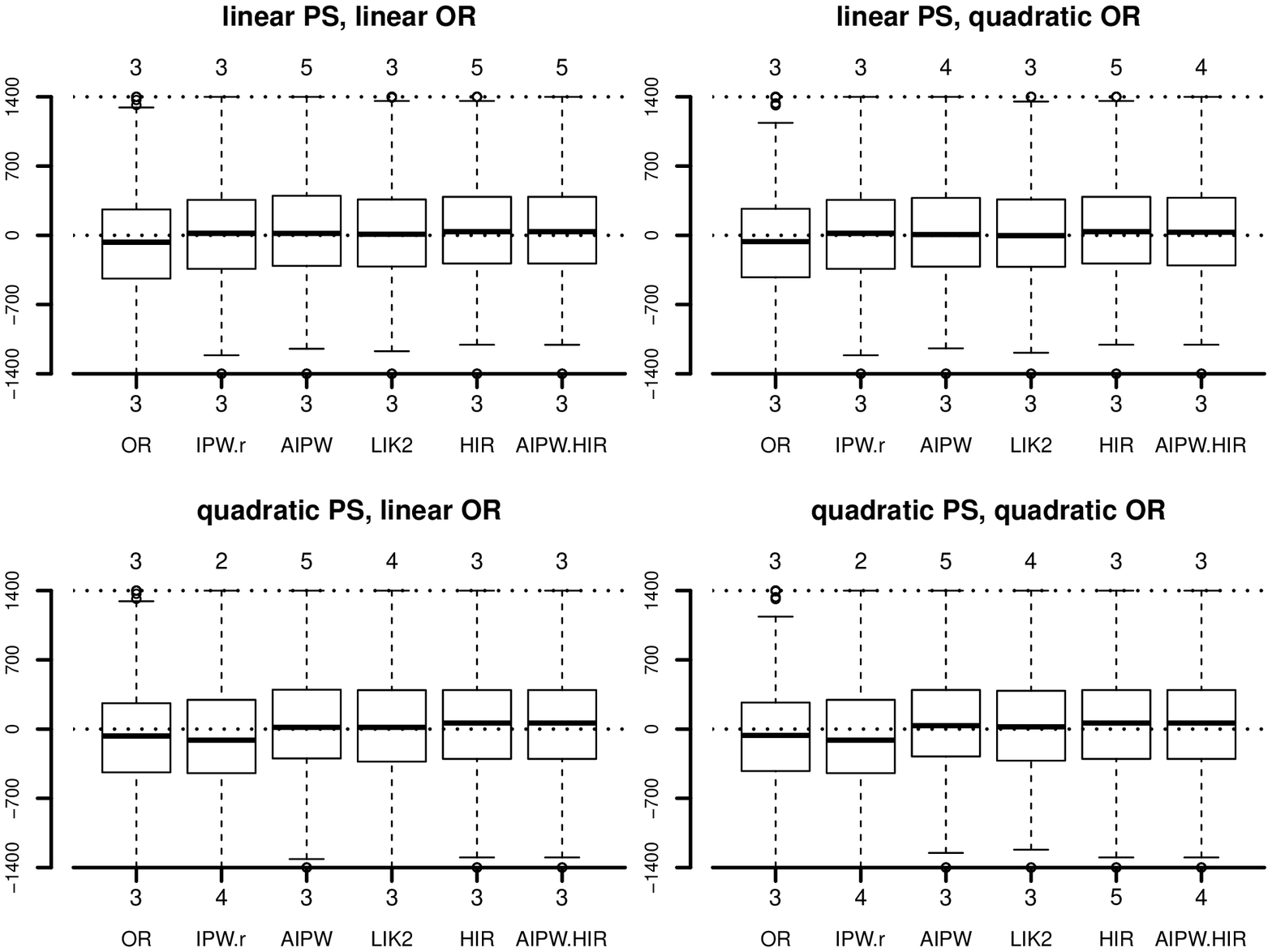}
\end{tabular}
\caption{\small Bootstrap boxplots of differences of bias estimates from Analyses (i) and (ii) on NSW$+$CPS composite data.}
\label{plot:CPS}
\vspace{-.1in}
\end{figure}

Table~\ref{table:CPS} and Figure~\ref{plot:CPS} present the results from Analyses (i) and (ii) for various estimators as listed in Section \ref{simulation},
based on 500 bootstrap samples of the NSW$+$CPS composite data.
There are much smaller differences between the performances of the estimators than when the NSW$+$PSID composite data are analyzed.
Another feature worthy of note is that none of the estimators lead to effect estimates close to the experimental benchmark \$886
or bias estimates close to 0, even though the differences between effect and bias estimates
are all roughly close to \$886.

\begin{table}[H]
\caption{Bootstrap results from Analyses (i) and (ii) on NSW$+$CPS composite data}
\scriptsize
\label{table:CPS}
\begin{center}
\def\arraystretch{0.8}%
\begin{tabular}{lcccccccc}
\hline\hline
\noalign{\medskip}
          &           & OR    & IPW.ratio & AIPW  & LIK2  & HIR   & AIPW.HIR \\
           \noalign{\medskip}\hline\noalign{\medskip}
Linear PS, Linear OR  & Treatment Effect     & -800  & -451      & -308  & -380  & -503  & -388     \\
                    &          & (475) & (518)     & (526) & (518) & (520) & (520)    \\
         \noalign{\medskip}
           & Evaluation Bias       & -1709 & -1336     & -1333 & -1364 & -1414 & -1413    \\
           &            & (374) & (414)     & (428) & (420) & (422) & (422)    \\
          \noalign{\medskip}
           & Difference & 803   & 885       & 903   & 880   & 910   & 910      \\
           &            & (527) & (518)     & (532) & (526) & (531) & (531)   \\
\noalign{\medskip}\hline\noalign{\medskip}
Linear PS, Quadratic OR   & Treatment Effect     & -800  & -451  & -308  & -380  & -503  & -388  \\
                            &            & (475) & (518) & (522) & (516) & (520) & (517) \\
            \noalign{\medskip}
             & Evaluation Bias       & -1611 & -1336 & -1196 & -1254 & -1414 & -1295 \\
             &            & (379) & (414) & (436) & (430) & (422) & (429) \\
             \noalign{\medskip}
             & Difference & 811   & 885   & 888   & 874   & 910   & 907   \\
             &            & (527) & (518) & (529) & (523) & (531) & (529)\\
\noalign{\medskip}\hline \noalign{\medskip}
Quadratic PS, Linear OR & Treatment Effect     & -906  & -427  & -421  & -424  & -465  & -465  \\
                        &            & (475) & (561) & (561) & (561) & (557) & (557) \\
               \noalign{\medskip}
              & Evaluation Bias       & -1709 & -1207 & -1335 & -1297 & -1383 & -1383 \\
              &            & (374) & (529) & (533) & (514) & (507) & (507) \\
               \noalign{\medskip}
              & Difference & 803   & 780   & 914   & 873   & 919   & 919   \\
              &            & (527) & (547) & (544) & (538) & (532) & (532)\\
\noalign{\medskip}\hline \noalign{\medskip}
Quadratic PS, Quadratic OR & Treatment Effect     & -800  & -427  & -465  & -438  & -432  & -465  \\
                 &            & (475) & (561) & (557) & (563) & (557) & (557) \\
              \noalign{\medskip}
              & Evaluation Bias       & -1611 & -1207 & -1383 & -1364 & -1313 & -1383 \\
              &            & (379) & (529) & (507) & (535) & (514) & (507) \\
               \noalign{\medskip}
              & Difference & 811   & 780   & 919   & 926   & 881   & 919   \\
              &            & (527) & (547) & (532) & (543) & (535) & (532)     \\
         \hline
\end{tabular}
\end{center}
Note: In the upper rows are the bootstrap means, and in the brackets are the corresponding bootstrap standard errors.
Treatment Effect is obtained from Analysis (i), and Evaluation Bias from Analysis (ii). The difference is to be
compared with the experimental benchmark \$886 with standard error \$488.
There was no issue of non-convergence when computing estimates during bootstrapping, and hence
Principle Component Analysis is not needed.
\end{table}

\setlength{\bibsep}{4pt}

\bibliographystyleappend{myapa}
\bibliographyappend{thesis}

\end{document}